\documentclass[preprint]{emulateapj}

\usepackage{graphics}
\usepackage{epsfig}
\usepackage{graphicx}
\usepackage{color}
\usepackage{epstopdf}

\newcommand{\etal}{et~al.\ }
\newcommand{\Ha}{H$\alpha$ }

\newcommand{\magsec}{mag/arcsec$^2$}

\pdfoutput=1

\shorttitle{}
\shortauthors{Janowiecki \& Salzer}

\received{8 Apr 2014}
\accepted{6 Aug 2014}

\begin{document}

\title{The Unique Structural Parameters of the Underlying Host
  Galaxies in Blue Compact Dwarfs}

\author{Steven Janowiecki\altaffilmark{1} and 
  John J. Salzer\altaffilmark{1} \\
}
\email{sjanowie@astro.indiana.edu, slaz@astro.indiana.edu}

\altaffiltext{1}{Department of Astronomy, Indiana University, 727 E
  3rd St, Bloomington, IN 47405}



\begin{abstract}


The nature of possible evolutionary pathways between various types 
of dwarf galaxies are still not fully understood. Blue compact dwarf 
galaxies (BCDs) provide a unique window
into dwarf galaxy formation and evolution and are often thought of as
an evolutionary stage between different classes of dwarf galaxies.
In this study we use deep optical and near-infrared observations of
the underlying hosts of BCDs in order to study the structural
differences between different types of dwarf galaxies.
When compared with dwarf irregular galaxies of similar luminosities,
we find that the \emph{underlying hosts} of BCDs have significantly
more concentrated light distributions, with smaller scale lengths and
brighter central surface brightnesses.  We demonstrate here that the 
underlying hosts of BCDs are distinct from the broad continuum of
typical dwarf irregular galaxies, and that it is unlikely that most
dwarf irregular galaxies can transform into a BCD, or vice versa.
Furthermore, we find that the starburst in a BCD only brightens it on 
average by $\sim0.8$ mag (factor of 2), in agreement with other studies.
It appears that BCDs are a long-lived and distinct type of dwarf galaxy
which exhibit an exceptionally concentrated matter distribution. We
suggest that it is this compact mass distribution that enables the
strong star-formation events that characterize this class of dwarf
galaxy, that the compactness of the underlying host can be used as a
distinguishing parameter between BCDs and other dwarf galaxies, and
that it can also be used to identify BCDs which are not currently
experiencing an intense starburst event.

\end{abstract}


\section{Introduction}












With the exception of mergers and interactions between galaxies, star
formation 
is the most transformative process that a galaxy can undergo. Star
formation 
in large gas-rich spiral galaxies is a very complex process that
depends on 
many internal and environmental factors. Alternatively, gas-rich dwarf
irregular galaxies (dIs) provide a 
much simpler laboratory for studying star formation processes in
galaxies. In particular, Blue Compact Dwarf galaxies (BCDs) are
especially unique objects, as they are currently experiencing
some of the most intense bursts of star formation in the local
universe. 

The term ``Blue Compact Dwarf'' is used {in the literature} to
describe a wide variety of galaxies. BCDs were first
identified by 
Sargent \& Searle (1970) as small galaxies with emission lines in
their spectra, giving
the appearance of ``extragalactic HII regions''. It is clear that BCDs
are relatively rare galaxies undergoing a significant starburst
and that they have characteristically low metallicities. However,
conclusions 
about the origin and nature of BCDs depend strongly on sample
selection. Indeed, there are many different 
definitions of BCDs in the literature (e.g., Thuan \& Martin 1981, Gil
de Paz \etal 2003).


Soon after their discovery, the very blue colors of BCDs were
found to be consistent with bursting late type dwarfs (Searle,
Sargent, \& Bagnuolo 1973). The combination of these blue colors with
their very low metallicities led some to propose that BCDs were
young galaxies possibly undergoing their first episode of star
formation (Searle \& Sargent 1972). Indeed, even modern
observations (e.g., I Zw 18, Izotov \& Thuan 2004) have concluded that
some BCDs are intrinsically young (i.e., $<500$ Myr old)
objects. However, whenever more sensitive 
observations are made (e.g., Aloisi \etal 2007), evolved stars are
detected and it is clear that all of the known BCDs have an underlying
old stellar population. Indeed, the discovery of an underlying lower
surface brightness component (Loose \& Thuan 1986) in BCDs was
another 
confirmation that old stars are an important component in BCDs,
meaning that they are not young objects or purely HII regions. While
suggestions have been made that there may
be a small fraction of intrinsically young galaxies (Thuan \etal
1999), 
the most recent searches have revealed older, underlying stellar hosts
in all BCDs (Papaderos \etal 2008). 

Since its discovery, 
the lower surface brightness component in BCDs has
been studied in the search for evolutionary connections between dwarf
galaxies. Lin \& Faber (1983) suggested that dwarf
spheroidal galaxies (dSph) could form when a dI loses its gas from ram
pressure  
stripping near a large galaxy. However, Thuan (1985) found that the
infrared colors of dwarf elliptical galaxies (dEs) and dIs imply
metallicities that are quite 
different; simple gas stripping is not enough to transform
a dI into a dE. Adding to these differences, Papaderos \etal 1996
found that the underlying host 
structure of BCDs was substantially different from dIs and dEs, which
means that the older stellar 
component must undergo serious structural changes if there are
evolutionary connections between types of dwarf galaxies. This
structural distinction in the underlying stellar component of dIs and
BCDs has also been 
noted by Marlowe \etal (1999), Salzer \& Norton (1999), and Doublier
\etal (1999).

\begin{deluxetable*}{lcccccccccc}
\tablewidth{7in}
\tablecaption{The BCD Sample\label{sample}}
\tablehead{
\colhead{Name} & \colhead{$\alpha$} & \colhead{$\delta$} & \colhead{$v$   } & \colhead{$D$  } & \colhead{$M_B$} & \colhead{$M_H$}  & \colhead{$Z$        } & \colhead{$t_B$} & \colhead{$t_{H\alpha}$} & \colhead{$t_H$}   \\
\colhead{    } & \colhead{(J2000) } & \colhead{(J2000) } & \colhead{[km/s]} & \colhead{[Mpc]} & \colhead{[Mag]}  & \colhead{[Mag]}  & \colhead{           } & \colhead{[s]  }   & \colhead{[s]        } & \colhead{[s]  }  \\
(1) & (2) & (3) & (4) & (5) & (6) & (7) & (8) & (9) & (10) & (11) 
}
\startdata
%
%
UM 323  & 01 26 46.6 & $-$00 38 46 & 1913 & 25.6 & -16.10 & -17.90 & 7.96$^1$ & 1800 & 1440 & 3600 \\
UM 408  & 02 11 23.4 & $+$02 20 30 & 3598 & 47.5 & -16.04 & -17.72 & 7.74$^1$ & 1800 & 1440 & 4680 \\
Mk 600  & 02 51 04.6 & $+$04 27 14 & 1008 & 13.6 & -15.55 & -17.30 & 7.94$^1$ & 1800 & 1440 & 3600 \\
Mk 5    & 06 42 15.5 & $+$75 37 33 &  792 & 15.3 & -15.18 & -17.74 & 8.06$^1$ & 3600 & 1440 & 2160 \\
CG 10   & 09 12 51.7 & $+$31 40 51 & 1902 & 30.7 & -15.01 & -17.47 & -        & 1800 & 2160 & 2160 \\
I Zw 18 & 09 34 02.0 & $+$55 14 28 &  751 & 14.6 & -14.68 & -15.19 & 7.18$^1$ & 1800 & 1440 & 3600 \\
Was 5   & 10 10 32.8 & $+$22 00 39 & 1259 & 23.1 & -15.05 & -16.86 & 7.85$^3$ & 1800 & 1440 & 2161 \\
Mk 36   & 11 04 58.5 & $+$29 08 22 &  646 &  8.4 & -14.23 & -16.03 & 7.82$^1$ & 1800 & 1440 & 2880 \\
UM 439  & 11 36 36.8 & $+$00 48 58 & 1099 & 15.9 & -16.11 & -18.23 & 8.08$^1$ & 1800 & 2160 & 2160 \\
Mk 750  & 11 50 02.6 & $+$15 01 23 &  749 &  5.2 & -13.42 & -      & 8.18$^3$ & 1800 & 2160 & -    \\
UM 461  & 11 51 33.3 & $-$02 22 22 & 1039 & 12.7 & -14.40 & -15.87 & 7.81$^4$ & 1800 & 1440 & 2160 \\
UM 462  & 11 52 37.3 & $-$02 28 10 & 1057 & 13.5 & -16.18 & -18.07 & 7.80$^4$ & 1800 & 1440 & 2160 \\
Mk 67   & 13 41 56.5 & $+$30 31 10 &  932 & 18.7 & -15.16 & -17.42 & 8.08$^1$ & 1800 & 1440 & 2880 \\
Mk 475  & 14 39 05.4 & $+$36 48 22 &  583 & 11.9 & -14.28 & -16.25 & 7.93$^1$ & 1800 & 1440 & 2160 \\
Mk 900  & 21 29 59.6 & $+$02 24 51 & 1152 & 18.9 & -17.16 & -19.85 & 8.07$^1$ & 1800 & 2160 & 2160 \\
Mk 324  & 23 26 32.8 & $+$18 15 59 & 1600 & 23.2 & -16.70 & -19.08 & 8.18$^2$ & 1800 & 1440 & 2160 \\
Mk 328  & 23 37 39.5 & $+$30 07 46 & 1379 & 20.6 & -16.57 & -19.57 & 8.64$^2$ & 1800 & 1440 & 4320 \\
\enddata
\tablecomments{
  Column 5: distances come from flow models;
  Column 6: B absolute magnitude;
  Column 7: H absolute magnitude;
  Column 8: gas phase abundance Z = 12 + log (O/H), $^1$: $T_e$
  abundances from Zhao \etal 2010; $^2$: $N2$ abundance from Zhao \etal
  2010; $^3$: Brinchmann \etal 2008; $^4$: Izotov \etal 2007.
  Columns 9,10,11: exposure times for each filter
}
\end{deluxetable*}

More recently, these types of structural decompositions have
been carried out on more types of galaxies, in order to find
evolutionary similarities at other mass ranges. In particular,
Luminous Blue Compact 
Galaxies (LBCGs, Werk \etal 2004, Garland \etal 2004, Salzer \etal
2009) have been fitted with a variety of profile shapes
(Micheve \etal 
2013, Amor\'{i}n \etal 2009), as have the so-called Green Pea Galaxies
(Cardamone \etal 2009). In general, the LBCGs and BCDs are both
dominated in their appearance by recent and intense star formation,
and the LBCGs may be higher-mass analogues to BCDs. If the
transformative effects of star formation can have a substantial impact
in the more massive LBCGs, the effects of recent star formation in
BCDs will be even more significant.

To understand the evolutionary status of galaxies currently hosting
BCD-like starbursts, we have acquired
observations of a sample of BCDs with high resolution and in many
wavelengths. In this paper we compare the isophotal structures of BCDs
with other types of dwarf galaxies to shed light on possible
evolutionary connections. In a subsequent paper we fit the Spectral
Energy Distributions (SEDs) of these BCDs with models to derive star
formation histories and further constrain evolutionary pathways.

{Many of the BCDs in our sample have been observed with space-
  and ground-based telescopes. Indeed, the literature is rich with
  high quality surface brightness profiles, some with even better
  resolution and depth than the observations presented
  here. However, most studies of the structural parameters of BCDs are
  carried out in great detail but on only a handful of objects. The
  structural parameters obtained from fitting surface brightness
  profiles can be affected by the methods used or even the
  observations themselves. In order to avoid systematic uncertainties
  inherent in a heterogeneous sample, we will use only our own
  observations to determine structural parameters for our sample of
  BCDs.
}

This paper is organized as follows: In Section \ref{sampleobs} we
present our BCD sample and the observations we have obtained. Our
surface photometry methods and fits are described in Section
\ref{sbfits} and the results of those fits are compared with other
dwarf galaxies in Section \ref{results}. Finally, we discuss the
implications of these results and summarize our conclusions in Section
\ref{discussion}.


\section{Sample and Observations}\label{sampleobs}

Our sample of 21 BCDs represents a variety of BCDs and BCD-like
galaxies, some of which are archetypal BCDs (e.g., I Zw 18) and some of
which may be similar to dEs but possess a strong central starburst 
(e.g., Mk 900). The sample includes a wide range of BCD morphologies, 
including BCDs 
with offset starbursts (e.g., Mk 36, Mk 750), BCDs with two primary
starburst regions (UM 461, Mk 600), BCDs with cometary shapes in their
outer isophotes (e.g., Mk 5), and BCDs with a large number of
active star formation sites (e.g., UM 439, UM 462, UM 323).

Our BCDs are all less luminous than $M_B = -18$ (typically around
$M_B \sim -16$), have blue colors with $B$-$V \sim 0$-$0.5$, and are
less than $50$ Mpc distant (their median distance is $16$ Mpc). Gas
phase abundances ($Z = 12 + \log (O/H)$) show that our sample is
characteristically metal poor, with a range between $Z=7.18$ (I Zw 18)
and $Z=8.64$ (Mk 328), although the vast majority have metallicity
between $7.7$ and $8.2$; the median value is $Z = 7.96$.

Table \ref{sample} gives the details of our BCD sample. Coordinates on
the sky, observed recession velocities, and flow-model distances all
come from 
NED\footnotemark \, (NASA Extragalactic Database). Our absolute
magnitudes come from aperture photometry of the galaxies, using
photometric zeropoints from stars in the field measured by SDSS and
2MASS (see Section \ref{BSDSS} and Section \ref{H2MASS}), and are
corrected for 
Galactic extinction. Also shown are the total integration times from
our broadband optical, near-infrared, and narrow band
observations. We varied the integration times on each target depending
on their expected brightness and sky conditions. Mk 750 was not
observed in near-infrared.

\footnotetext{This research has made use of the NASA/IPAC Extragalactic
  Database (NED) which is operated by the Jet Propulsion Laboratory,
  California Institute of Technology, under contract with the National
  Aeronautics and Space Administration.}

\subsection{Observations}
\label{observations}

In order to measure the structural parameters of the underlying host
galaxies of our BCDs, we need very deep images that are sensitive to
their faint outer reaches. Many of the BCDs in our sample
have been studied before, but previous surface photometry lacked the
necessary depth and sensitivity to measure the structure of the host
galaxies. We find that our understanding of the evolutionary context
of BCDs requires detailed knowledge about the structure of their faint
underlying host galaxies.

If BCDs are typical dwarf galaxies undergoing a recent and localized
burst of star formation, their underlying old stellar population
should be similar to normal dwarf galaxies. It is easier to separate
this old population from the recently-formed stars by considering only
the underlying host galaxies of the BCDs. We restrict our surface
brightness 
profile fits to this underlying host light in the outskirts of the
BCDs. While the high surface brightness regions of the BCDs
typically have clumpy and irregular morphologies, the outskirts have
more regular shapes and smoother light distributions. It is this
underlying host galaxy that we fit in order to probe the BCD as it
was before the current burst of star formation.

Given the very blue optical colors of our BCDs, we chose to observe
them in the Johnson B filter in order to achieve the greatest
sensitivity to the faint 
stellar populations of the underlying host galaxy. While this is the
optimal filter choice for a metal-poor population, some of the blue
light in the BCDs also comes from their substantial ongoing star
formation. This recent star formation can be identified by bright \Ha
emission from the ionized gas around young massive stars. We observe
each BCD in a narrow-band \Ha filter to locate regions with
active star formation. We use these \Ha images to mask out regions of
the B image involved in current star formation and keep only light
from the underlying host galaxy.

A more direct way to study the underlying host 
galaxy light is to use a much redder filter which is mainly sensitive
to the light from evolved stars and which is much less affected by
recently formed stars. Toward this end, we use the near-infrared (NIR)
H filter 
at $\lambda \sim 1.7\mu m$ to directly observe the old stellar
populations of the host galaxies. Very young ($\sim 5$ Myr) stellar
populations can have B-H $\sim 0$, but old ($\sim 10$ Gyr) stellar
populations have B-H $\sim 4$ {(Bressan \etal 2012, Girardi
  \etal 2010, c.f. Carter \etal 2009)}. 
Although the BCDs are much fainter and harder to observe in H than in
B, the NIR light is more representative of the old underlying
component.

We observed our sample of BCDs using the Wisconsin Yale Indiana NOAO
(WIYN\footnotemark \footnotetext{The WIYN Observatory is a joint
  facility of the University of Wisconsin-Madison, Indiana University,
  Yale University, and the National Optical Astronomy Observatory.})
3.5m telescope at Kitt Peak National Observatory\footnotemark
\footnotetext{Kitt Peak National Observatory, National Optical
  Astronomy Observatory, which is operated by the Association of
  Universities for Research in Astronomy (AURA) under cooperative
  agreement with the National Science Foundation.}, between November
2008 and April 2010. The Minimosaic and OPTIC (Orthogonal Parallel
Transfer Imaging Camera) imagers were used 
to obtain deep images in broadband B and narrow-band \Ha filters, and
WHIRC (WIYN High Resolution Infrared Camera, Meixner \etal 2010) was
used to obtain deep near-infrared images in the H filter. The
{good} native seeing of the WIYN site provides images with an
average of $1''$ seeing. 
The following subsections describe the individual observations and
processing.

\subsection{Deep B Imaging with WIYN}
\label{BSDSS}

The B images of the BCDs were observed with the Minimosaic and OPTIC
imagers. Minimosaic and OPTIC both consist of two side-by-side 2048 x
4096 pixel CCDs, where the pixels are 0.141$''$ on a side, and the
field-of-view is 10$'$. Targets were placed on the cosmetically
cleanest area of both cameras while avoiding the small chip
gaps. Two 900s (or three 600s) exposures were taken in B to aid in
cosmic ray rejection. One galaxy (Mk 5) was observed twice as long to
compensate for poor sky transparency during the observations.

These images were reduced with the standard
IRAF\footnotemark \, \textsc{ccdproc} and \textsc{mscred} packages,
including an overscan 
subtraction, bias image correction, and sensitivity correction from
dome flat fields. Laplacian cosmic ray rejection was applied to
individual images with LACOS (van Dokkum 2001), a world coordinate
system (WCS) was applied using the USNO-B1 catalog, and
\textsc{mscimage} was used to create single-extension images from the
multi-extension images. Finally, the individual images of each BCD
were combined to create a single, deep image by averaging and
using standard bad pixel masks.

\footnotetext{IRAF is distributed by the National Optical Astronomy
  Observatory, which is operated by the Association of Universities
  for Research in Astronomy (AURA) under cooperative agreement with
  the National Science Foundation.}

Many of the B observations were not taken under photometric
conditions, so we 
use a boot-strap calibration from the Sloan Digital Sky Survey (SDSS)
Data Release 9 (Ahn \etal 2012). All but one of the BCDs (Mk 5) are
within the survey footprint. Using broadband observations from another
observing run with Minimosaic during photometric conditions where
standard stars 
were observed (Haurberg \etal in preparation), we derived our own
empirical conversions between SDSS ugriz photometry and Minimosaic BVR 
photometry, and found the following relationship for B as a function
of g and r:

\vspace{0.5cm}

$B = g + (0.266 \pm 0.036) (g-r) + (0.258 \pm 0.021)$.

\vspace{0.5cm}

\noindent
We measure the brightnesses of foreground stars on the same chip and
amplifier as the 
BCDs in our $B$ images and find a zero-point for each image based on
the difference between the SDSS catalog photometry and our measured
photometry. We selected unsaturated stars with $r<22$ mag, and
with photometric measurements better than $\sim<0.05$ mag. Typical
uncertainties (standard deviations of multiple stars in each image) on
these computed zero-points are $0.03 - 0.04$ mag. We use these
boot-strap calibrations for all $B$ photometry throughout this work for
consistency. The final calibrated $B$ images are shown in the
left-most column of Figure \ref{images}.


We also find a small systematic offset between this aperture
photometry and earlier $B$ photometry (Salzer, private communication),
such that the new observations are $\sim 0.16$ magnitudes brighter,
with a standard deviation of only $\sim0.10$ magnitudes. This
systematic difference could be related to the increased depth of our
new observations which may have detected previously unseen low surface
brightness light in the outskirts of the BCDs.

\subsection{Deep \Ha Imaging with WIYN}
\label{H2MASS}

The narrow-band \Ha images of the BCDs were obtained with Minimosaic,
placing the targets on the cosmetically optimal portion of
the detector, as 
with the B images. We used two narrow-band \Ha filters appropriate to
the redshift of the targets ($90\%$ widths: $6554-6600$\AA \, and
$6596-6646$\AA). As with the B
images, in order to effectively reject cosmic rays, two 720s
narrow-band exposures were taken of each target. A short (180s)
image in the R filter (encompassing the narrow-band filters in
wavelength range) was taken between the pairs of narrow-band exposures
to provide an image for subtraction of the stellar continuum. The same
image reduction methods were applied as with the B images.

We remove the continuum from our \Ha images by re-scaling the short
$R$ filter exposure, following a well-documented procedure (Van
Sistine \etal in preparation). We first align the three images to match each
other and apply a WCS using 2MASS stars in the field. Next we measure
the full width at half-maximum 
(FWHM) of the profile of many stars in the three images to identify
any mismatches in image quality through the sequence. In order to have
effective continuum-subtraction, the images must be at approximately
the same resolution. If the average FWHM differs between images by
more than 0.2$''$, the image with the smaller FWHM is 
smoothed with a Gaussian kernel to match the larger FWHM. Finally,
stellar photometry is carried out in all three images, and the images
are multiplicatively scaled so that stars, which we assume have no
significant \Ha emission on average, have the same flux in all
images. To produce the final continuum-subtracted image, the scaled R
image is subtracted from the average of the two \Ha images.

Due to non-photometric conditions, standard stars were not
observed during our \Ha 
observations. However, we again use the calibrated photometric
narrow-band Minimosaic observations from Haurberg \etal (in prep) to
calculate a typical offset between photometry of stars in the R and
\Ha images. We find that the two filters have a consistent zero-point
offset of $3.39 \pm 0.03$ magnitudes. Finally, we compare the
photometry of well-measured stars from the images to derive a
relationship between calibrated SDSS magnitudes and Minimosaic R and
find the following relationship:

\vspace{.5cm}

$R = r + (-0.142 \pm 0.038) (g-r) + (-0.108 \pm 0.023).$

\vspace{.5cm}

\noindent
We adopt the assumption that the narrow-band images here have
no significant color term. Thus, we use this relation to derive
zero-points for our R images, and the standard offset between R and
\Ha zeropoints to calibrate our \Ha images. These \Ha images are
reprojected using IRAF's \textsc{mscimage} task to be matched
precisely on the same pixel scale as the B images.  The final
calibrated \Ha images are shown in the right-most column of Figure
\ref{images}.

\subsection{Deep H Imaging}

The H images of the BCDs were observed between Nov 2008 and Sep 2009
with the WHIRC imager. WHIRC has a $2048$x$2048$ HgCdTe array with
$0.1''$ pixels and a $3.3'$x$3.3'$ field-of-view. Total exposure times
were between 2200s and 4600s per object.

The most important step in reducing ground-based NIR observations is
the sky subtraction. Atmospheric OH emission contributes heavily
to the bright ($\sim 14$ \magsec) near-infrared sky. With this high
sky background in mind, we used dithered sequences of 3-minute
exposures on each target. Since the BCDs are much smaller than WHIRC's
field-of-view, most of the image is sky and a good sky subtraction can
be obtained without separate dedicated sky observations. We dithered
in a 2x2 box 50$''$ on a side, and offset subsequent dither boxes by
10$''$. Most targets had 12 three-minute exposures (3 separate 2x2
dithers), but we spent more time (up to 26 three-minute exposures) on
the fainter targets.

These images were reduced with standard IRAF routines, but required
some special attention and efforts, described in detail in Appendix
A. In brief, the WHIRC task \textsc{wprep} is run to trim and
linearize all of the raw images. Dome flat fields were created by
subtracting an average of 10 lamps-off flat fields from an average of
10 lamps-on flat fields, to remove the very high background {
  (Alam \& Predina 1999) }. These
dome flats were also used to create a bad pixel mask. Later on in the
reduction process we created sky flat fields that flattened the
observations significantly better than the dome flats. Finally, we
took darks at every exposure time we planned to use, since the dark
level on WHIRC is not negligible.

To achieve a successful sky subtraction we used an iterative
subtraction method with object masking. {On each iteration, we
  use the combined images to mask out objects from the individual
  dithered frames, generate an improved sky subtraction, and then
  create better combined images. We also found it necessary to
  characterize and remove a dark ``palm-print'' feature from some
  images, which resulted from an issue with the amplifier
  electronics. The complete details of the sky subtraction and image
  processing are detailed in Appendix A.
}

After the final images are created we measured the brightnesses of all
of the 2MASS stars in our frames to determine a photometric zero-point
for each image. The final calibrated
images are re-projected using \textsc{mscimage} to match the pixels in
the B images (0.14'') for consistent surface photometry. The final
calibrated $H$ images are shown in the left-center column of Figure
\ref{images}, using 2x2 pixel binning (0.28'' pixels) to increase the
signal-to-noise ratio in each pixel.

\subsection{SDSS image processing}
\label{sdss}

We also download calibrated $ugriz$ images from SDSS DR9 (Ahn \etal
2012) for the BCDs in our sample (excluding Mk 5, which has not been
observed by SDSS). These images have average FWHM of 1.3$''$ but are
not as deep as our observations. Using IRAF's \textsc{mscimage} task
and the compatible WCS on the SDSS images and our own, we
re-project each SDSS filter image to match the pixels in our $B$
image. While less deep than our observations, we use these
$ugriz$ images to construct surface brightness profiles in additional
colors across the high surface brightness regions of the BCDs. This
variety of colors is useful to compare with our deep $B-H$ color
profiles. In particular, we convert the $g-r$ surface color
profile into a $B-V$ profile.

\subsection{WISE catalog photometry}
\label{wise}

In order to more directly probe the intrinsic stellar mass of the
BCDs, we use infrared observations from the Wide-field Infrared Survey
Explorer (WISE, Wright \etal 2010) catalog. WISE surveyed the sky in 4
photometric bands between 3.4$\mu$m ($w1$ band) and 22 $\mu$m ($w4$
band). The recent AllWISE Data Release
({http://wise2.ipac.caltech.edu/docs/release/allwise/})
contains the most reliable photometric catalog of sources released to
date. In particular, we have obtained catalog aperture photometry for
all of our BCDs by finding the nearest WISE source within 5$''$ of our
BCDs. These  infrared luminosities are easily converted into stellar
masses via the reliable mass-to-light ratio of McGaugh \& Schombert
(2013), and will be even less affected by the recent star formation
than the near-infrared observations.

%

\subsection{Multi-wavelength images}

Our observations in $B$, $H$, and \Ha for each BCD are shown in
Figure \ref{images}. { Most images are shown using a
  logarithmic intensity scaling to adequately show the high dynamic
  range between both the high and low surface brightness portions of
  each galaxy.}  The $B$ and $H$ 
images are shown in units of calibrated surface brightness (magnitudes
per square arcsecond, \magsec) and 
visibly demonstrate the range of morphologies of BCDs in our
sample. For most of our sample, we also produce $B-H$ color maps,
which are shown in the third column of Figure \ref{images}. For Mk 750 
we do not have an $H$ image and thus cannot create a $B-H$ color
map. We also do not show a $B-H$ color map for UM 461 because our $H$
image of UM 461 is not sensitive enough.

While the $B$ and $H$ images look generally similar to each other, the
$B-H$ color maps show that there are significant color differences
within  
the BCDs. In particular, many BCDs (c.f. UM 439) have regions of
significantly bluer ($B-H$) color which also correspond to regions of
stronger \Ha emission, likely indicating the locations of HII regions
with hot young stars ionizing the low-metallicity gas that 
surrounds them. Across most of the sample there is a ($B-H$) color
trend for the galaxies to have somewhat bluer centers and somewhat
redder outskirts, which likely indicates the different stellar
populations present across the galaxies. 

{
While many of our BCDs have already been observed by other researchers
and have published surface brightness profiles, we restrict our analysis
to those profiles which we have observed and processed with our simple
and consistent method. One of the largest sample of modern high
quality observations is Micheva \etal 2013, which includes
multi-wavelength ($UBVRIHK$) surface photometry of 24 Blue Compact
Galaxies, three of which are common to our study (UM 439, UM 461, \&
UM 462). Gil de Paz \& Madore (2005) present $BRH$ surface
photometry of 114 BCDs, with many in common to our study. Hunter \&
Elmegreen (2006) study the profiles of $\sim 100$ dIs and BCDs with
deep $UBV$ and some $JHK$ photometry. Finally, Mk 36, UM 408, and UM
461 have been observed in $JHK$ with higher resolution and greater
depth (Lagos \etal 2011) than in our work. 
}

\begin{figure*}
\figurenum{1}
\centering
\label{images}
\epsscale{0.9}
\plotone{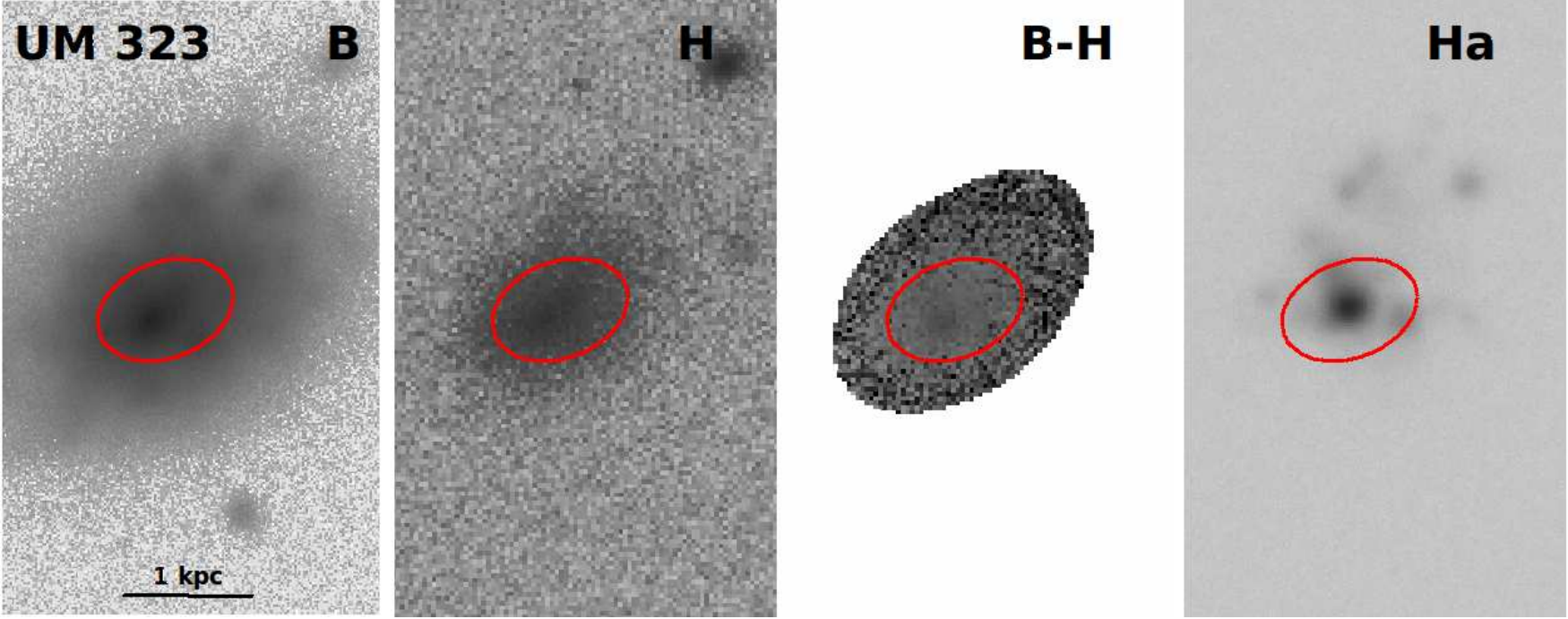} 
\plotone{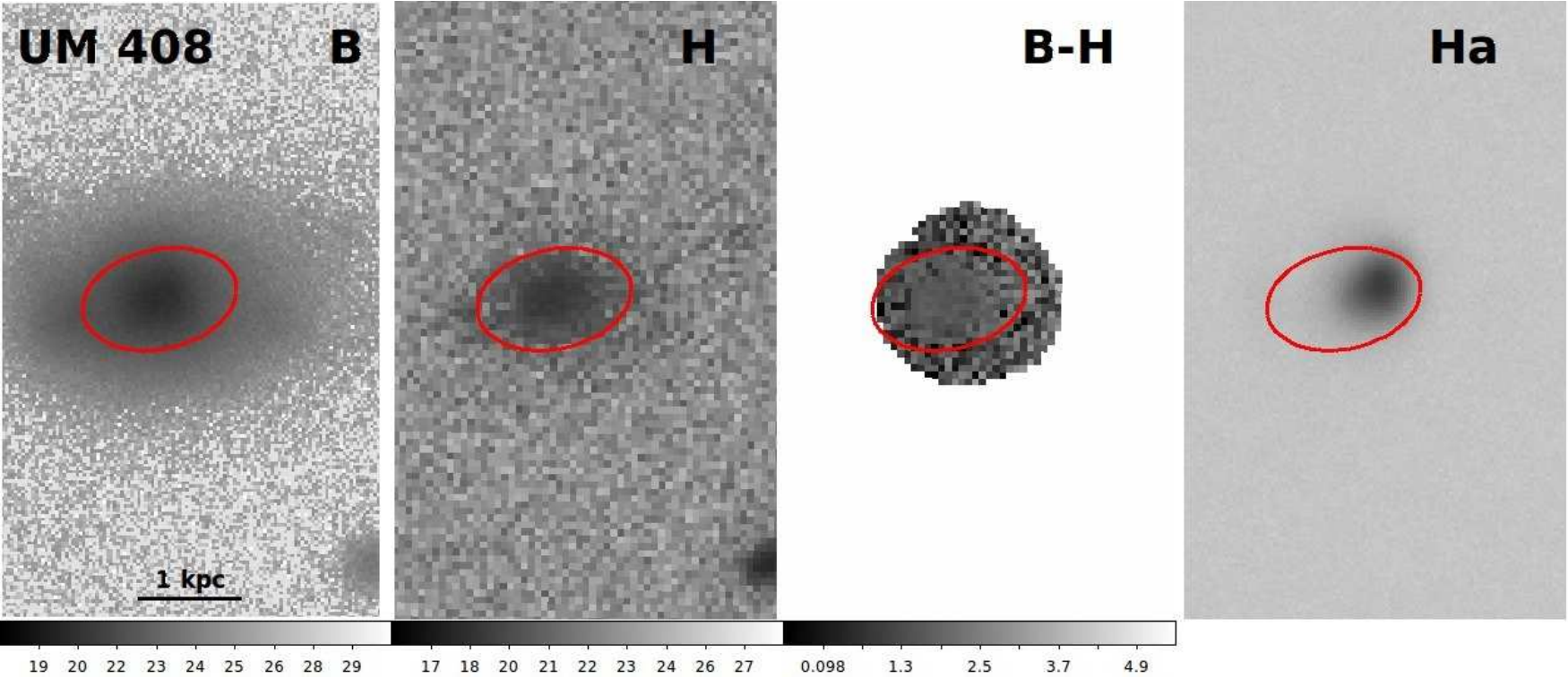} 
\caption{Final images of the BCDs. Shown here are the primary
  observations of each BCD in our sample, all matched to the $B$ image.
  The left-most panel shows an inverted image in the B filter with
    0.14'' pixels, logarithmic intensity scaling, and a scale bar 1
    kiloparsec in length. The scale at the bottom shows the calibrated
    surface brightness in \magsec.
  The left-center panel shows an inverted image in the H filter binned
    2x2 (resulting in 0.28'' pixels) with logarithmic intensity
    scaling. Again the scale at the bottom shows the calibrated
    surface brightness in \magsec.
  The right-center panel shows a B-H color map with 2x2 binning
    (0.28'' pixels) where darker grey shading corresponds to bluer
    colors, as shown in the
    scale bar at the bottom. The color map is truncated beyond where
    the signal-to-noise of the B and H images is too low.
  The right-most panel shows an inverted continuum-subtracted \Ha
    image with a logarithmic intensity scaling, and 0.14'' pixels.
  On all images, an ellipse marks the isophote inside of which more
    than half of the pixels in each isophote are masked. Inside of
    that ellipse, our isophotal fits have their geometric parameters
    fixed.
}
\end{figure*}

\begin{figure*}
\figurenum{1}
\centering
\epsscale{0.99}
\plotone{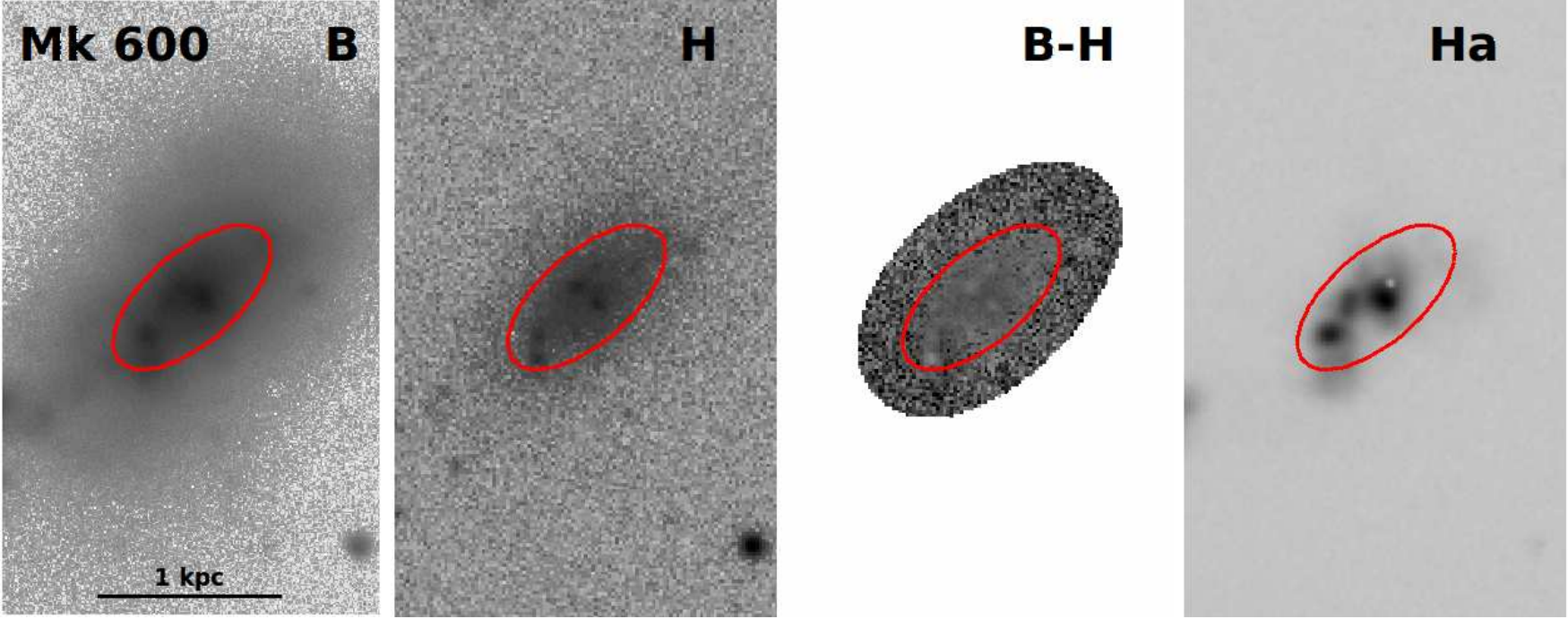} 
\plotone{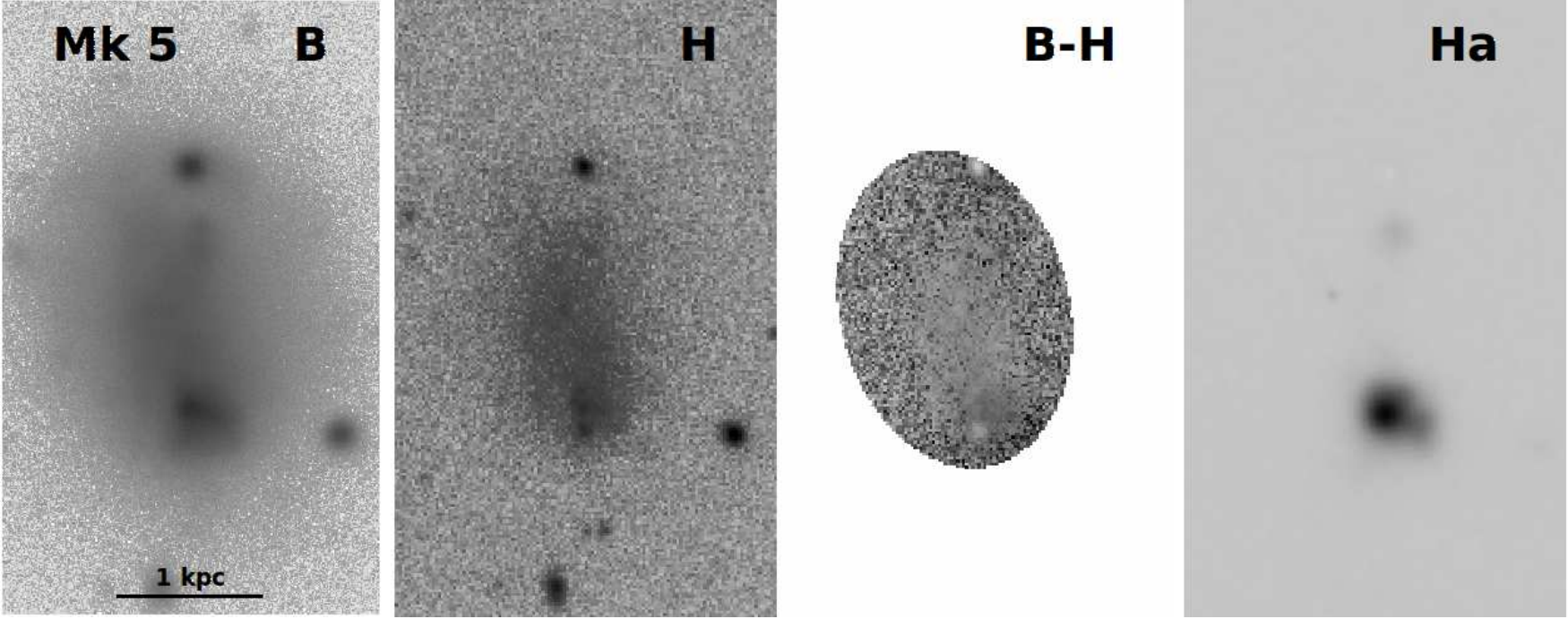} 
\plotone{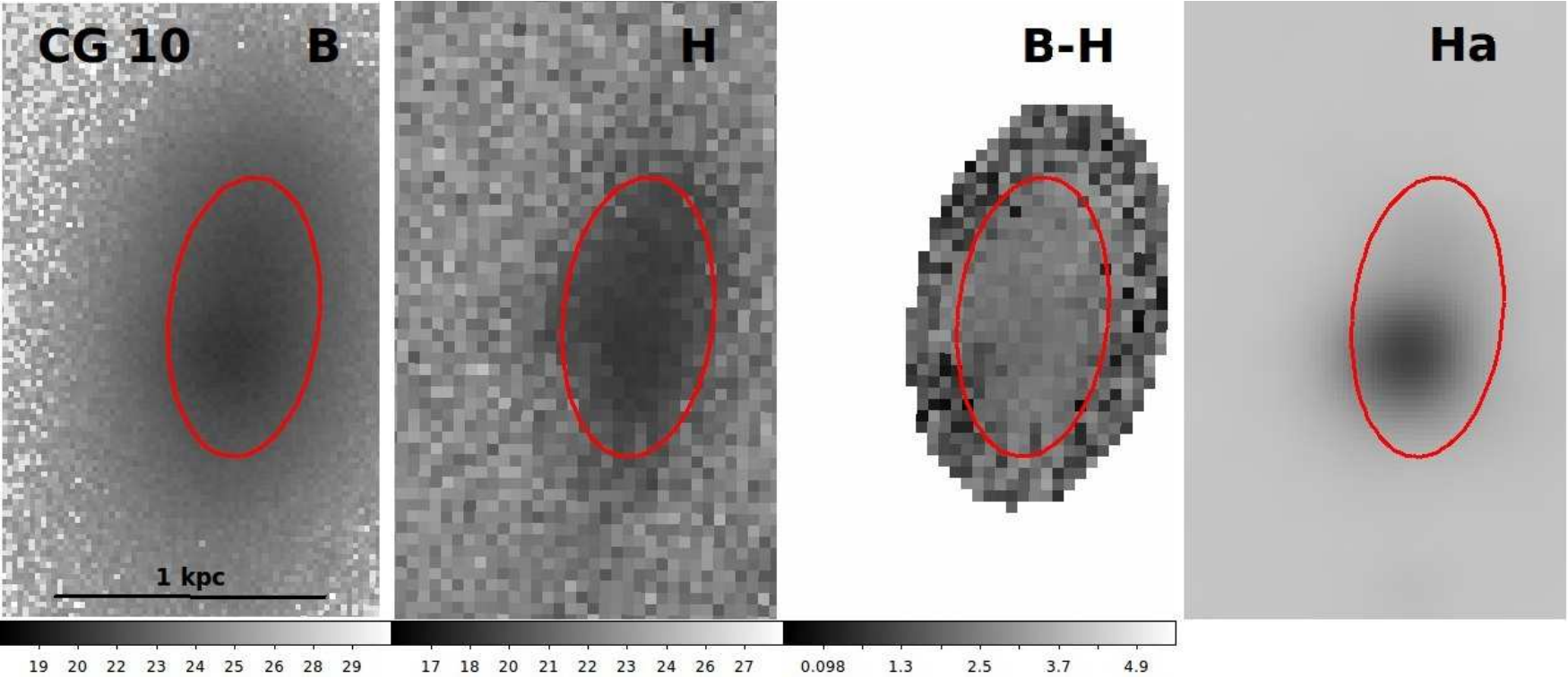} 
\caption{continued}
\end{figure*}

\begin{figure*}
\figurenum{1}
\centering
\epsscale{0.99}
\plotone{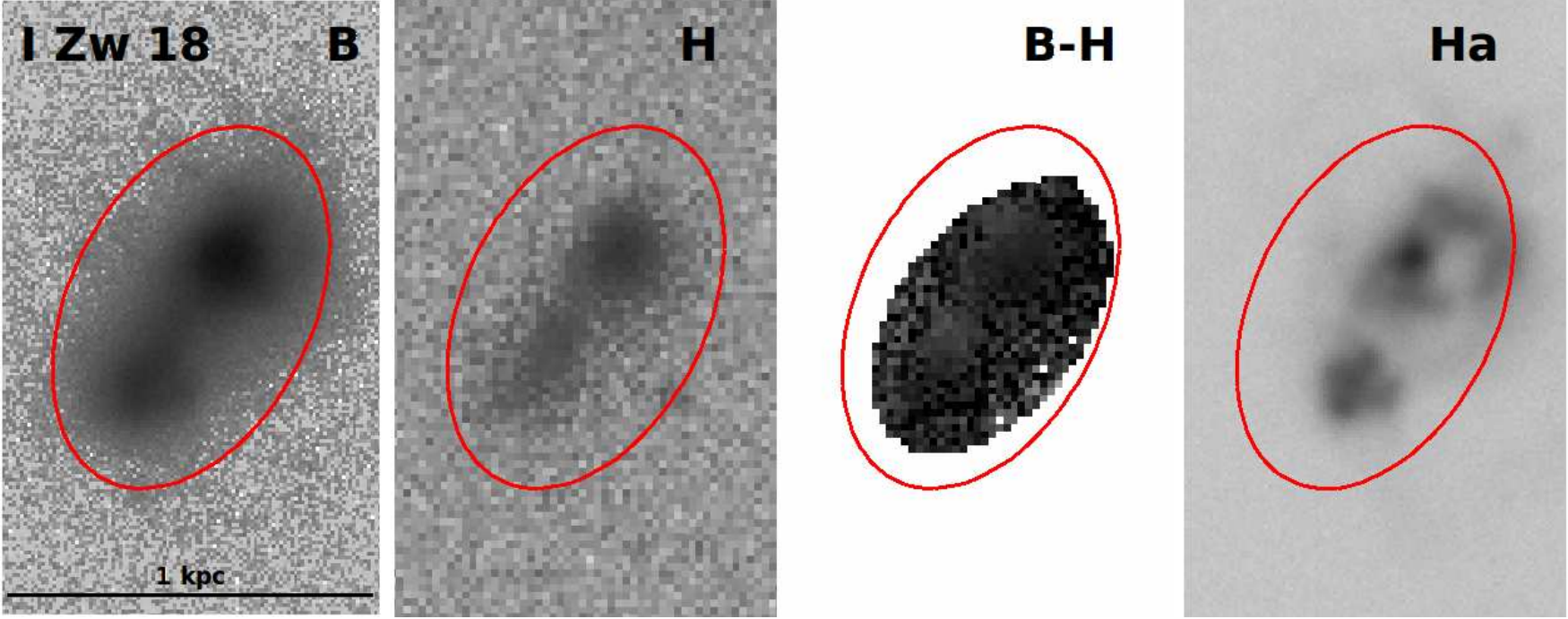} 
\plotone{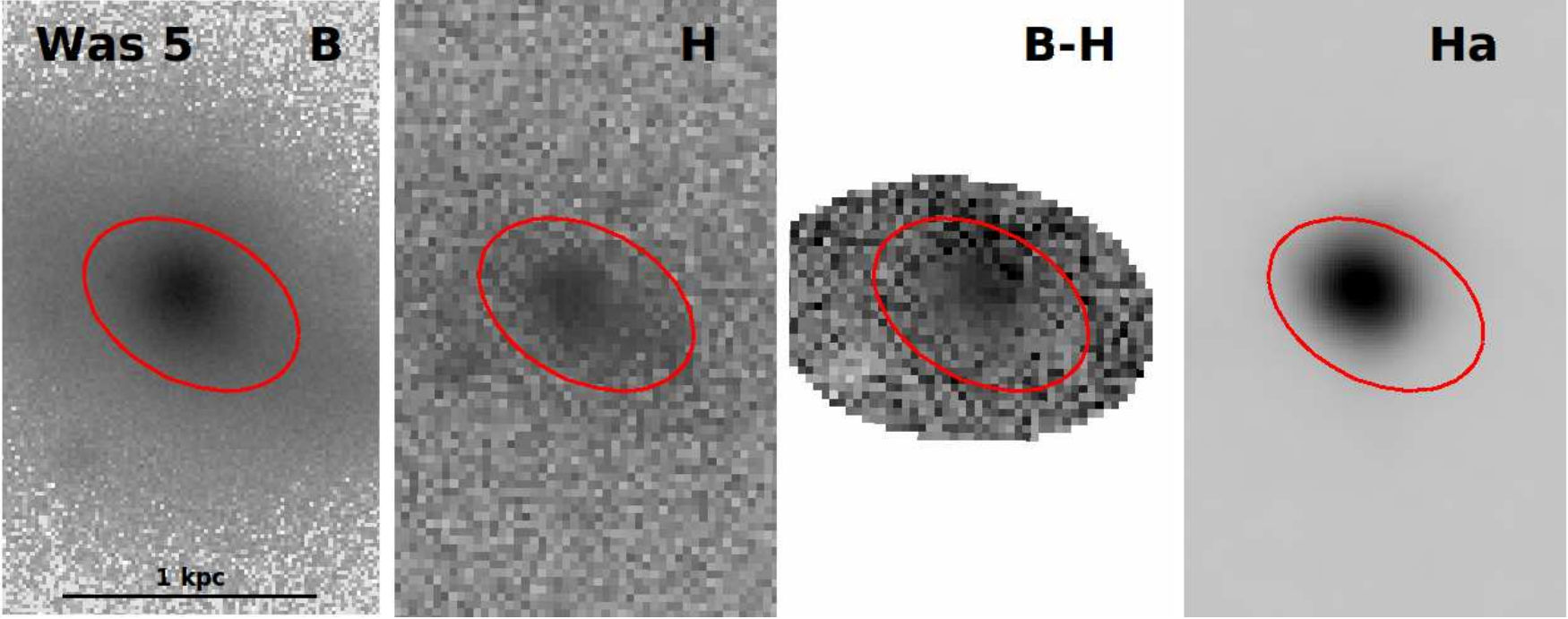} 
\plotone{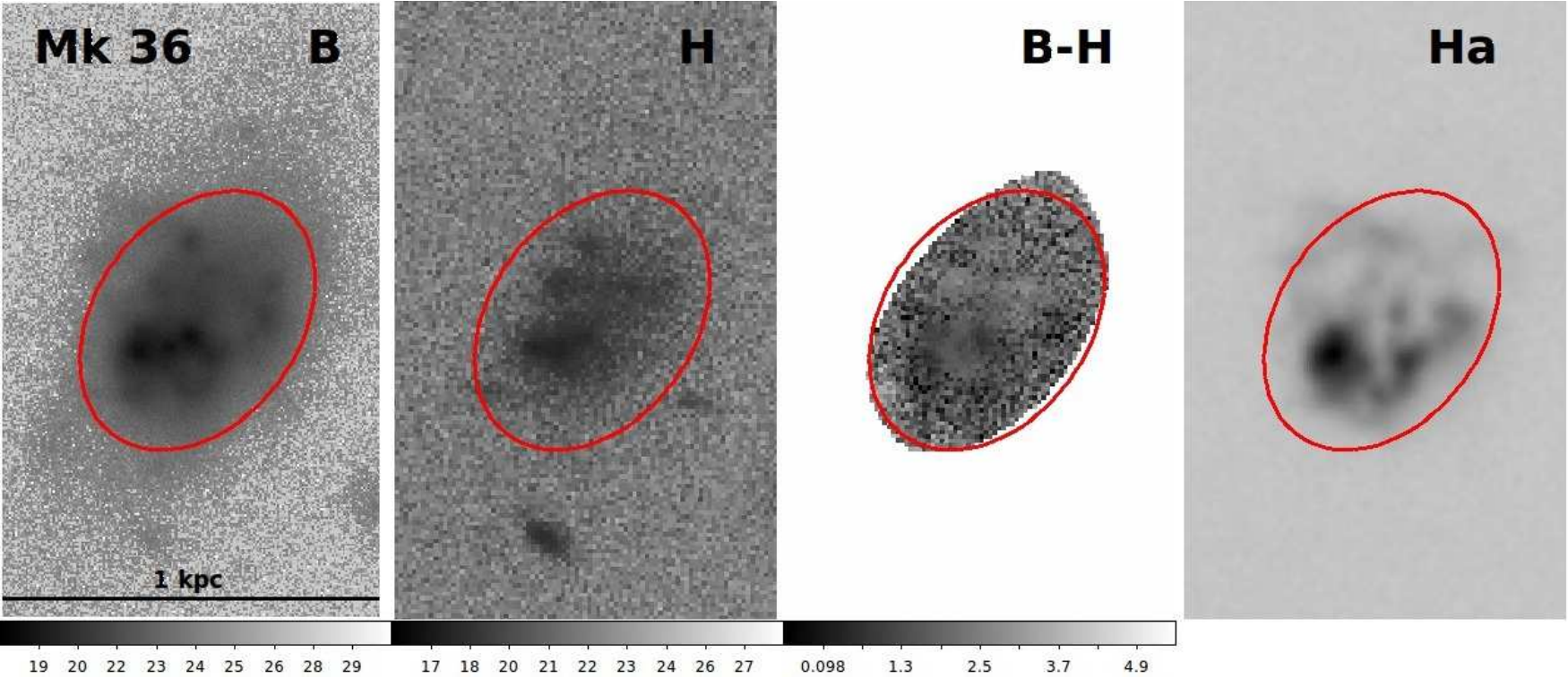} 
\caption{continued}
\end{figure*}

\begin{figure*}
\figurenum{1}
\centering
\epsscale{0.99}
\plotone{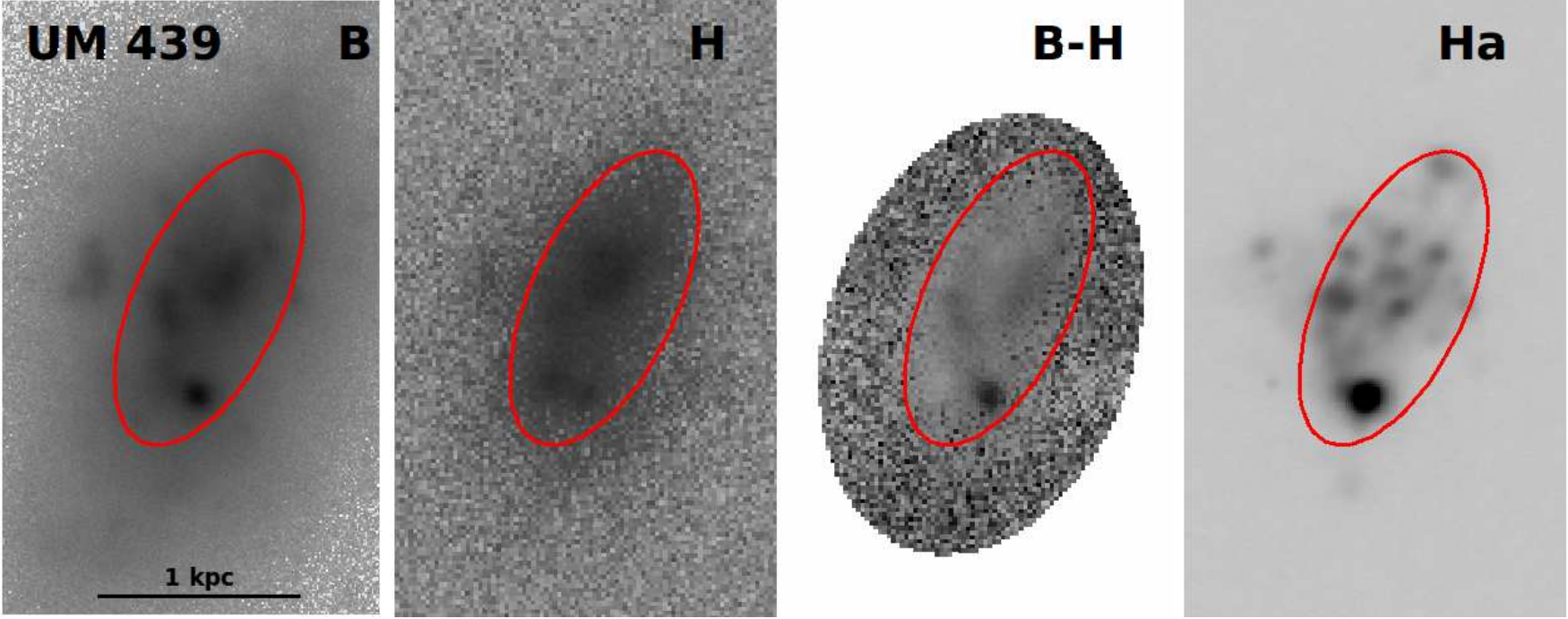} 
\plotone{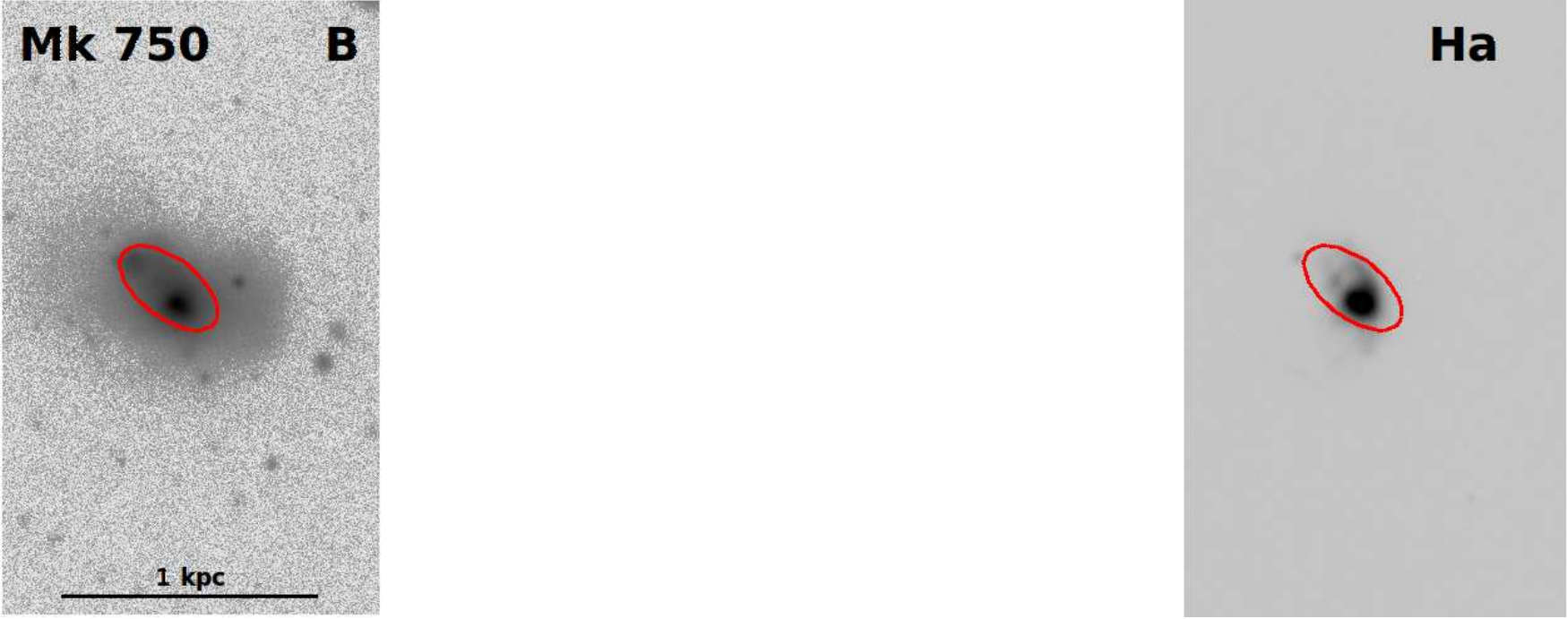} 
\plotone{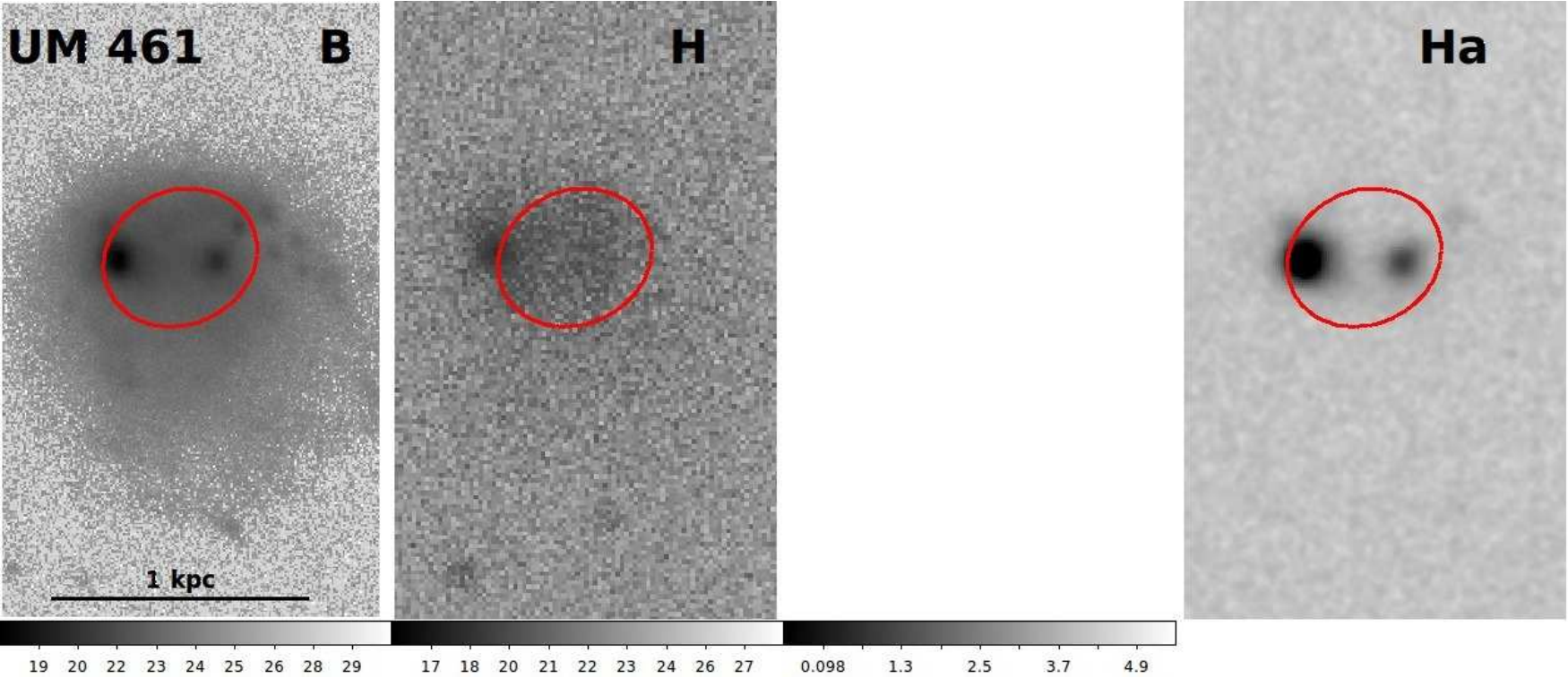} 
\caption{continued}
\end{figure*}

\begin{figure*}
\figurenum{1}
\centering
\epsscale{0.99}
\plotone{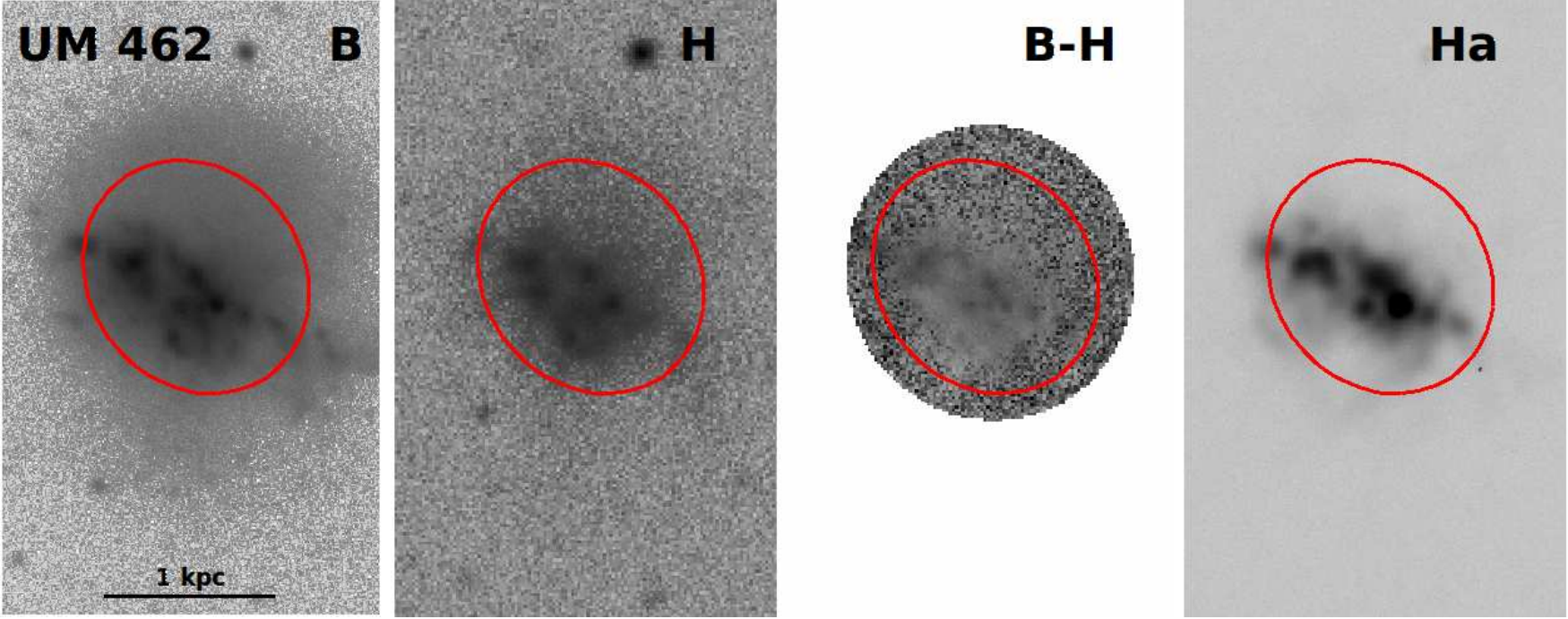} 
\plotone{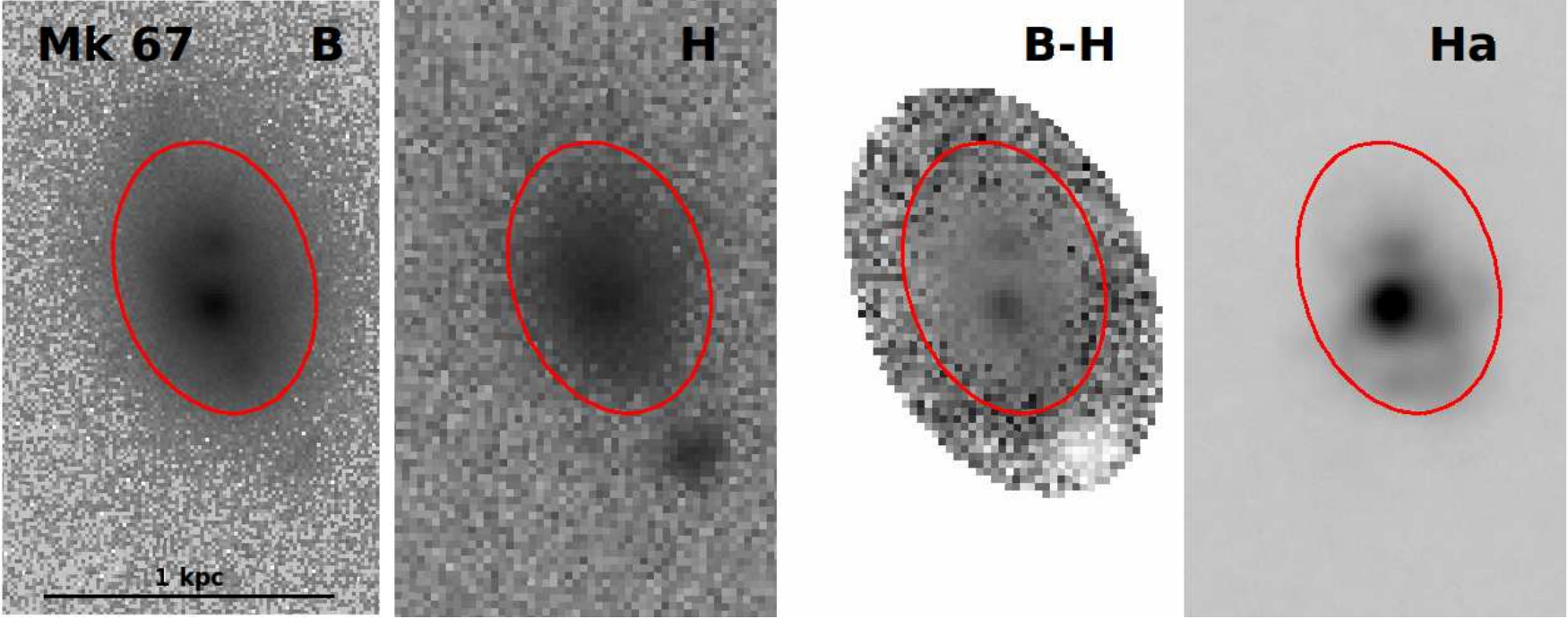} 
\plotone{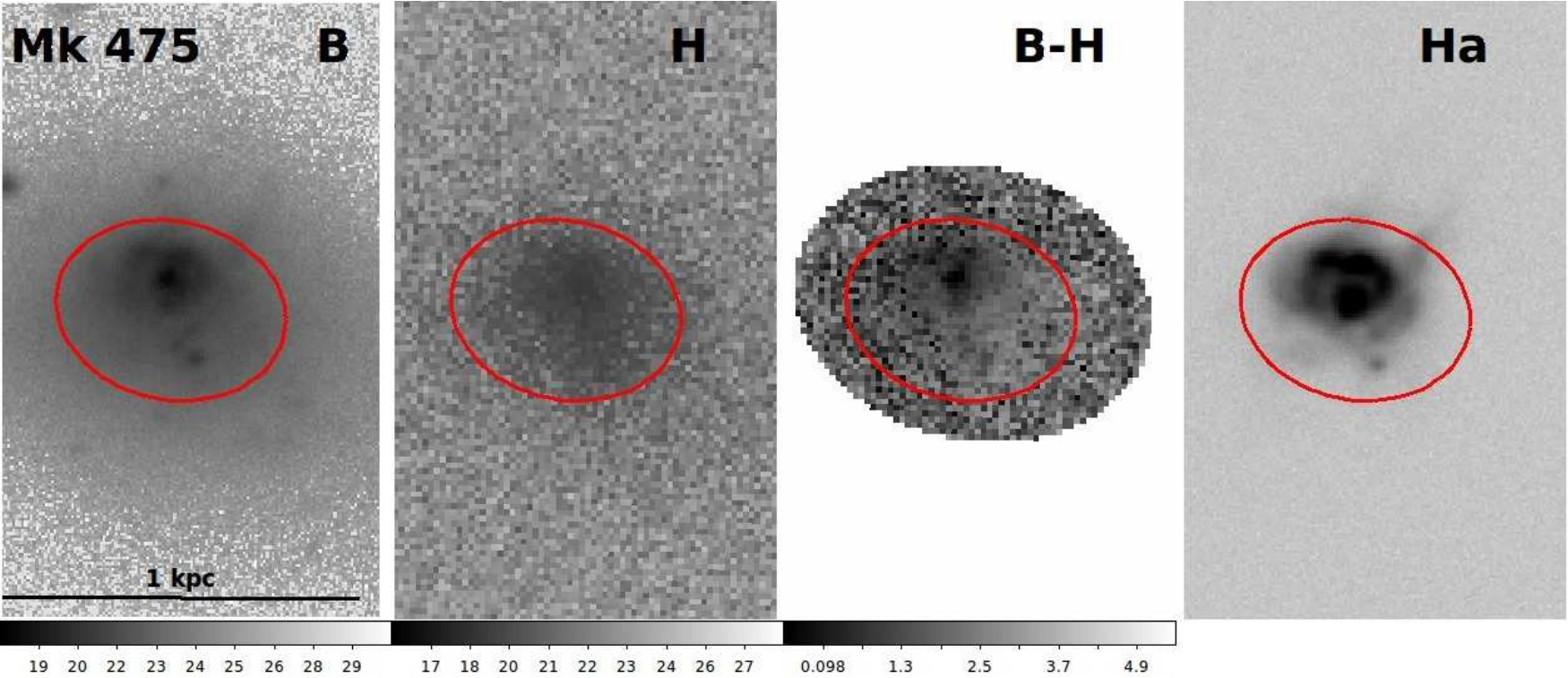} 
\caption{continued}
\end{figure*}

\begin{figure*}
\figurenum{1}
\centering
\epsscale{0.99}
\plotone{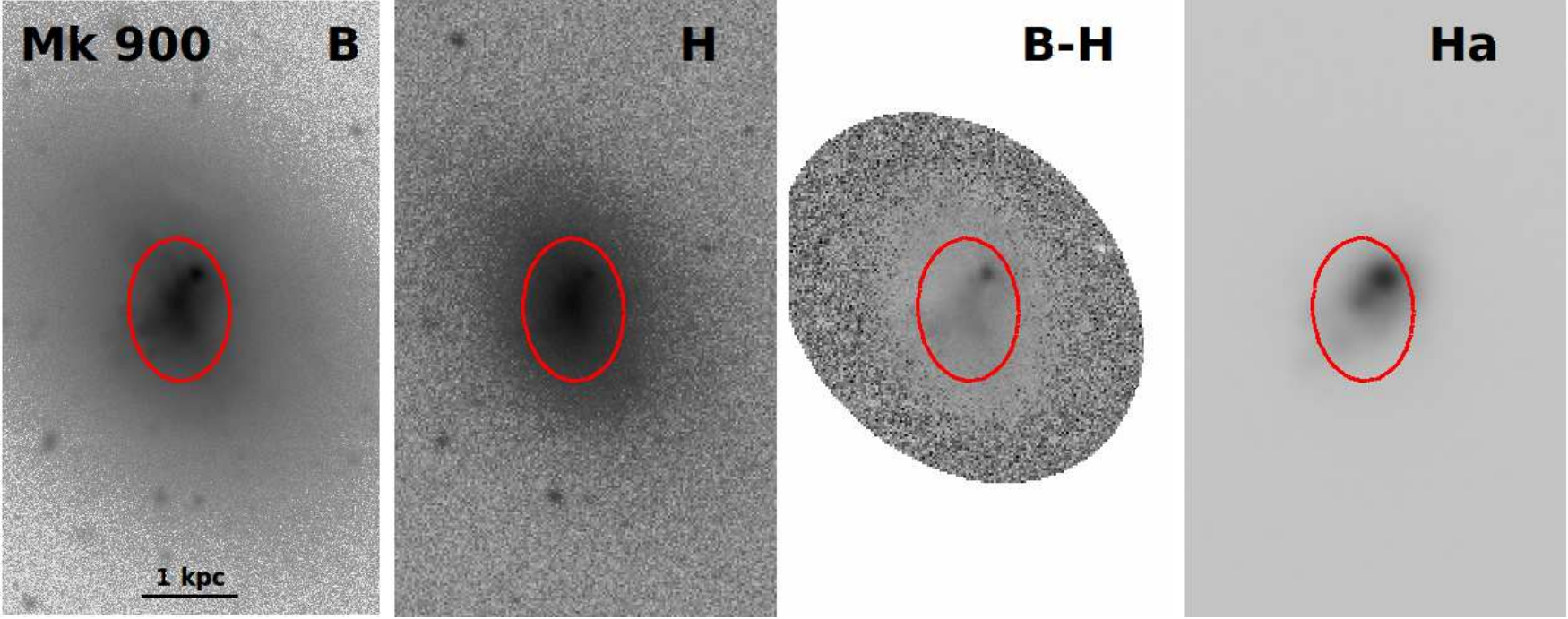} 
\plotone{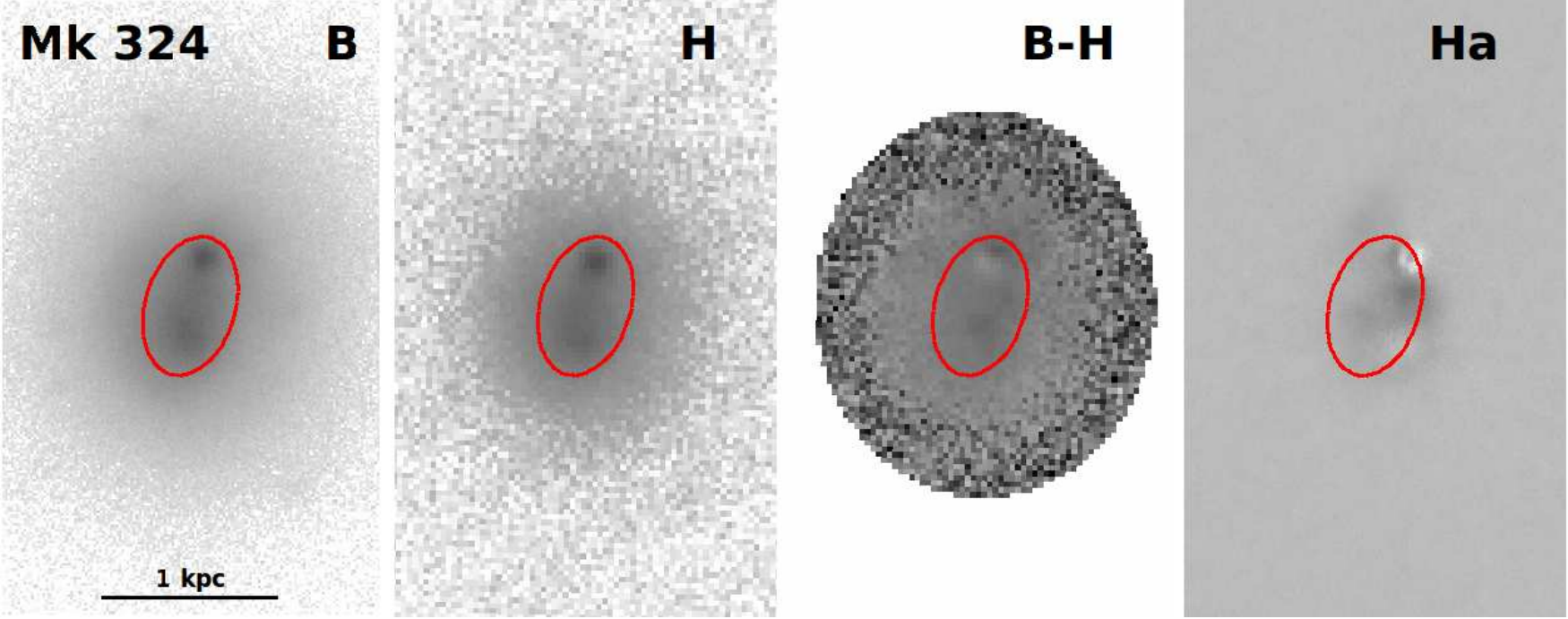} 
\plotone{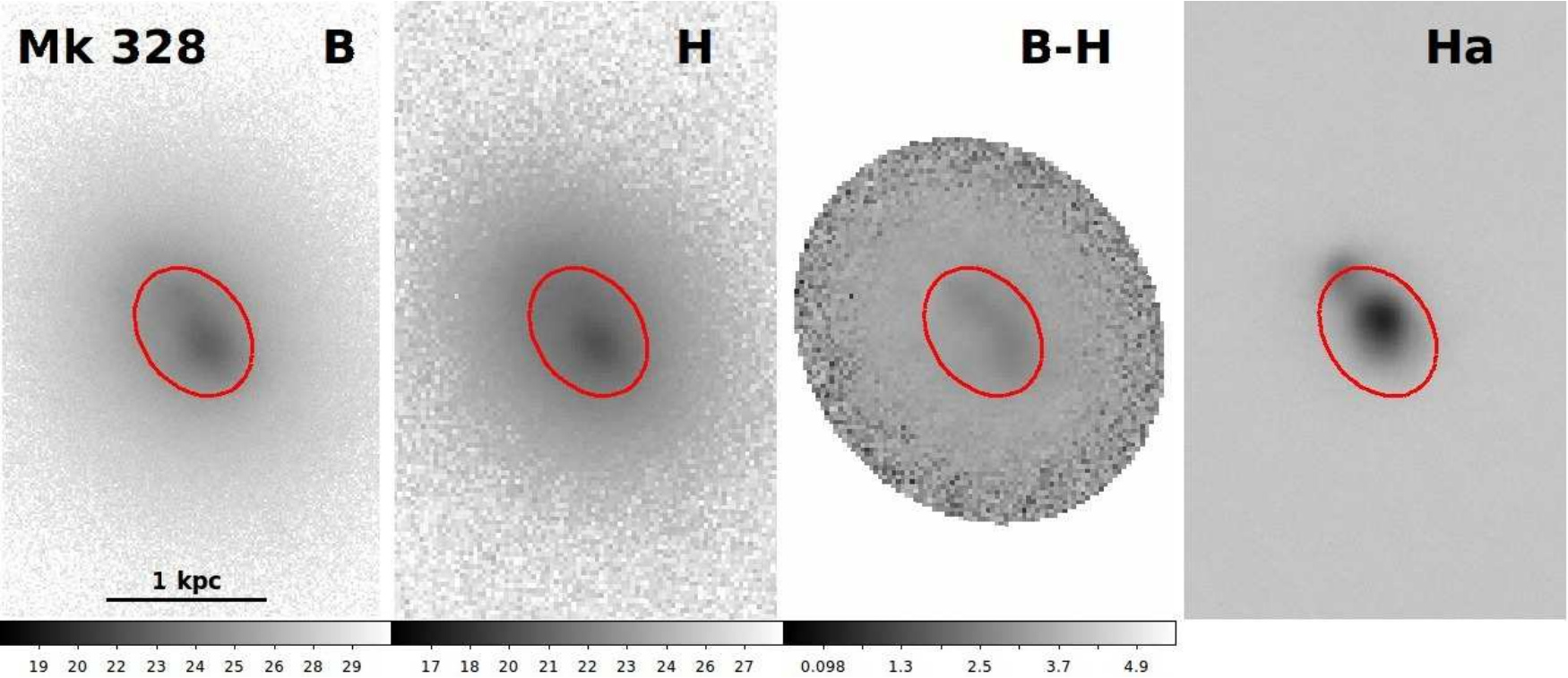} 
\caption{continued}
\end{figure*}

\section{Surface Photometry and Fitting}\label{sbfits}

Our objective is to measure the properties and structure of the $B$
and $H$ light of 
the underlying host in each of our BCDs. While some regions of each
BCD are dominated by the light from the recent starburst, the
outskirts are dominated by the light from old stars. 
We use our calibrated \Ha images to identify and mask areas of
active star formation in our B images. The B-filter light comes mostly
from stellar emission, particularly from young, recently-formed
stars. 
{In order to compare BCD hosts with other types of dwarf 
galaxies, independent of recent star formation we use the \Ha
threshold masks to ensure that we only fit the light from 
the old stars and that our surface brightness profile fit represents
the light from the underlying host galaxy. }


We use the \textsc{galphot} package and
\textsc{ellipse} (Jedrzejewski 1987) task in IRAF to fit
elliptical isophotes to the $B$ images.
Starting from an initial isophote, the
fit proceeds toward the center of the galaxy until it runs out of
pixels to fit, and then proceeds toward the outskirts of the
galaxy. At each isophote, the algorithm varies the geometric
parameters (center, ellipticity, position angle) of subsequent
ellipses until it best fits the galaxy at that isophotal level. Below
a certain signal-to-noise threshold, there is not enough galaxy light
to accurately determine the shape of the isophotes, so the parameters
from the last well-determined ellipse are held fixed and all larger
ellipses have the same ellipticity and position angle.

The $B$ images are fit both with and without the \Ha masks, and we
find that the mask has a negligible effect outside of the bright,
central region of the galaxy. As our profile
fitting will be limited to the outskirts of the galaxies, these \Ha
masks will not drive our results but are a good precaution. However,
to display sensible $B$ surface brightness profiles inside the
regions excluded by the \Ha masks, we need to make a composite
profile. When fitting inwardly on the unmasked image, we hold the
shape parameters fixed within the region covered by the mask. In
detail, we determine the innermost isophote with at least half of its
pixels unmasked, and use that as the transition point between an
unconstrained shape fit, and holding the shape fixed all the way to
the center.

The resulting isophotal ellipse fits obtained from the B-filter
surface photometry are then applied to the
$H$ images to generate a comparable surface brightness profile in
$H$. We found very small differences in the $H$ surface brightness
profile when it was determined using the $B$ annuli compared to when
the $H$ profile was determined from an
unconstrained fit to the $H$ image itself. However, due to the
shallower nature of 
the $H$ images, the $B$ annuli are able to trace the $H$ profiles to
fainter levels than if the $H$ images had been fit with independent
isophotes.

These same $B$ annuli are also used on each of
the SDSS $ugriz$ images to determine surface brightness profiles in
each filter. The use of spatially-matched images and common isophotes
for all wavelengths allows us to create surface brightness profiles
where each isophotal annulus measures the same
physical part of the galaxy in each filter, enabling us to also create
\textit{surface color} profiles.
We also apply a correction in each filter for Galactic extinction
obtained through NED, 
from the recently re-calibrated dust maps of Schlafly \& Finkbeiner (2011).


Dwarf galaxy surface brightness profiles can be fit with a
wide variety of functions across their radial extent. A wide variety
of surface brightness profile breaks are observed in dwarf galaxies
(Herrmann \etal 2013), but each component is still fitted
with 
exponential profiles (Hunter \& Elmegreen 2006), consistent with
stellar disks. As motivated in Section \ref{observations}, in order to
probe the structure of the underlying old stellar populations in the
outskirts of the BCDs, we fit only the outer portion of each BCD's
surface brightness profile, where the shape is well described by a
single exponential profile. This exponential decline likely indicates
the presence of a somewhat regular underlying stellar disk.

The inner limit of the fit to the underlying host galaxy is set at the
point where the contamination 
by light from recent or ongoing star formation is no longer
significant and the profile has a simple exponential decline. The
outer limit of the fit is set by the signal-to-noise 
of the profile above the sky background, in terms of how bright the
galaxy is relative to the standard deviation ($\sigma$) of the
measured sky level. Sometimes these fit limits are different between
the $B$ and $H$ filter, owing usually to the shallower nature of the
$H$ images. In 
a few cases the $H$ images are not deep enough to reach a region where
the $H$ surface brightness profile follows an exponential decline, and
no $H$ profile fitting is possible. For this reason, we do not fit
$H$ profiles to Was 5, Mk 36, or UM 461 { (c.f. Hunter \&
  Elmegreen 2006, Lagos \etal 2011, and Micheva \etal 2013, for deeper
  NIR profiles) }.

We use the SDSS $g$ and $r$ profiles to construct a $B-V$ profile
with the photometric conversions from Jester \etal 2005. While the
$g$ and $r$ 
images are less deep than our observations, the additional color
information is valuable even if it is only available in the higher
surface brightness regions of the BCDs. The $B-V$ and $B-H$
colors typically flatten to a constant value in the region of the
outskirts where we fit exponentials to the profiles, and we compute
the average color within this region as well.

Our linear least-squares fit to the surface brightness profiles of the
underlying host galaxy gives
a best-fit value of the slope (related to the disk scale length) and
intercept (related to the central surface brightness), but these
observed quantities must be converted to intrinsic
quantities. First we need to geometrically correct the observed
central surface brightness $\mu$ to the central surface brightness for
a face-on orientation $\mu^0$ using a determination of the inclination
angle, $i$, assuming an optically thin disk, with:

\vspace{0.5cm}

$\mu^0 = \mu - 2.5 \textrm{log} \cos i$.

\vspace{0.5cm}

\noindent
Following the method of Staveley-Smith \etal (1992), we consider the
galaxy as an oblate spheroid with thickness to length ratio $q$, which
on average for dwarf galaxies is $\left< q \right> =
0.57$. Staveley-Smith \etal (1992) derived the following
relationship between semi-minor to semi-major axis ratio $b/a$, and
the thickness-to-length ratio $q$: 

\vspace{0.5cm}

$q = 0.65(b/a) - 0.072(b/a)^{3.9}$.

\vspace{0.5cm}

\noindent
For each galaxy we determine the mean ellipticity ($\epsilon = 1 -
b/a$) within the outskirts of the galaxy where the underlying surface
brightness profile is fit. We use this ellipticity to determine an
estimate of the inclination 
angle, and correct all central surface brightnesses to face-on. We
also use distances derived from flow models (Mould \etal 2000) to
convert the scale radius into physical units.
All of these best-fit parameters are listed
in Table \ref{structparms}, and are plotted in the figures that follow 
(absolute magnitudes are corrected for Galactic extinction).

The surface brightness profiles, ellipticity profiles, color profiles,
and fits are all 
shown in Figure \ref{sb}. For each surface brightness profile we show
the $H$ and $B$ profiles on the same scale and also plot the masked
and unmasked $B$ profile fits together to show the minimal
differences. Because of the mask, the masked profile does not continue
to the center of the galaxy. Instead, we show a profile of fixed-shape
isophotes which takes the shape of the inner-most well-determined
isophote using the mask, and extending that shape inward. Using a
fixed-shape ellipse in the inner regions of the BCDs has largely a
cosmetic effect on the surface brightness profile, preventing the
ellipse shape from varying wildly as it tries to match the irregular
structures in the inner regions of most of the BCDs. The fixed-shape
profile is not necessarily physically meaningful, but it is
representative of the azimuthally-averaged light distribution in the
complex central regions of the BCDs.

We also show the ellipticity profile at the top of each graph in
Figure \ref{sb}. As our profiles are all displayed in terms of
semi-major axis, the ellipticity profile is necessary in order to
interpret changes and shapes in the surface brightness profile. A few
galaxies show dramatic changes in ellipticity in their fits. Notably,
Mk 600 has $\epsilon \sim 0.5$ across most of its profile, but in the
inner regions, the ellipticity rapidly decreases as the isophotes
become much more circular. That sudden change in $\epsilon$ causes a
break 
in the surface brightness profile, which is evident in the graph as
well. In general it is safer and simpler to use the semi-major axis
for surface brightness profiles, but in cases where the ellipticity
(or position angle) change rapidly, the resulting profiles are more
difficult to interpret. In our BCDs, the underlying host profiles are
all fit in regions of relatively stable ellipticity, so that our fits
are unaffected by the irregular central regions where isophotal shapes
may change more rapidly.

The complex inner regions of the BCDs have less bearing on this work
than do the outskirts, as we are most interested in the underlying
host light. There are some notable shapes and features in the profiles 
interior to the portions we fit. Some BCDs have significant curvature
in their inner profiles (Was 5, Mk 475, Mk 900, Mk 324, Mk 328), which
looks similar to a deVaucouleurs profile, suggesting that these
galaxies may have bulge-like centers or be similar to dEs. Other BCDs
have irregular bumps and jumps in their inner profiles (Mk 5, I Zw 18,
Mk 36, Mk 750), which show the difficulty of fitting smooth ellipses
to inherently more complex structures. In particular, the central
regions of UM 461 and UM 462 have a bright and nearly uniform level of
intensity, punctuated with sites of active star formation, which makes
for a very complex environment. Finally, some BCDs have profiles which
remain almost exponential over their entire extent (UM 323, UM 408, CG
10, UM 439), which may indicate a somewhat simpler recent history of
star formation and activity.

The shapes of a few galaxies were particularly difficult to
fit with our simple method, due to their significant
irregularities. The cometary shape of Mk 5 creates a significant
positional offset between the center of the bright star-forming region
and the center of the faint outer isophotes. This rapid shifting of
successive isophotes leads to a poorly determined fit in the central
regions, and so we do not display the inner part of the fit. The fit
to the smooth, regular outer isophotes is reliable, though, so we keep
this galaxy in our sample { (see also Section 6.1.1 in Amor\'{i}n
  \etal 2007  for a 2D treatment of Mk 5) }. In a similar way, I Zw 18
presents a 
challenging structure to be fit. Because of the visible and irregular
structures present throughout much of the galaxy, we caution that even
though our fit is restricted to the outermost regions of the galaxy,
we may not have reached the true underlying population of this faint,
unusual galaxy.

The graphs of $B-H$ and $B-V$ color in Figure \ref{sb} are also
powerful tools in understanding these complex galaxies. The shallower
depth of the SDSS images makes the $B-V$ profile become noisy and
erratic beyond the exponential fit region. Inside the fit region and
in the inner regions of the BCDs, however, the $B-V$ and $B-H$
profiles are mostly well behaved and have interesting features. As
noted in the $B-H$ color maps in Figure \ref{images}, the BCDs
tend to have bluer centers and redder outskirts. Some BCDs have
particularly strong color gradients over most of their radial extent
(CG 10, I Zw 18, Was 5, Mk 750, UM 461, Mk 67, Mk 475) while others
are surprisingly flat (UM 323, UM 408, Mk 600, UM 439, Mk 324). The
surface colors ($B-V$ and $B-H$), averaged within the region of the
profile where we fit an exponential, are given in Table
\ref{structparms}, and are fairly red. A more detailed analysis of
both the surface colors and the color gradients throughout the BCDs is
discussed in Section \ref{ccp} and shown in Figure \ref{BVBH}.

\begin{deluxetable*}{cccccccccc}
\tablewidth{7in}
\tablecaption{Structural Parameters of Underlying Host Galaxies of BCDs
  \label{structparms} }
\tablehead{
\colhead{Galaxy} & \colhead{$M_{B,tot}$} & \colhead{$\mu^0_{B,0}$} &
  \colhead{$\alpha_{B,exp}$} & 
\colhead{$M_{H,tot}$} & \colhead{$\mu^0_{H,0}$} &
  \colhead{$\alpha_{H,exp}$} & 
\colhead{b/a} & \colhead{$<$B-V$>$} & \colhead{$<$B-H$>$} \\ 
\colhead{}       & \colhead{[mag]} & \colhead{[\magsec]}   &
  \colhead{[kpc]}   & \colhead{[mag]} & \colhead{[\magsec]}   &
  \colhead{[kpc]}                 
& \colhead{}  & \colhead{(env.)} & \colhead{(env.)} }
\startdata
  UM323 & $ -16.10 $ & $  19.85 $ & $  0.30 $ & $ -17.90 $ & $ 18.49 $ & $  0.41 $ & $ 0.69 $ & $  0.44 $ & $  2.09 $  \\ 
  UM408 & $ -16.04 $ & $  20.40 $ & $  0.36 $ & $ -17.72 $ & $ 19.28 $ & $  0.49 $ & $ 0.65 $ & $  0.46 $ & $  1.96 $  \\ 
  Mk600 & $ -15.55 $ & $  20.96 $ & $  0.38 $ & $ -17.30 $ & $ 18.84 $ & $  0.35 $ & $ 0.48 $ & $  0.46 $ & $  1.83 $  \\ 
    Mk5 & $ -15.18 $ & $  19.79 $ & $  0.25 $ & $ -17.74 $ & $ 17.64 $ & $  0.31 $ & $ 0.59 $ & $  -     $ & $  2.76 $  \\ 
   CG10 & $ -15.01 $ & $  20.10 $ & $  0.23 $ & $ -17.47 $ & $ 18.11 $ & $  0.27 $ & $ 0.58 $ & $  0.47 $ & $  2.36 $  \\ 
  IZw18 & $ -14.68 $ & $  19.25 $ & $  0.10 $ & $ -15.19 $ & $ 17.31 $ & $  0.10 $ & $ 0.66 $ & $  0.45 $ & $  0.97 $  \\ 
   Was5 & $ -15.05 $ & $  20.16 $ & $  0.22 $ & $ -16.86 $ & $ -     $ & $  -     $ & $ 0.68 $ & $  0.62 $ & $  2.65 $  \\ 
   Mk36 & $ -14.23 $ & $  22.17 $ & $  0.22 $ & $ -16.03 $ & $ -     $ & $  -     $ & $ 0.62 $ & $  0.65 $ & $  3.03 $  \\ 
  UM439 & $ -16.11 $ & $  20.73 $ & $  0.43 $ & $ -18.23 $ & $ 17.84 $ & $  0.32 $ & $ 0.58 $ & $  0.46 $ & $  2.27 $  \\ 
  Mk750 & $ -13.42 $ & $  21.20 $ & $  0.12 $ & $  -     $ & $ -     $ & $  -     $ & $ 0.61 $ & $  0.61 $ & $  -     $  \\ 
  UM461 & $ -14.40 $ & $  21.28 $ & $  0.17 $ & $ -15.87 $ & $ -     $ & $  -     $ & $ 0.88 $ & $  0.41 $ & $  1.44 $  \\ 
  UM462 & $ -16.18 $ & $  20.25 $ & $  0.24 $ & $ -18.07 $ & $ 17.95 $ & $  0.24 $ & $ 0.89 $ & $  0.45 $ & $  2.42 $  \\ 
   Mk67 & $ -15.16 $ & $  20.16 $ & $  0.17 $ & $ -17.42 $ & $ 19.52 $ & $  0.63 $ & $ 0.70 $ & $  0.54 $ & $  2.71 $  \\ 
  Mk475 & $ -14.28 $ & $  22.16 $ & $  0.24 $ & $ -16.25 $ & $ 20.04 $ & $  0.33 $ & $ 0.74 $ & $  0.56 $ & $  2.65 $  \\ 
  Mk900 & $ -17.16 $ & $  21.45 $ & $  0.73 $ & $ -19.85 $ & $ 18.32 $ & $  0.70 $ & $ 0.72 $ & $  0.80 $ & $  3.12 $  \\ 
  Mk324 & $ -16.70 $ & $  20.40 $ & $  0.28 $ & $ -19.08 $ & $ 17.76 $ & $  0.27 $ & $ 0.85 $ & $  0.65 $ & $  2.65 $  \\ 
  Mk328 & $ -16.57 $ & $  21.37 $ & $  0.39 $ & $ -19.57 $ & $ 17.81 $ & $  0.37 $ & $ 0.85 $ & $  0.90 $ & $  3.34 $  \\ 
\enddata
\tablecomments{Our H observations of Was 5, Mk 36, and UM 461 were not
  sensitive enough for surface brightness profile fitting. The $b/a$
  values are averages determined for the underlying host galaxy,
  within the same range of SMA as the exponential fit.}
\end{deluxetable*}


\begin{figure*}
\figurenum{2}
\label{sb}
\centering
\epsscale{0.95}
\plottwo{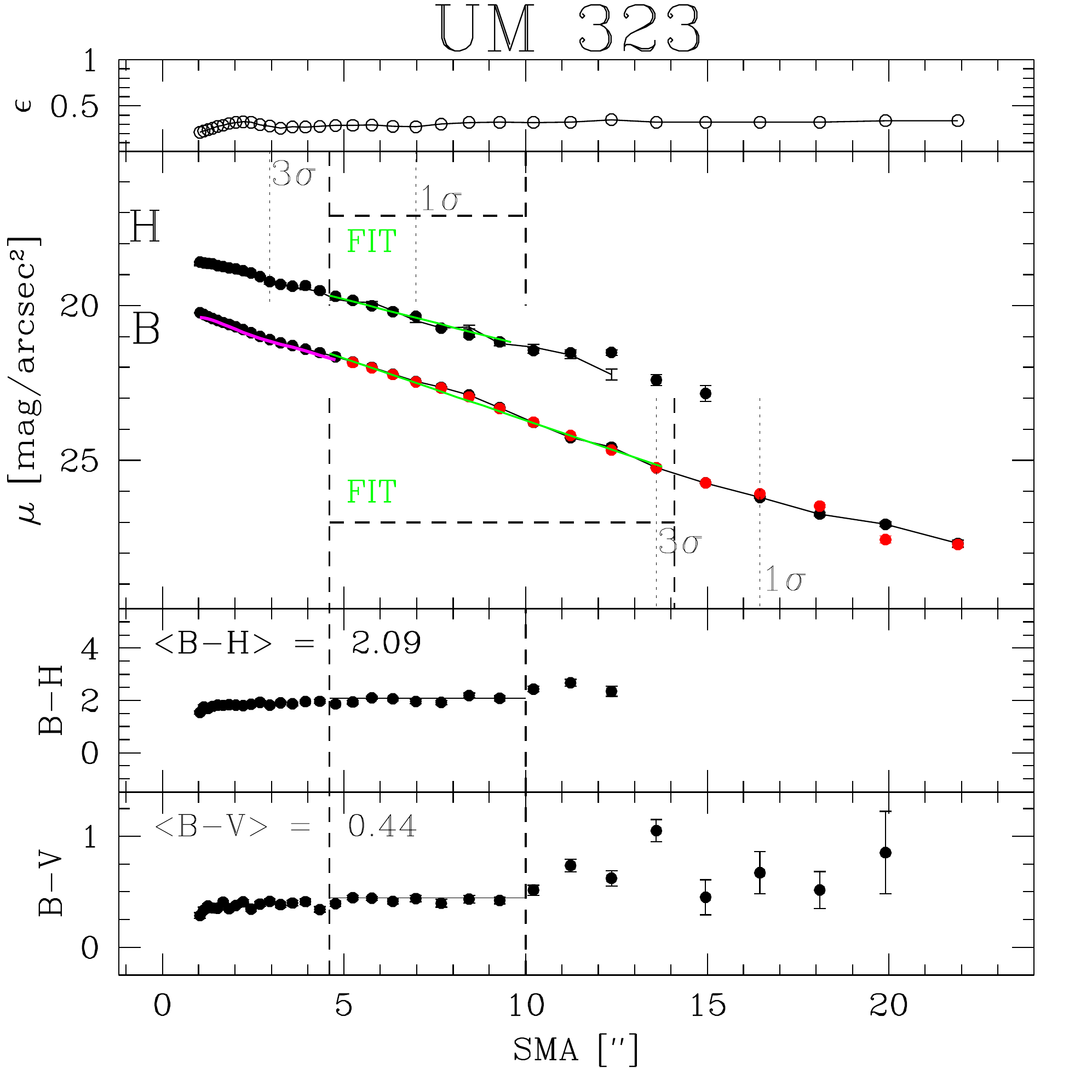}{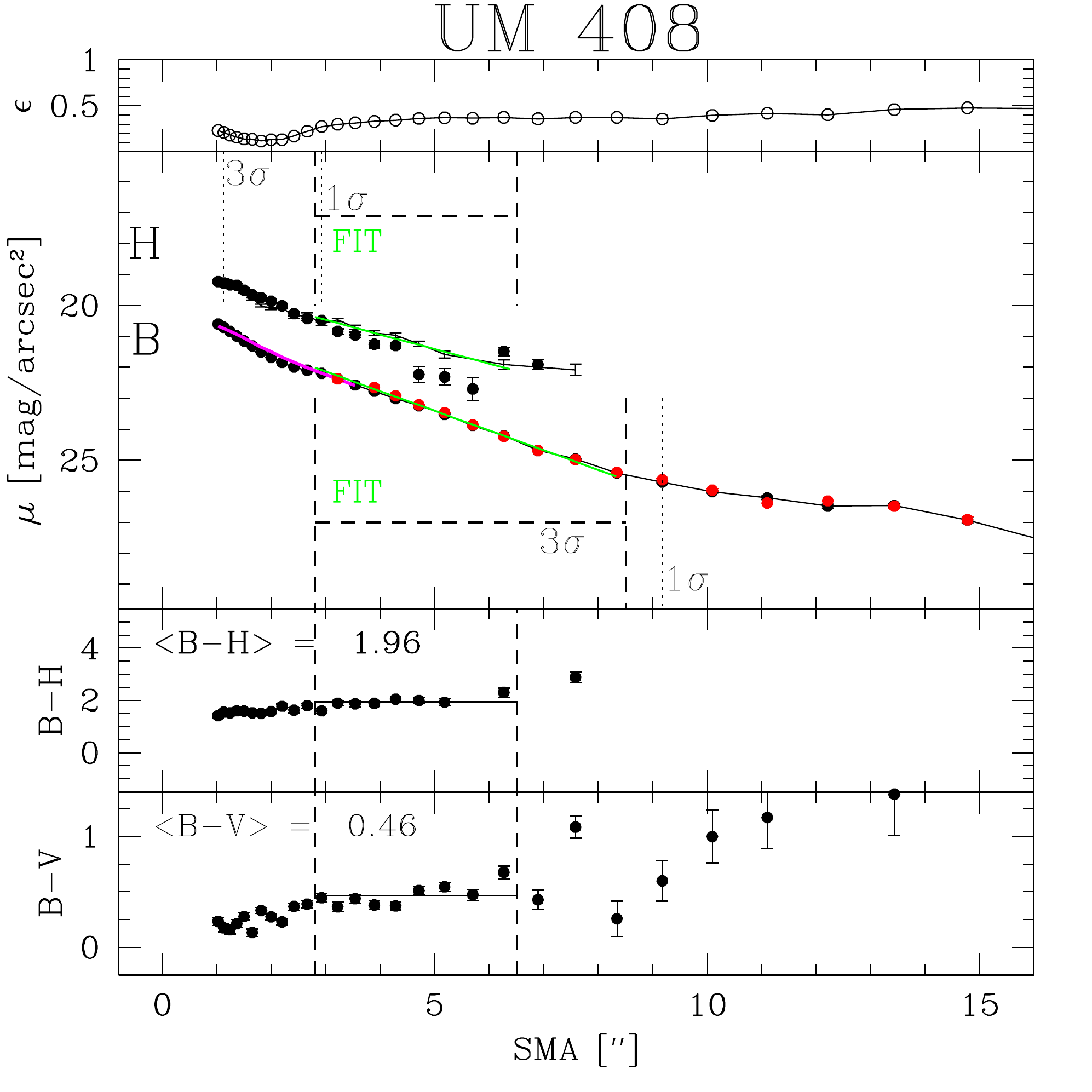} 
\plottwo{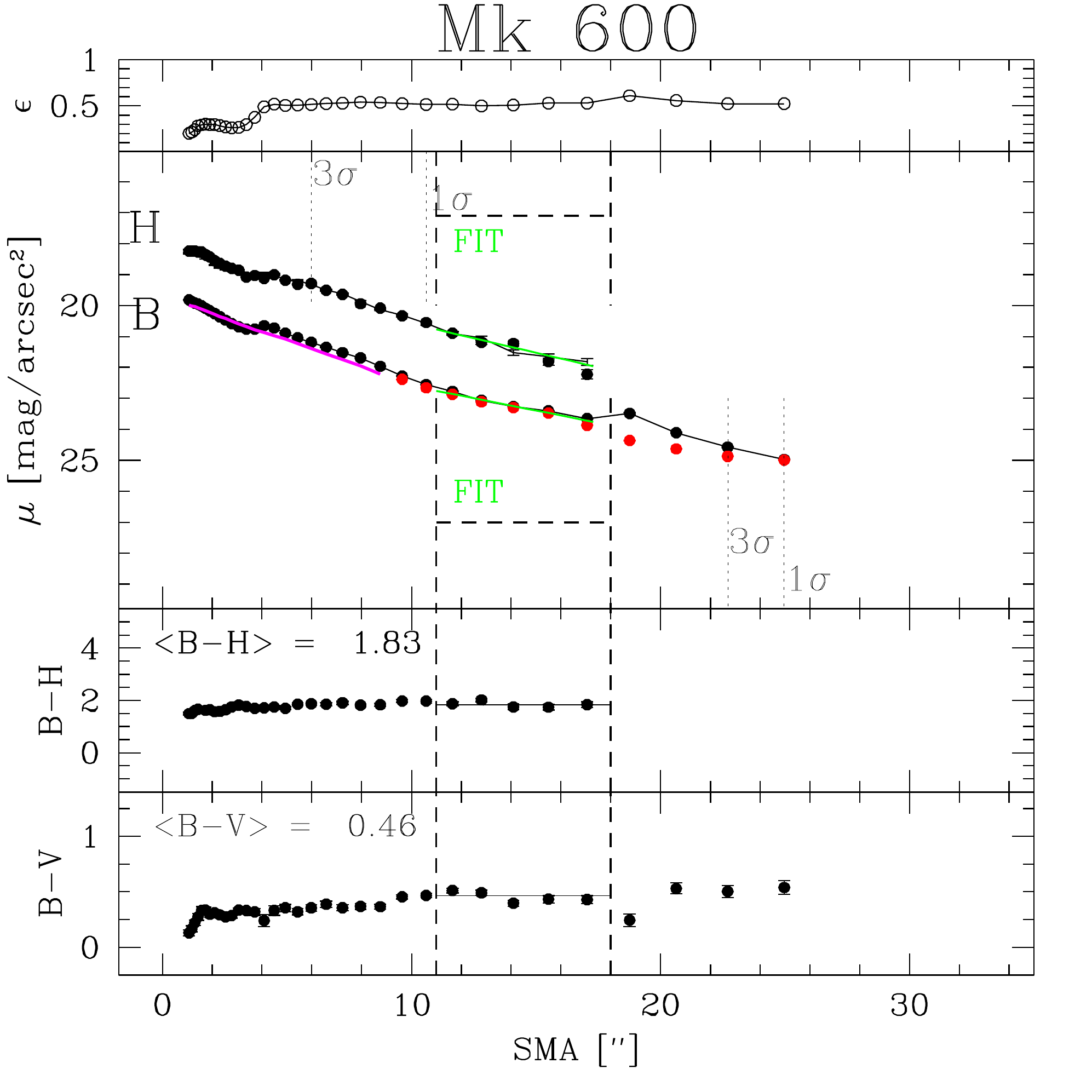}{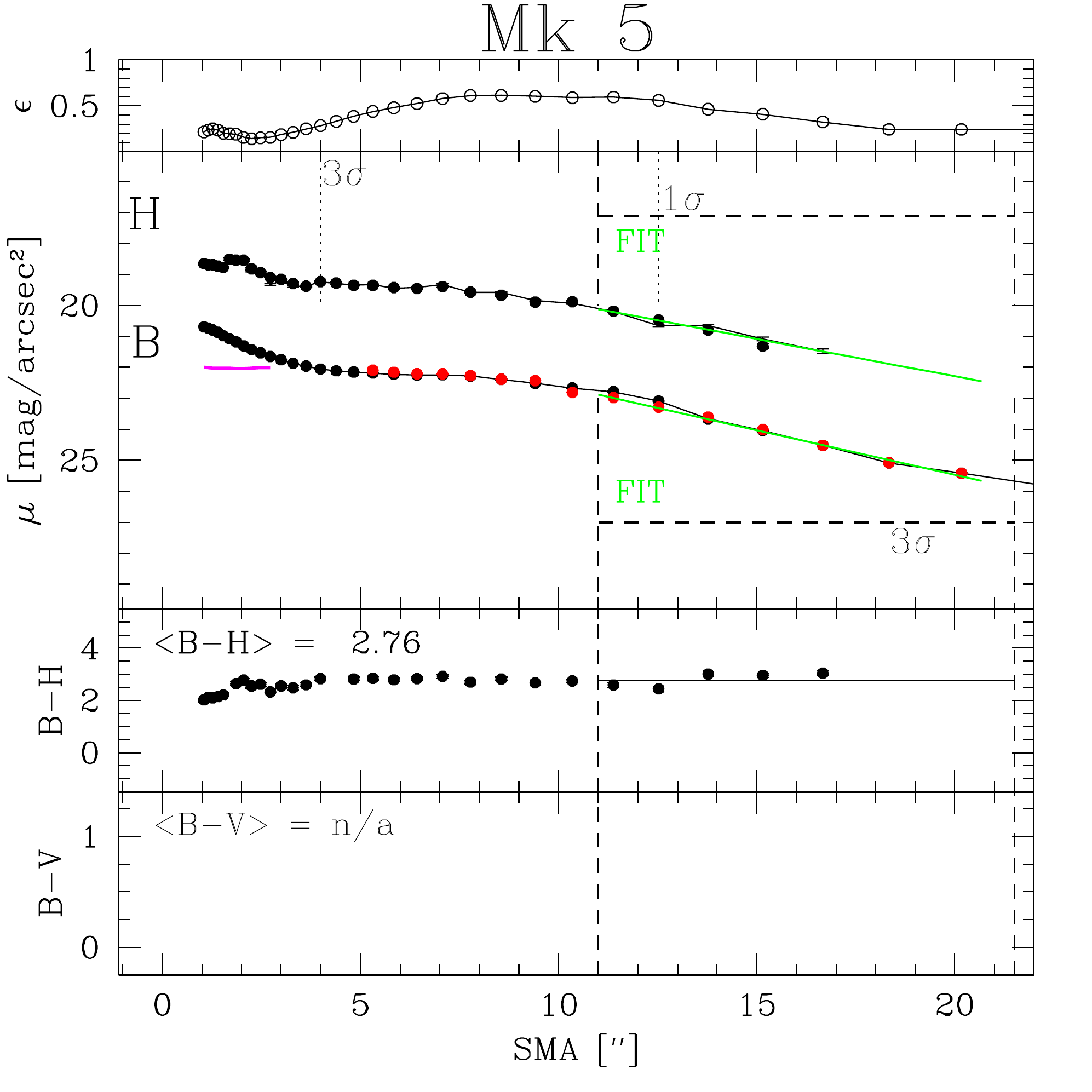} 
\caption{Surface brightness profiles and fits. Each graph shows the
  surface brightness profiles, color profiles, and fits for each BCD,
  plotted as a function of semi-major axis (SMA), in arcseconds. The
  large main panel shows the unmasked $B$ surface brightness profile
  (connected black dots), the masked $B$ surface brightness profile
  (red dots), the inner region of fixed-shape fit (magenta line), the
  $H$ surface brightness profile from an unconstrained fit 
  (black dots), and the $H$ surface brightness profile as constructed
  from the best-fitting $B$ ellipses (black line). Vertical dotted
  lines show the thresholds where the surface brightness profiles
  reach $3\sigma_{sky}$ and $1\sigma_{sky}$, and connected vertical
  dashed lines
  show the SMA range used to fit the profile. The exponential fits to
  the underlying host galaxies are shown as green lines, over the
  range of SMA where they were determined.
  The upper panel shows the ellipticity of the best-fit unmasked
  isophotes as a function of $SMA$.
  The lower middle panel 
  shows the $B-H$ profile as calculated within the matched ellipses
  between the $B$ and $H$ images, and the average $B-H$ color within
  the common fit region from both filters. The bottom panel shows the
  $B-V$ color as converted from the matched $g$ and $r$ profiles (see
  Section \ref{sbfits}).
}
\end{figure*}

\begin{figure*}
\figurenum{2}
\centering
\epsscale{0.95}
\plottwo{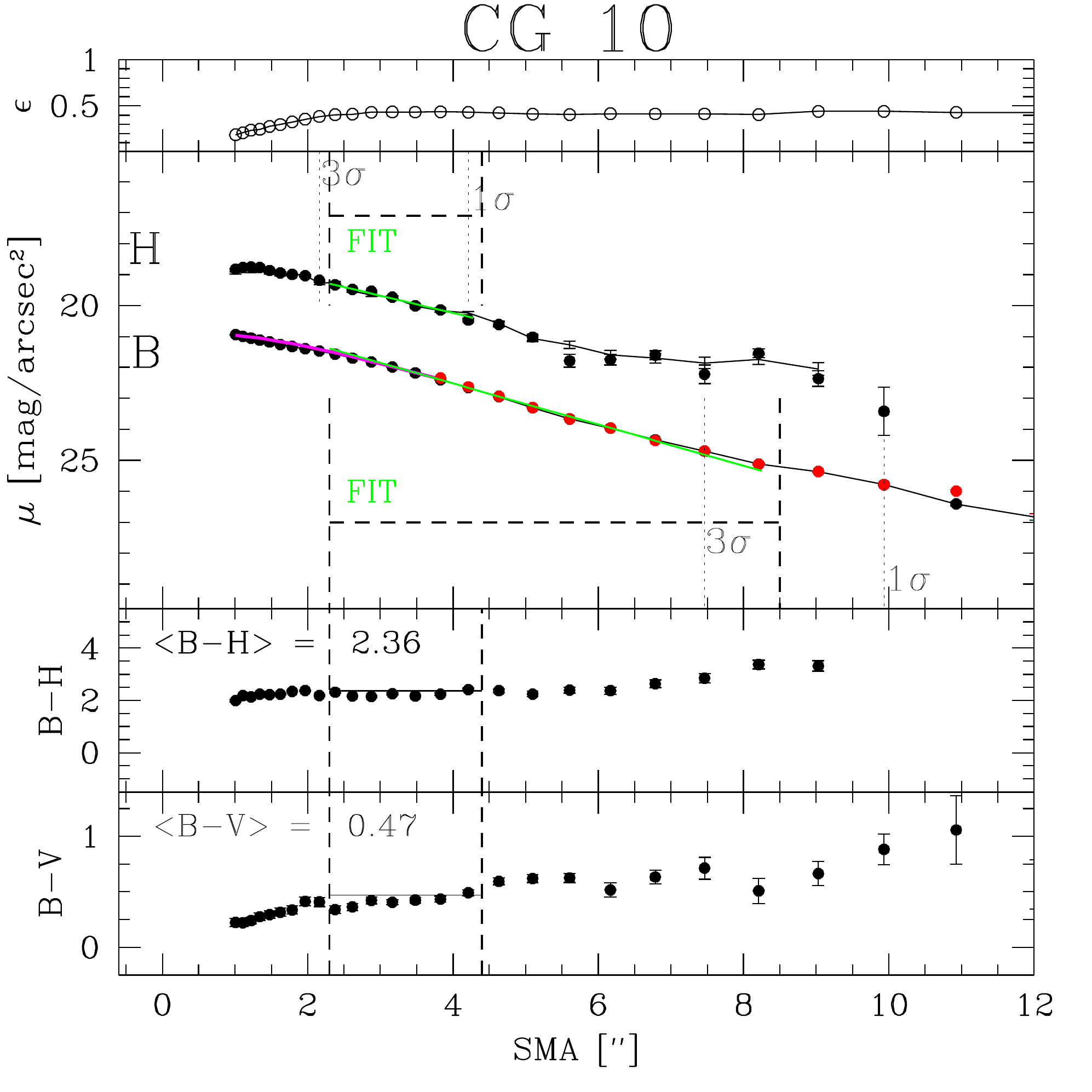}{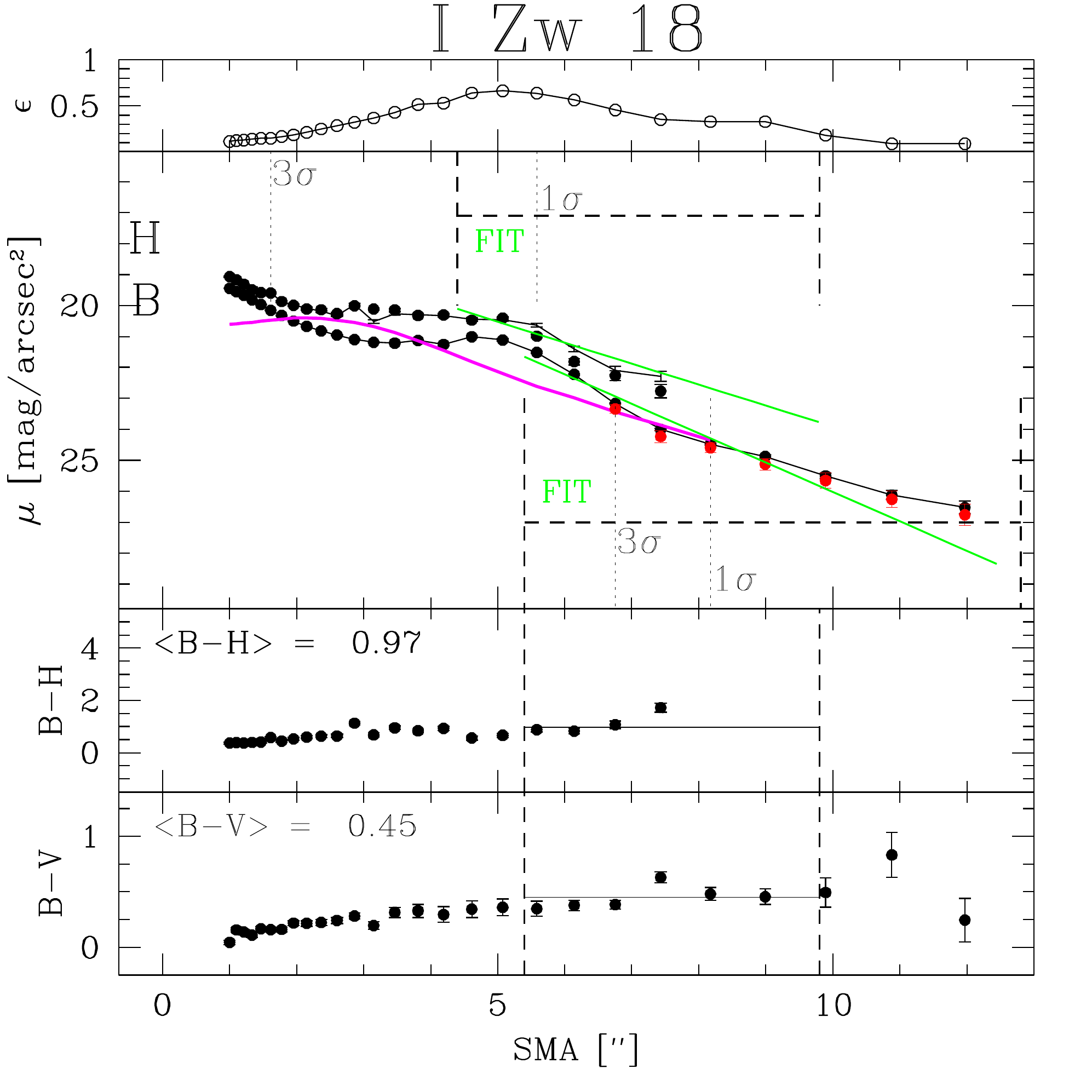}  
\plottwo{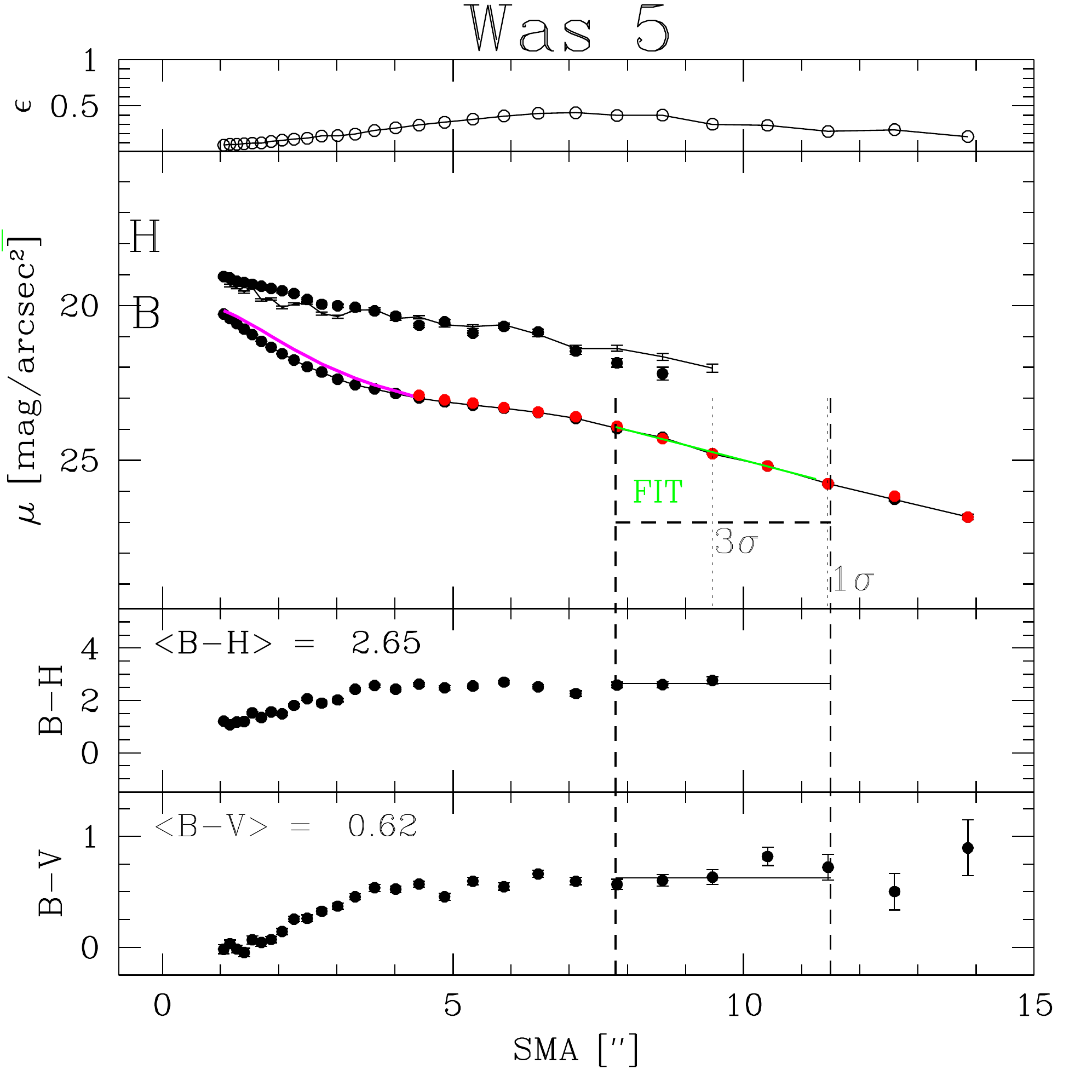}{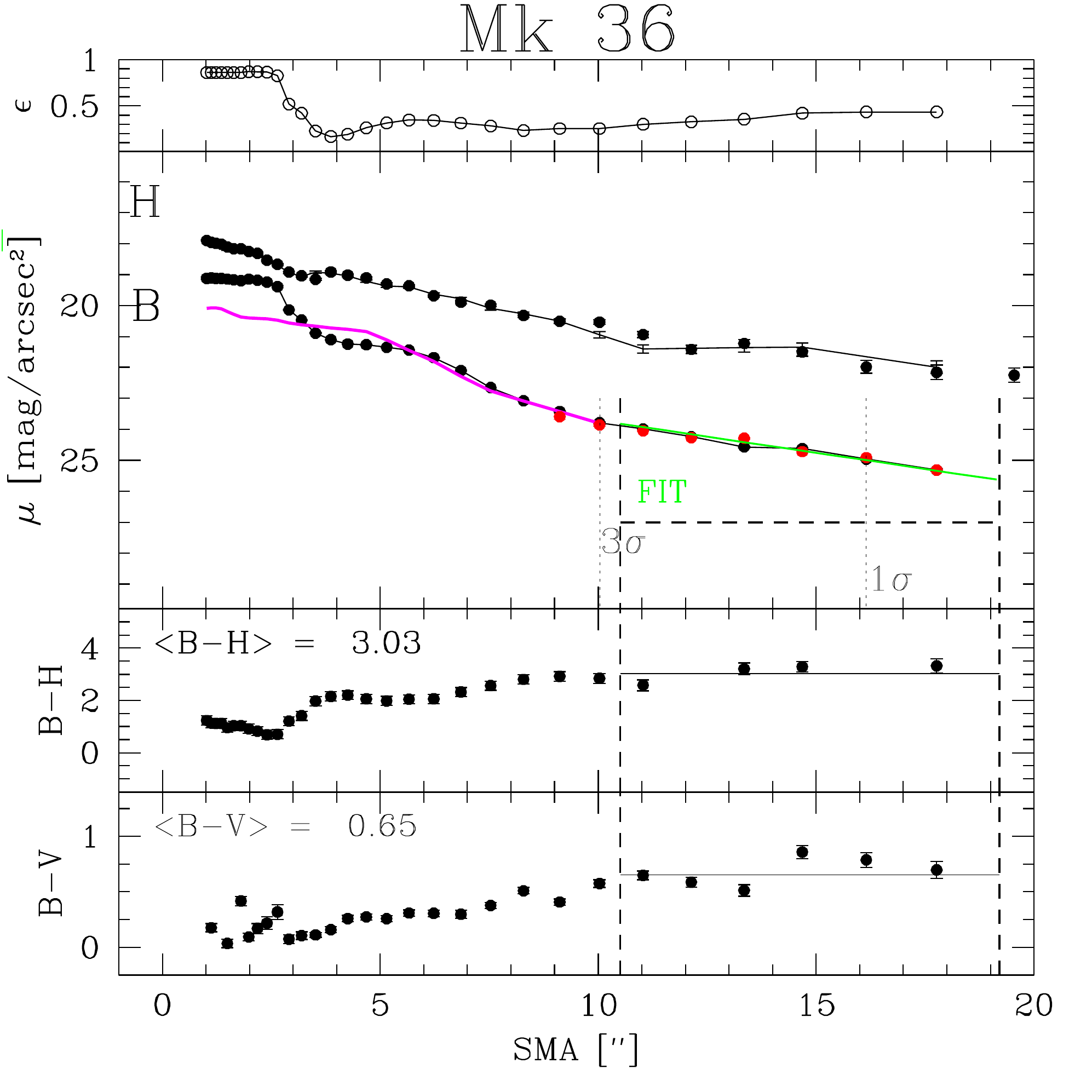}  
\plottwo{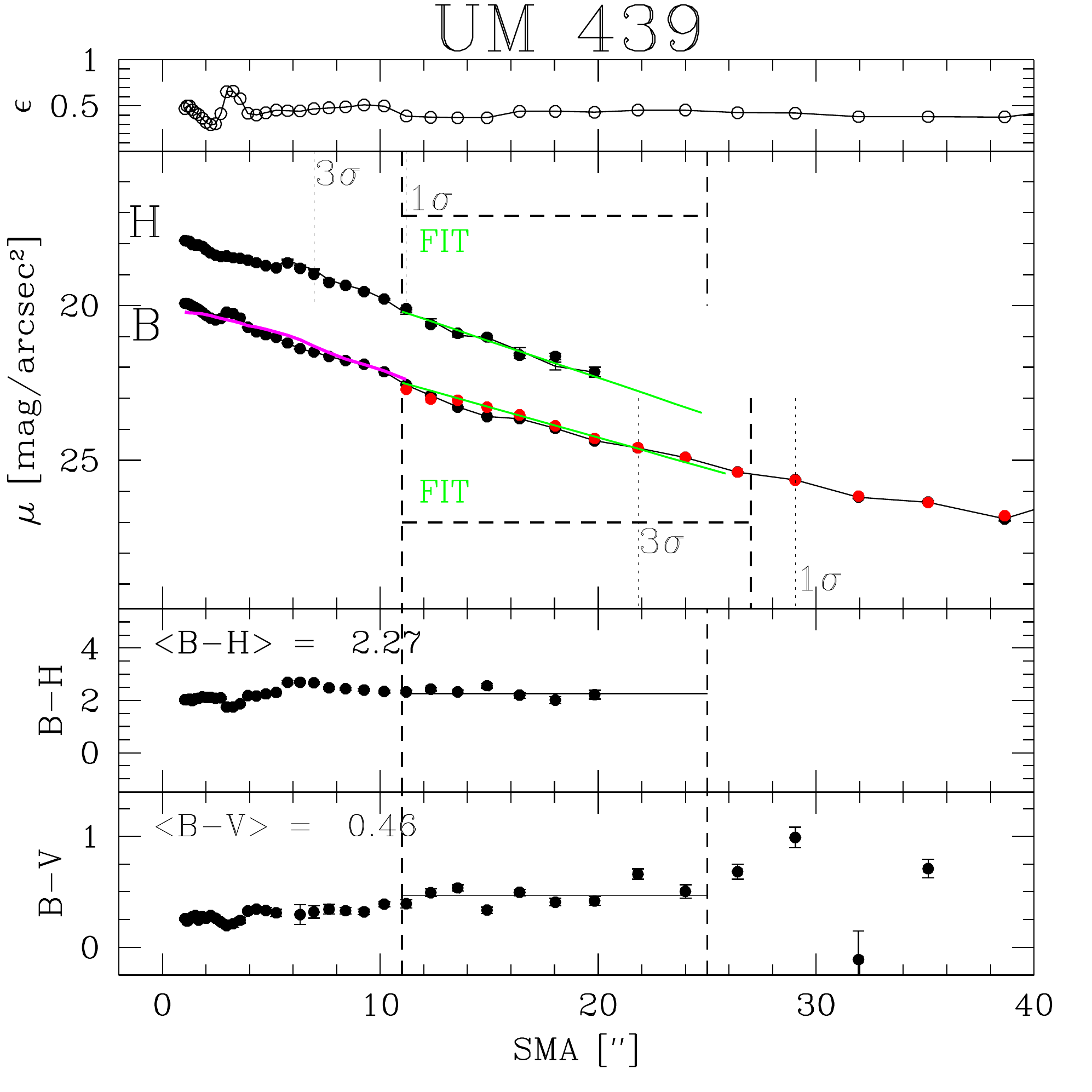}{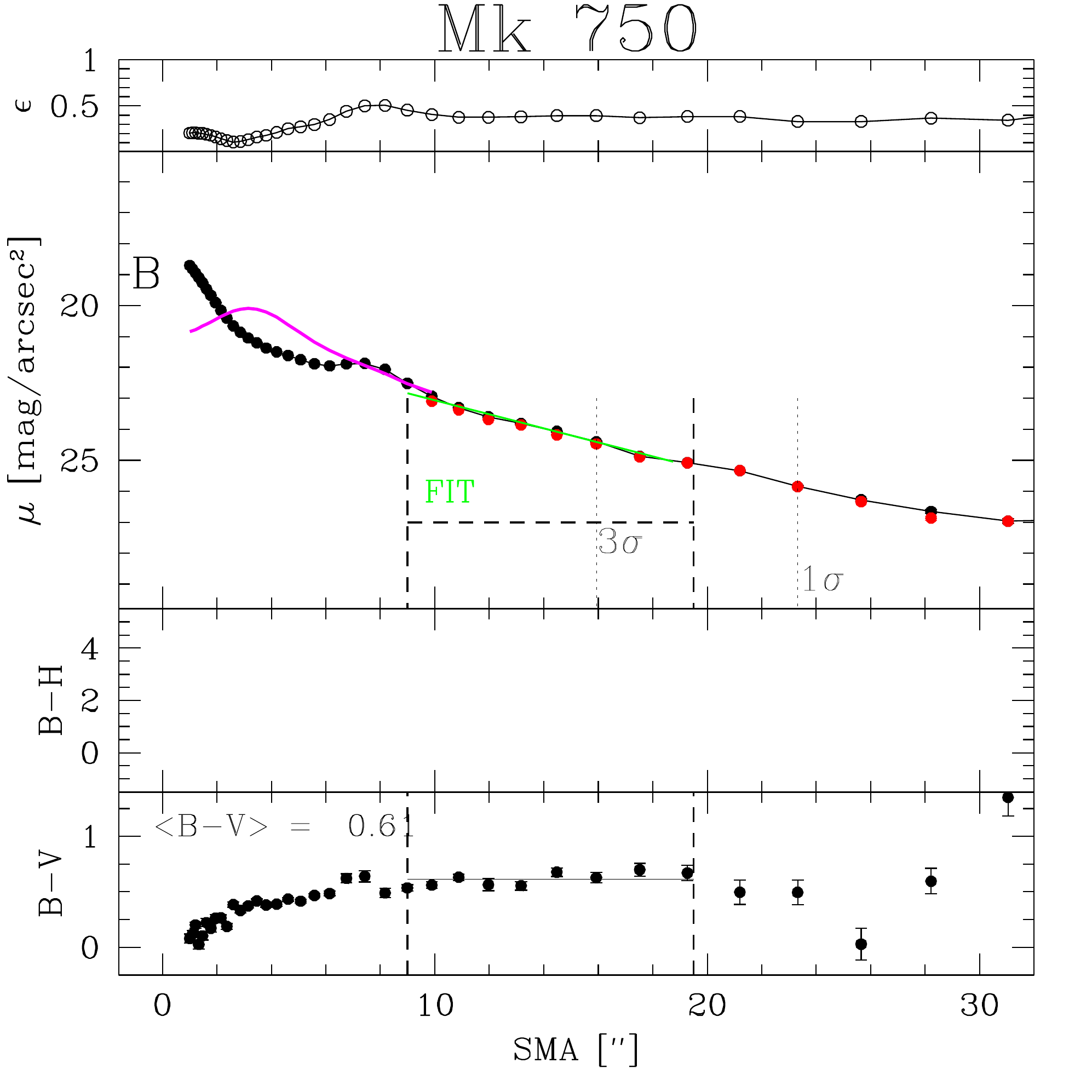} 
\caption{continued}
\end{figure*}

\begin{figure*}
\figurenum{2}
\centering
\epsscale{0.95}
\plottwo{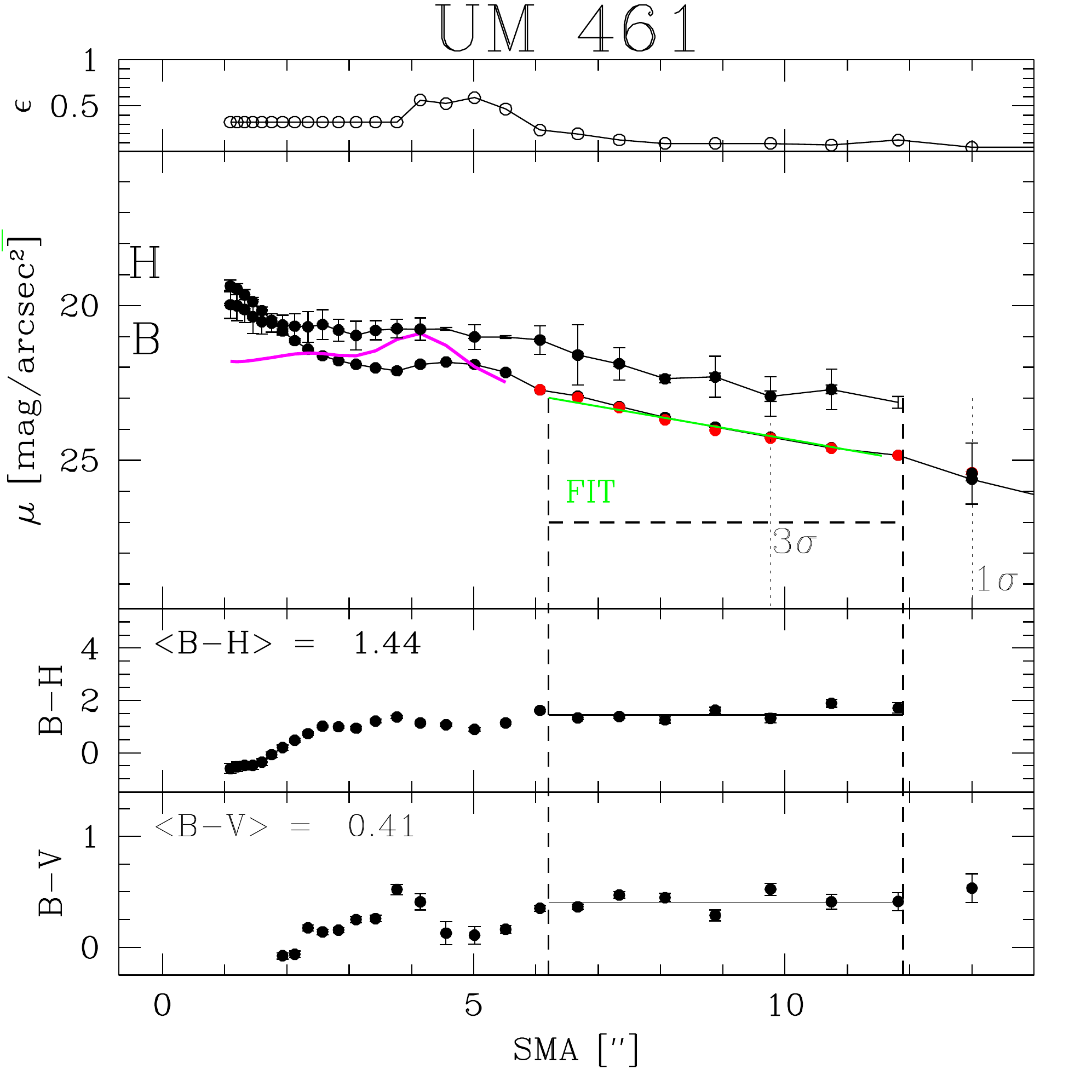}{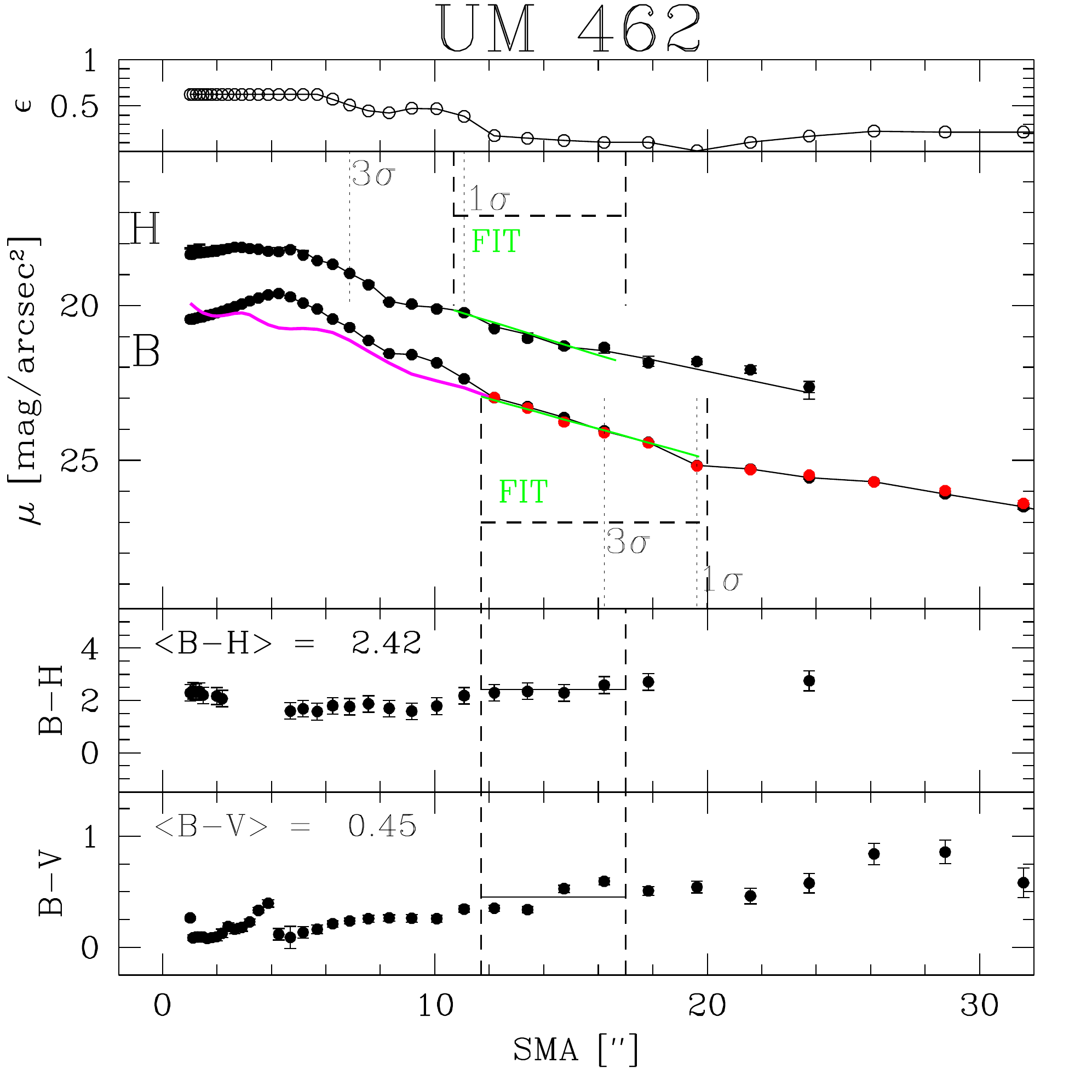} 
\plottwo{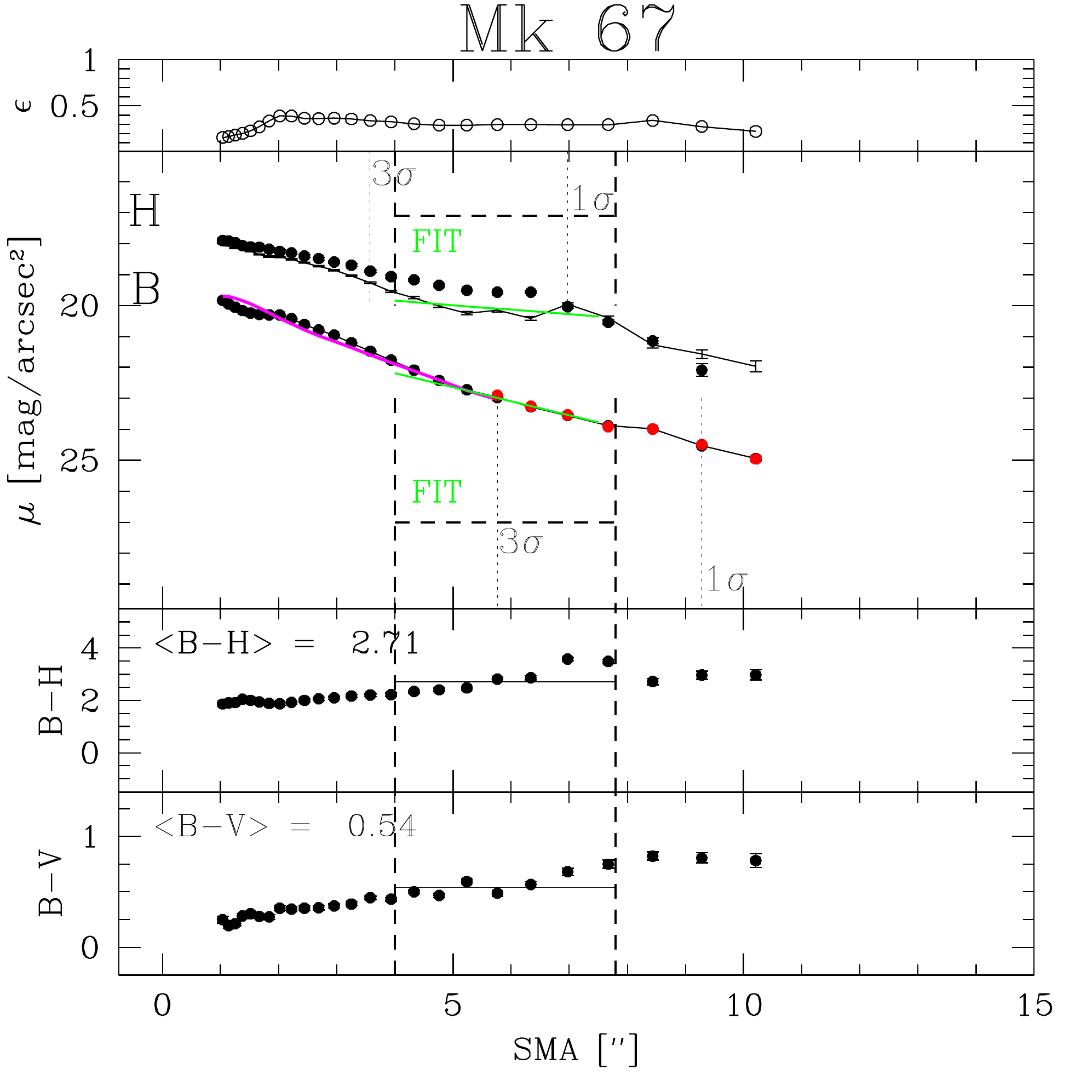}{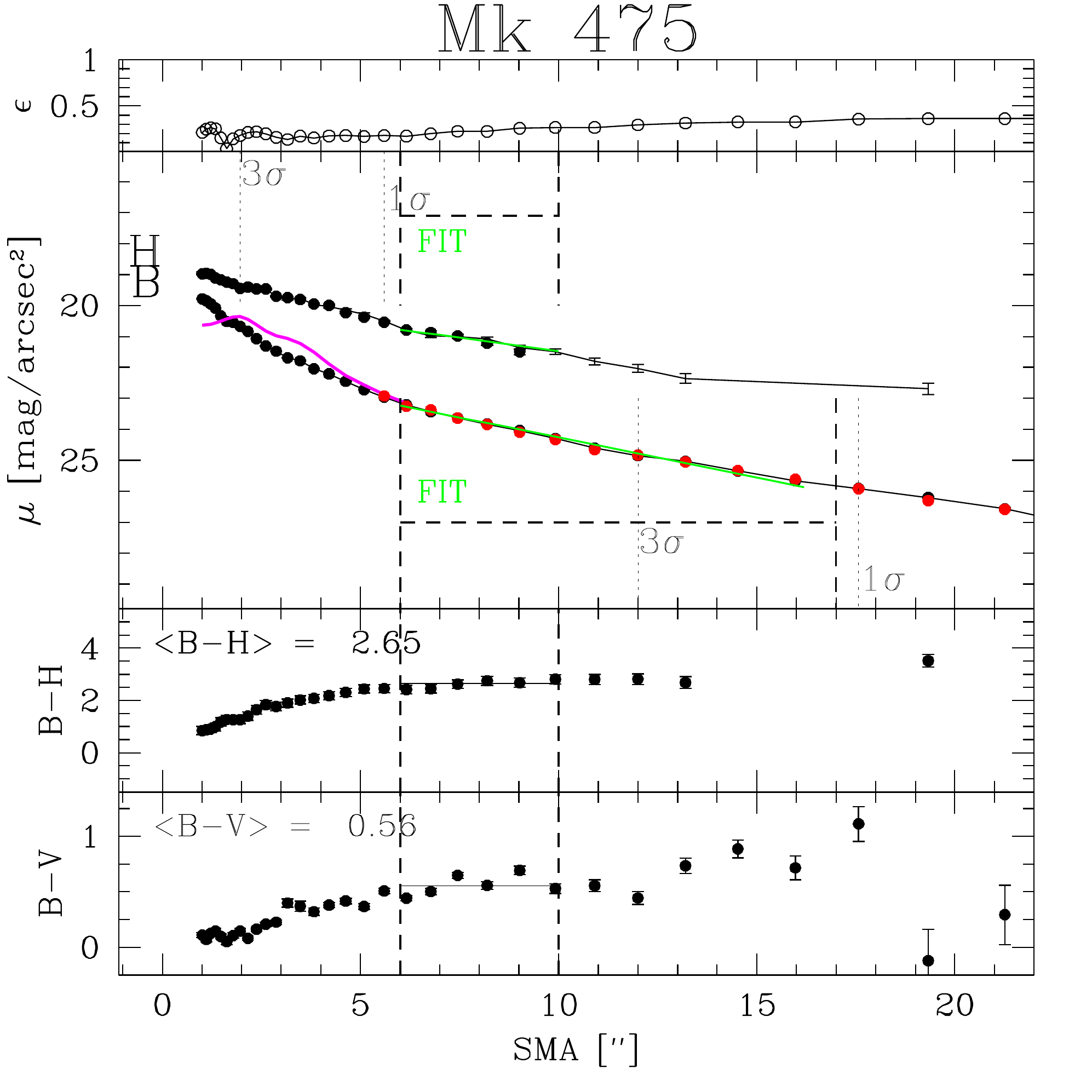} 
\plottwo{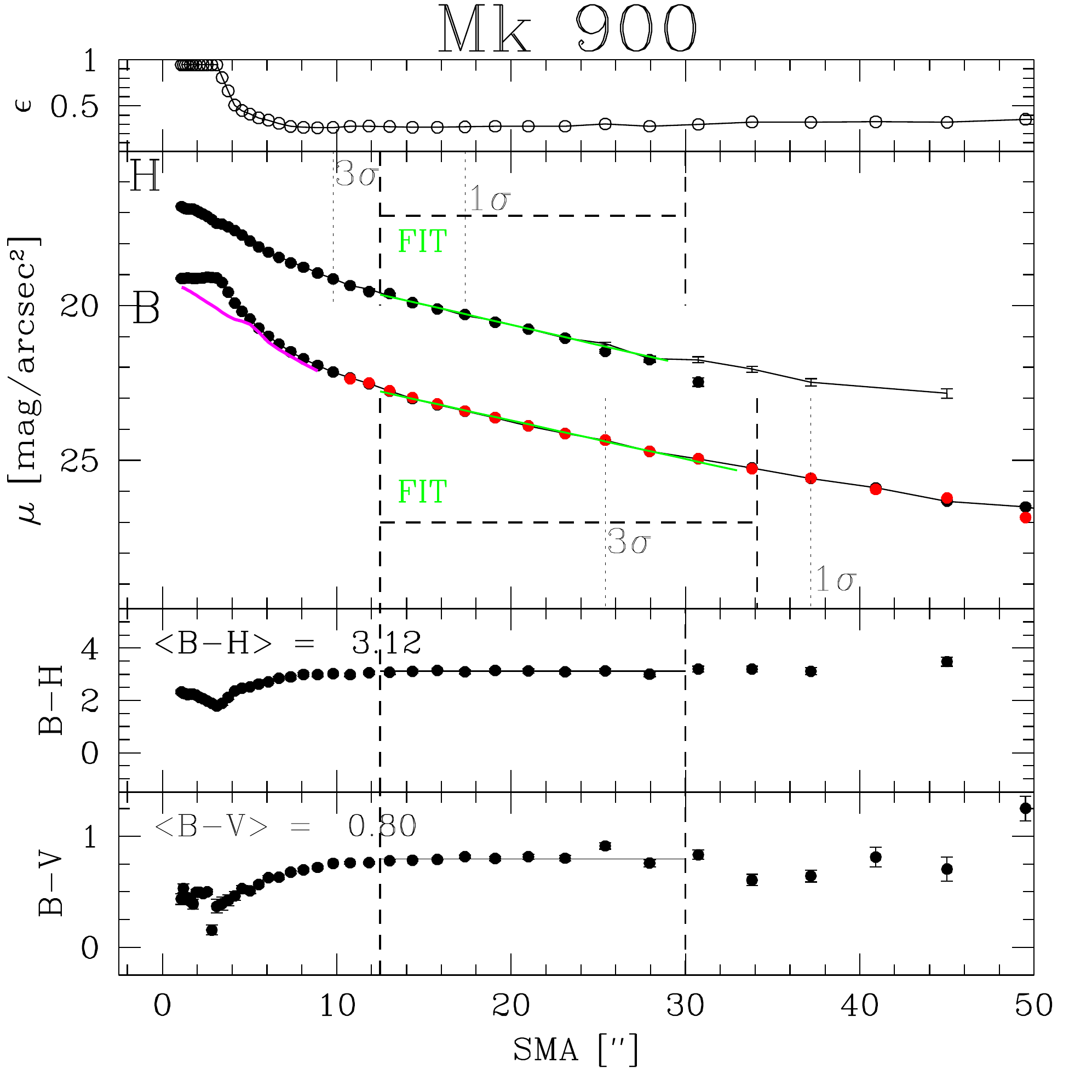}{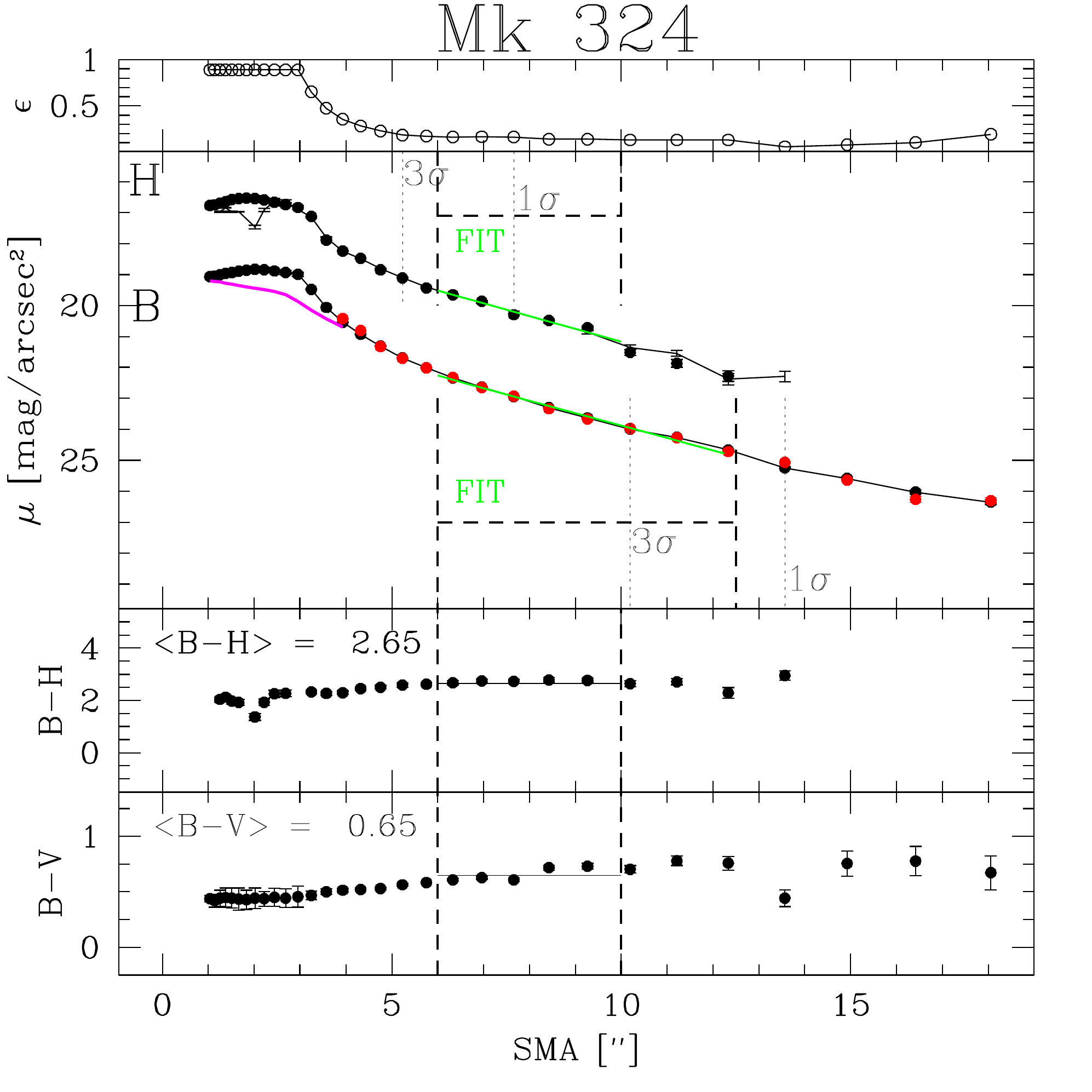} 
\caption{continued}
\end{figure*}

\begin{figure}[h]
\figurenum{2}
\centering
\epsscale{0.99}
\plotone{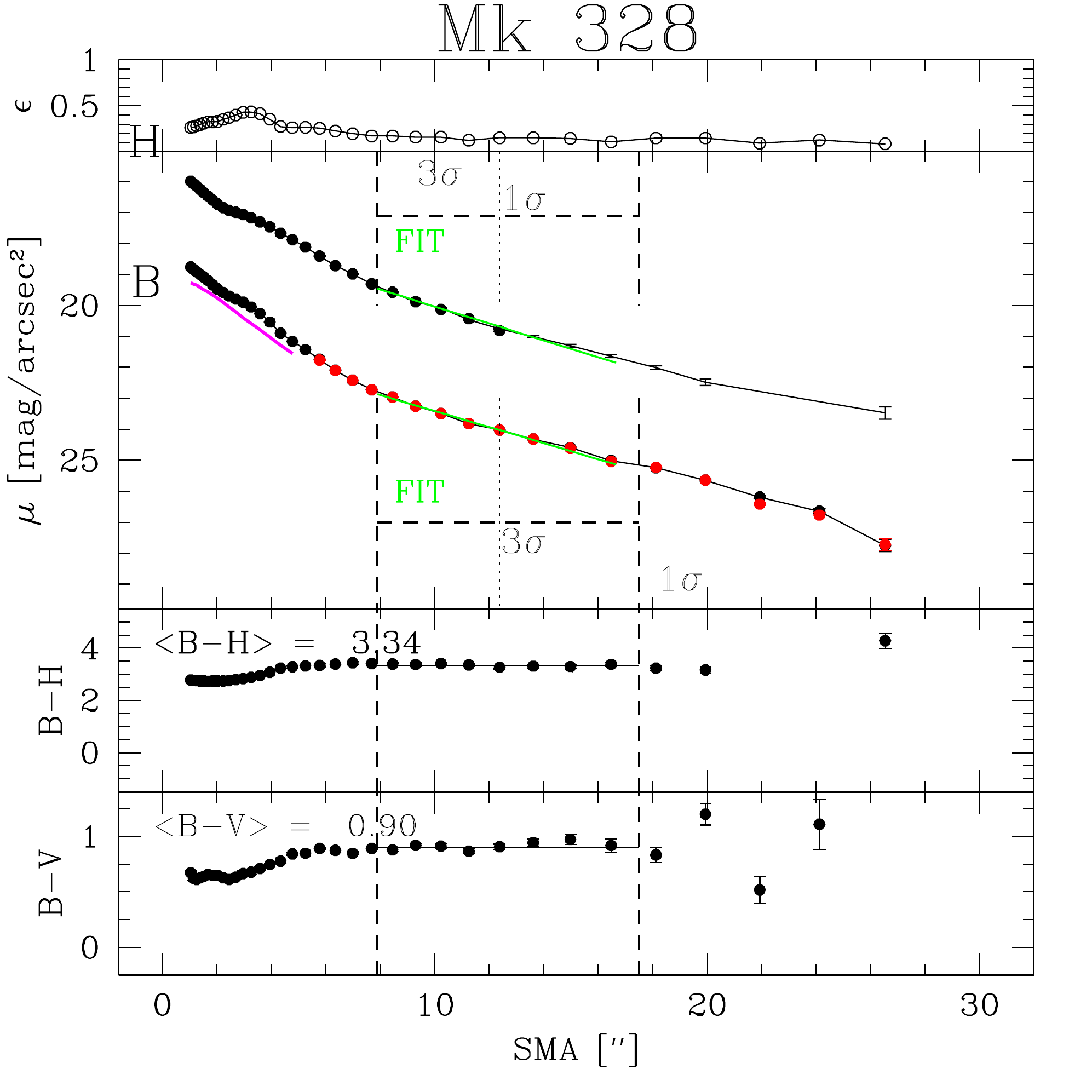} 
\caption{continued}
\end{figure}

\section{Results}\label{results}

\subsection{Comparison Samples}

To understand where BCDs fit in the context of dwarf galaxy evolution,
we select a broad set of dwarf galaxies as a comparison sample,
drawn from a variety of sources in the literature. It
is important that { we select only studies which used similar
  surface brightness profile fitting methods to ours, and which had
  similar goals in separating the old stellar population from recent
  star formation, so that the structural parameters represent the
  underlying old stellar population.}

Our primary optical comparison sample comes from Salzer \& Norton
(1999, SN99) who observed some of these same BCDs and also 11 dIs in
$B$ and \Ha 
filters and used this same method to fit exponential profiles to the
outskirts of the 
surface brightness profiles. We supplement these comparison dIs with
other samples from the literature, each of which is briefly discussed
here.
Patterson \& Thuan (1996, PT96) studied 51 dIs and LSBs and reported
results from fits of the underlying exponential profiles in the $B$
filter. They fit the outer portions of the profile to avoid light from
star-forming regions.
Van Zee (2000, Z00)  observed a sample of isolated dIs in $UBV$ and
\Ha, and we include her structural parameters from exponential disk
fits to 50 galaxies.
Parodi \etal (2002, P+02) used observations of many late-type dwarf galaxies
to fit disk profiles in $B$ and $R$ filters. Out of the 72 dwarfs in
the sample, we use the structural parameters from the 48 that are
classified as Sm or Im.
Finally, we include structural parameters of 5 dwarf LSBs from
Pustilnik \etal (2011, P+11), and of 109 gas-rich field dwarfs from Pildis
\etal (1997, P+97).

It is also important to find an appropriate comparison sample of
near-infrared observations of dwarf galaxies. Surface photometry
from deep NIR observations of dwarf galaxies is relatively rare in the
literature. We use the large amalgamated sample of 66 dwarf galaxies
from McCall \etal 2012. These galaxies' 
surface brightness profiles are fitted with hyperbolic secant
($sech$) 
functions instead of the exponential functions used in this work,
so we carried out a simple Monte Carlo simulation to derive empirical
relations between the structural parameters of these two profile
types. We use the galaxies' best-fit 
$sech$ profile parameters to generate theoretical surface brightness
profiles with $\sim5\%$ noise, and re-fit the outskirts with 
exponential profiles. We recover the following relations between the
exponential parameters ($\mu_{0,exp}$,$\alpha_{exp}$) and the $sech$
parameters ($\mu_{0,sech}$,$\alpha_{sech}$), where $\mu_0$ is in
\magsec \, and $\alpha$ is in parsecs:

$\mu_{0,exp} = (0.996 \pm 0.001) * \mu_{0,sech} + (0.821 \pm 0.012)$

$\alpha_{exp} = (0.987 \pm 0.000) * \alpha_{sech} + (6.763 \pm 0.283)$

\noindent
Overall these fits show that it is possible to make a reliable
conversion between exponential and $sech$ fits. The slopes of the
relations are nearly unity, indicating no difference in the
relationships across the relevant parameter space. The offset of $\sim
0.8$ \magsec \, in surface brightness is expected, as the $sech$ profile
flattens out in its central regions significantly more than the
exponential 
profile. When plotted logarithmically, the $sech$ and exponential profiles
overlap at large radii, but at smaller radii the $sech$ profile
becomes increasingly fainter than the exponential profile. The overall rms
scatter in the relation for $\mu_0$ is $0.06$ \magsec, and in the
relation for $\alpha$ is $7.9$ pc. These conversions are robust
enough for this sample to be a useful comparison data set for the
structural parameters of the BCDs.

From this data set, we also correct the observed central surface
brightnesses to be face-on ($\mu_0^o$) using the inclination angle
given by McCall \etal (2012). Finally, since observations are in
$K_s$, we determine the $H$ filter equivalent of the the central
surface brightness and absolute magnitude using a color of $H-K =
0.1$, which is representative of these dwarf galaxies.
We also use the recent NIR observations from Noeske \etal (2005) who
measured optical and NIR structural parameters for a sample of 11
BCDs, 3 of which are common to our sample.

\subsection{Optical Structural Parameters}

We now consider relationships between the structural parameters
and total luminosities of the BCD hosts relative to our comparison
sample of dIs. Here the structural
parameters are determined in the outskirts of the BCDs and reflect the
structure of the underlying host galaxy, while the absolute magnitudes
are total magnitudes for the BCDs, including both host and starburst
light. The exponential fit to this underlying host galaxy has two
parameters: $\alpha$, the scale length in kiloparsecs, and $\mu_0$,
the central surface brightness corrected to face-on in magnitudes per
square arcsecond. These structural parameters only apply 
to the surface brightness profile in the outskirts, and are different
from the scale length of the overall galaxy light profile and
the central surface brightness that the overall profile reaches at the
center of the galaxy. We use 
$\alpha$ and $\mu_0$ as measures of the overall shape and structure of
the underlying host galaxy, and measure 
them on the $B$ image ($\alpha_B$ and $\mu_{0,B}$) and on the $H$
image ($\alpha_H$ and $\mu_{0,H}$).

\begin{figure*}[htb]
\figurenum{3}
\centering
\epsscale{1.1}
\plottwo{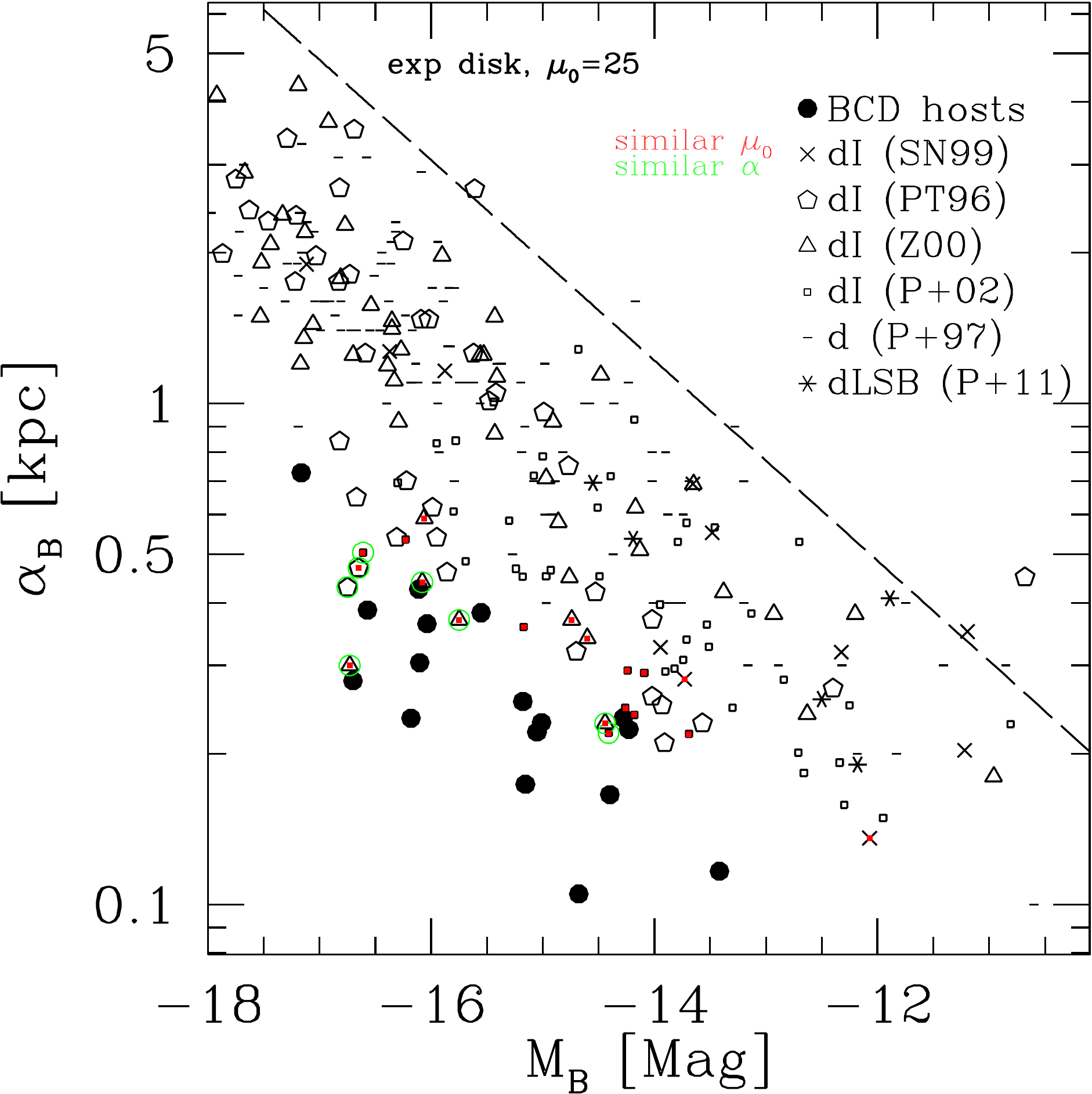}{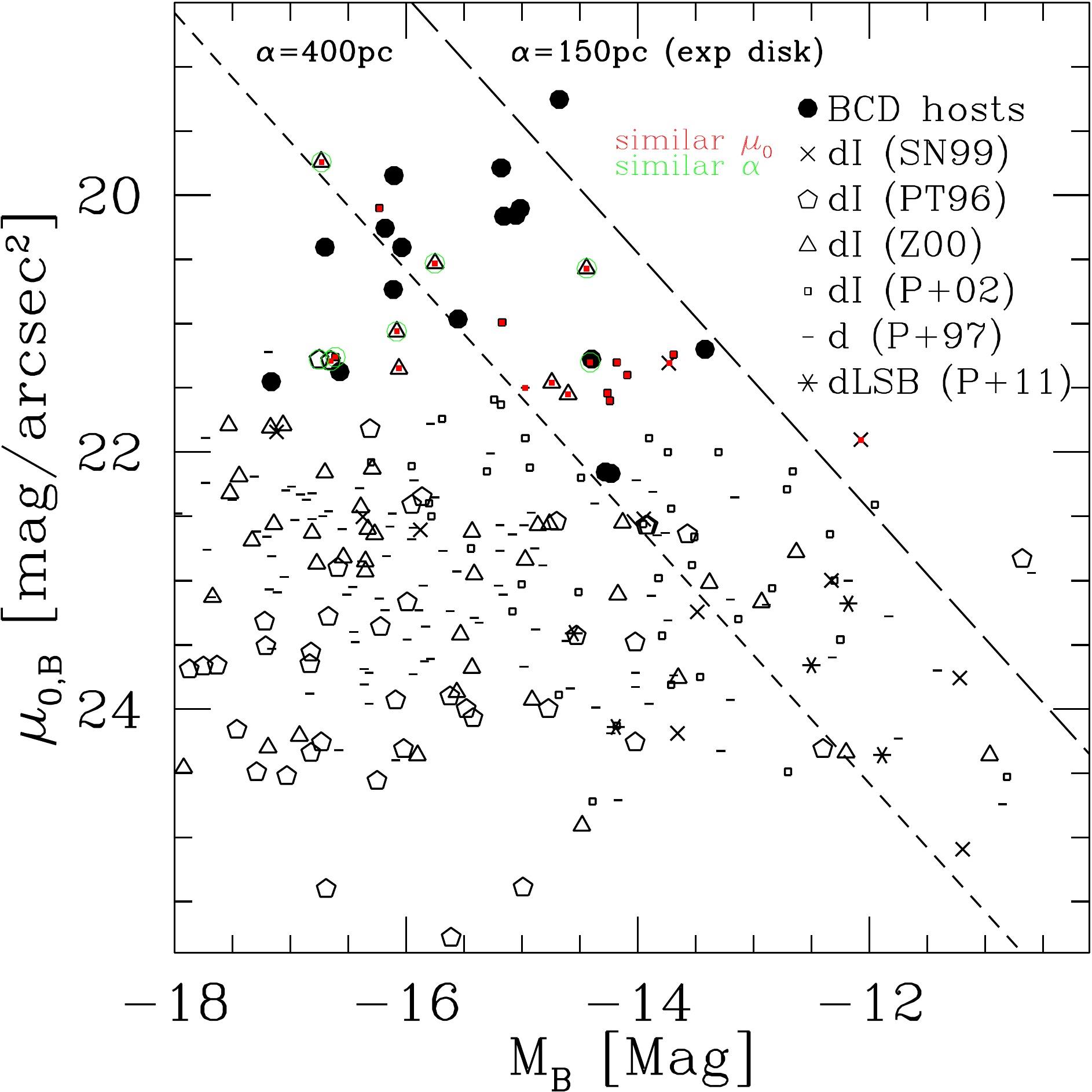}
\caption{Optical structural parameters ($B$ scale length $\alpha_B$
  and central surface brightness $\mu_{0,B}$) as a function of $M_B$
  for the underlying host galaxies of BCDs and the comparison
  samples. We 
  highlight the comparison galaxies which have similar $\alpha_B$ and
  similar $\mu_0$ to the BCD hosts with an open green circle if they
  have similar scale lengths, and an internal red dot if they have
  similar central surface brightnesses.
  Diagonal lines indicate the relationships between parameters for
  exponential disks with the given parameters.
  On the left panel, the long-dashed line shows the relationship
  between $\alpha_B$ and $M_B$ for a pure exponential disk with
  $\mu_{0,B} =  25$ \magsec.
  On the right panel, the long-dashed line shows the relationship
  between $\mu_{B,0}$ and $M_B$ for a pure exponential disk with
  $\alpha_B = 150$ pc, and the short-dashed line shows an exponential
  disk with $\alpha_B = 400$ pc.
  \label{opt}  }
\end{figure*}

Figure \ref{opt} shows the relationship for our BCDs and comparison
sample galaxies between the optical structural
parameters $\alpha_B$ and $\mu_{0,B}$ and the $B$ luminosity, $M_B$.
Many groups have used these types of graphs to study the structural
parameters of a wide variety of galaxies since these were first
introduced (Kormendy 1977). Graham (2013) summarizes the historical and
modern scaling relations for a variety of galaxies, and describes
these two relationships in particular.

The relationship between scale length and luminosity is not a simple
power law across the wide range of galaxy masses from dwarf to giant
(Binggeli \etal 1984), but when considering samples restricted to
dwarf galaxies, the relationship is well fit with a power law (Lisker
2009 for dEs, Sharina \etal 2013 for dIs). Our plot of $\alpha_B$ vs
$M_B$ shows a clear trend for the more luminous comparison dwarf
galaxies to have larger scale radii. The underlying host galaxies of
the BCDs have scale
radii which are unusually small for their luminosities, and seem to
occupy one edge of the parameter space defined by the comparison
galaxies. In fact, the underlying host galaxies of the BCDs are
displaced from the comparison 
sample in the sense that they have smaller scale lengths for their
luminosity, which, assuming an exponential disk profile, implies they
will also have brighter central surface brightnesses. The left panel
of Figure 
\ref{opt} also includes a dashed line showing the relationship between
$\alpha_B$ and $M_B$ for an exponential profile with $\mu_{0,B} = 25$
\magsec. This central surface brightness is typical for very low
surface brightness galaxies, and as they are difficult to
observe, the comparison sample contains very few galaxies in this
region. The underlying host galaxies of the BCDs are found on the
opposite side of this 
parameter space, indicating that they have exceptionally bright
central surface brightness.

The relationship between luminosity and central surface brightness has
been characterized with a single power law from $-12 > M_B > -20$
(Binggeli \etal 1984, Binggeli \& Cameron 1991), but appears to 
deviate brighter than $M_B \sim -20$. Our relationship 
between $\mu_{0,B}$ and $M_B$ for the comparison dwarf galaxies shows
a large scatter across much of parameter space. Even so, the
underlying host galaxies of the BCDs have central surface brightnesses
which are at and beyond 
the bright edge of the parameter space defined by the comparison
galaxies. As with the scale length relationship, there are
implicit restrictions on this parameter space. To show these limits,
the right panel of Figure \ref{opt} includes lines indicating the
relationships 
for exponential disks with particular scale lengths. The comparison
sample does not include more than a few galaxies with scale lengths
smaller than 
$\alpha_B \sim 150$ pc, as systems much smaller than that are not
likely to be classified as galaxies. Here again, the underlying host
galaxies of the BCDs are
near the edge of parameter space with their unusually bright central
surface brightnesses and small scale lengths.

Both of these relationships for the underlying host galaxies of BCDs
have been discussed 
previously in great detail, { but never with as simple and
  direct a method as we employ. Even with the variety of analysis and
  fitting methods used, previous works have reached similar
  conclusions to those of this work } (Papaderos \etal 1996,
Amor\'{i}n   \etal 2009, Gil de Paz \& Madore 2005). In 
particular, Papaderos \etal (1996) found that BCD hosts have
$\mu_{0,B}$ enhanced by $\sim1.5$ mag and have $\alpha_B$ which are a
factor of $\sim 2$ smaller than comparison dIs of similar $M_B$. Our
comparison sample of dIs in the same luminosity range as the BCDs
($M_B = -17$ to $M_B = -14$) has a median $\mu_{0,B} = 22.85$ \magsec
\, with a standard deviation of $1.03$ \magsec. The underlying hosts
of our BCD sample have a median that is nearly 2.5 magnitudes
brighter, at $\mu_{0,B} = 20.40$ \magsec \, with a standard deviation of
$0.83$ \magsec. In terms of scale length, our luminosity-matched
comparison sample has a median $\alpha = 1.01$ kpc with a standard
deviation of $0.68$ kpc, while the BCD hosts in our sample have a
median $\alpha = 0.24$ kpc, with standard deviation $0.15$
kpc. { However, these median statistics are only intended to
  roughly quantify the significant differences between the structure
  of the BCD hosts and the structure of the comparison galaxies. As is
  apparent from Figure \ref{opt}, there is a broad and continuous
  distribution of structural parameters with a wide variety of values
  for dwarf galaxies. The structural parameters of the BCD hosts are
  distinctly at one extreme of this continuum.}

\begin{deluxetable*}{ccccl}
\tablewidth{0pt}
\tablecaption{BCD-like comparison galaxies}
\tablehead{\colhead{ID} & \colhead{$M_B$ [mag]} & \colhead{$\alpha_B$ [kpc]} & \colhead{$\mu_{0,B}$ [\magsec]} & \colhead{Reference}}
\startdata
     UGC 1104   &   -16.08   &    21.06    &    0.44 & Z00  \\
 CGCG 007-025   &   -15.75   &    20.53    &    0.37 & Z00  \\
     UGC 5288   &   -14.44   &    20.57    &    0.23 & Z00  \\
     UGCA 439   &   -16.73   &    19.74    &    0.30 & Z00  \\
     UGC 2905   &   -14.41   &    21.30    &    0.22 & P02  \\
     NGC 2915   &   -16.61   &    21.26    &    0.50 & P02  \\
    UGC 00772   &   -16.65   &    21.29    &    0.47 & PT   \\
    UGC 03860   &   -18.72   &    19.67    &    0.60 & PT   \\
     D640-15    &   -19.09   &    20.43    &    1.60 & P+97 \\
\enddata
\label{similar}
\end{deluxetable*}
We also indicate in Figure \ref{opt}
the comparison galaxies that have similar structural parameters to
the underlying host galaxies of the BCDs. These ``BCD-like''
comparison galaxies have 
structural parameters within $1\sigma$ of the locus of BCD host points
on each graph, and we find 9 of them, as listed in Table
\ref{similar}. 
While a detailed followup of these 
galaxies is outside the scope of this work, we do note that they
typically have blue colors consistent with star forming dIs (van Zee
2001). These galaxies could represent candidates for post- or
pre-burst BCDs, which are not currently experiencing an intense
starburst, but which still have the unusually concentrated underlying
stellar distribution. {In this way, morphology alone is not
  enough to classify a dwarf galaxy as a BCD or as a dI. Without
  considering the compactness of the underlying light distribution,
  some BCDs may be classified as dIs and some dIs may be classified as
  BCDs.} 
It may be that a concentrated underlying host is
a necessary requirement for a galaxy to become a
BCD, and the concentrated distribution may persist after the BCD
starburst has aged and faded.

\subsection{Near-infrared Structural Parameters}

{
Next we look at the relationships between the NIR structural
parameters ($\alpha_H$ and $\mu_{0,H}$) and the overall $H$
luminosity, $M_H$. Figure \ref{nirH} shows these relationships for our
BCDs and comparison samples.
\begin{figure*}[hbt]
\figurenum{4}
\centering
\epsscale{1.1}
\plottwo{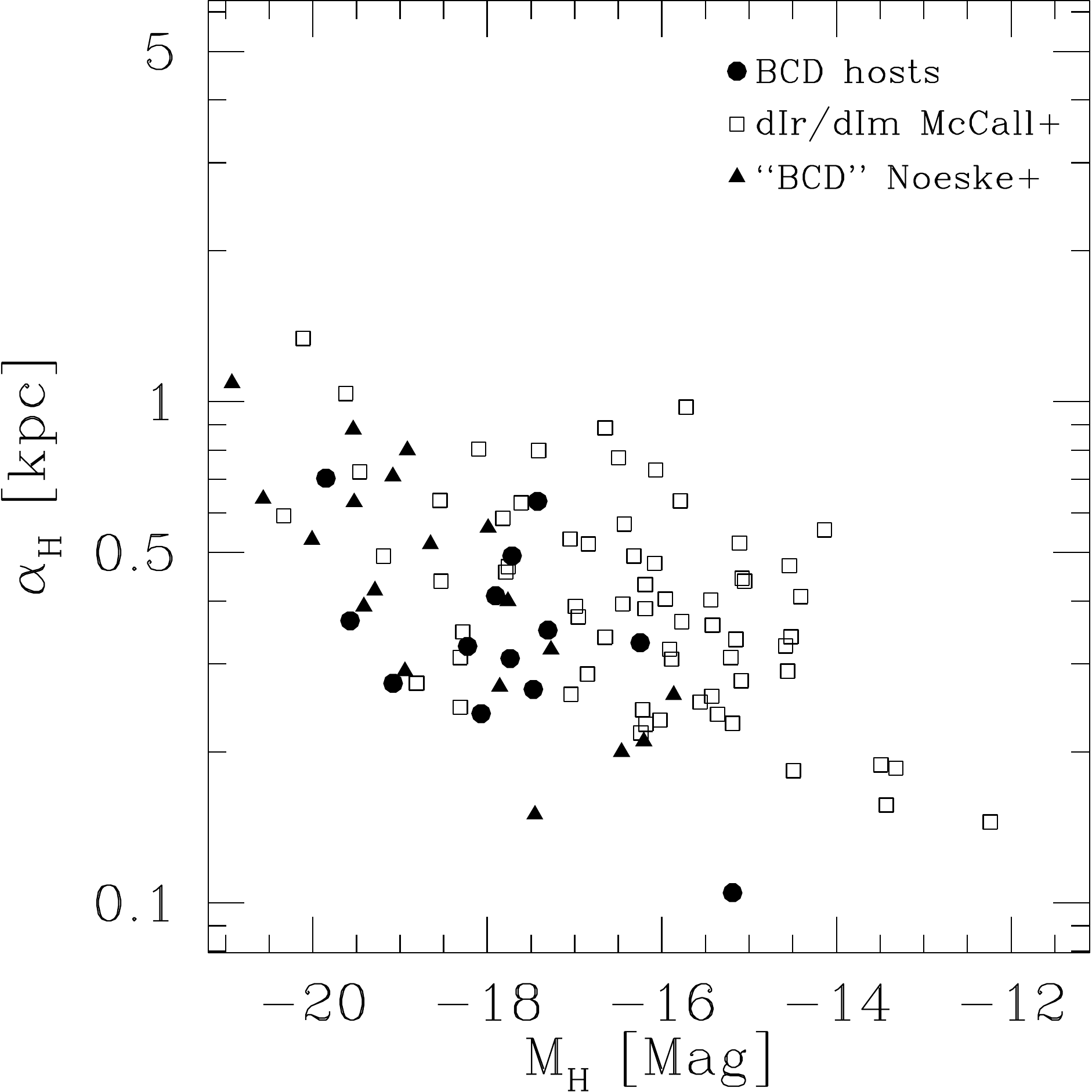}{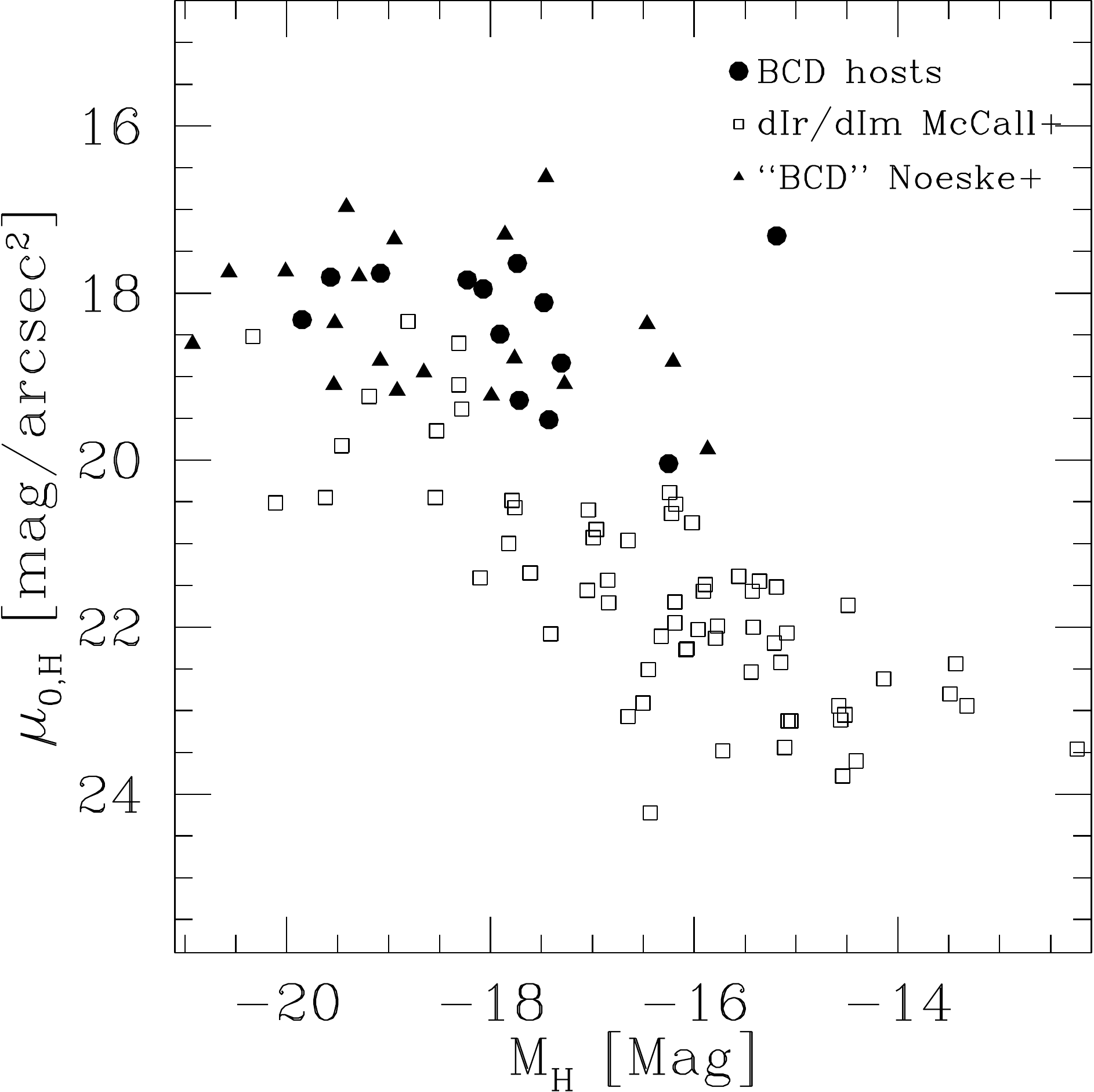}
\caption{NIR structural parameters ($H$ scale length $\alpha_H$ and
  central surface brightness $\mu_{0,H}$ as a function of $M_B$ for
  the underlying host galaxies of the BCDs and the comparison
  samples.
  \label{nirH}  }
\end{figure*}
The trends with NIR structural parameters are similar to those seen in
the optical structural parameters. The dIs form a sequence in their
NIR structural parameters such that BCDs at a given absolute $H$
magnitude typically have smaller $\alpha_H$ than the comparison
sample, although the slope 
is shallower than it was for the $\alpha_B$ relationship. The
$\mu_{0,H}$ of the dIs also gets fainter for fainter absolute $H $
magnitudes, but with a stronger trend than for the $\mu_{0,B}$
relationship. In both plots the BCD sample (both ours and those of
Noeske 
\etal 2003) exists at an extreme of parameter space. Just as seen with
$\alpha_B$, the BCD hosts have smaller $\alpha_H$ for a
given $M_B$ than the dIs, and the BCD hosts have 
brighter $\mu_{0,H}$ than the dIs. Even within the broad continuum of
NIR structural parameters, the BCD hosts are exceptionally compact.
}

%

\subsection{Optical and Infrared Structural Parameters}

\begin{figure*}[htb]
\figurenum{5}
\centering
\epsscale{1.0}
\plottwo{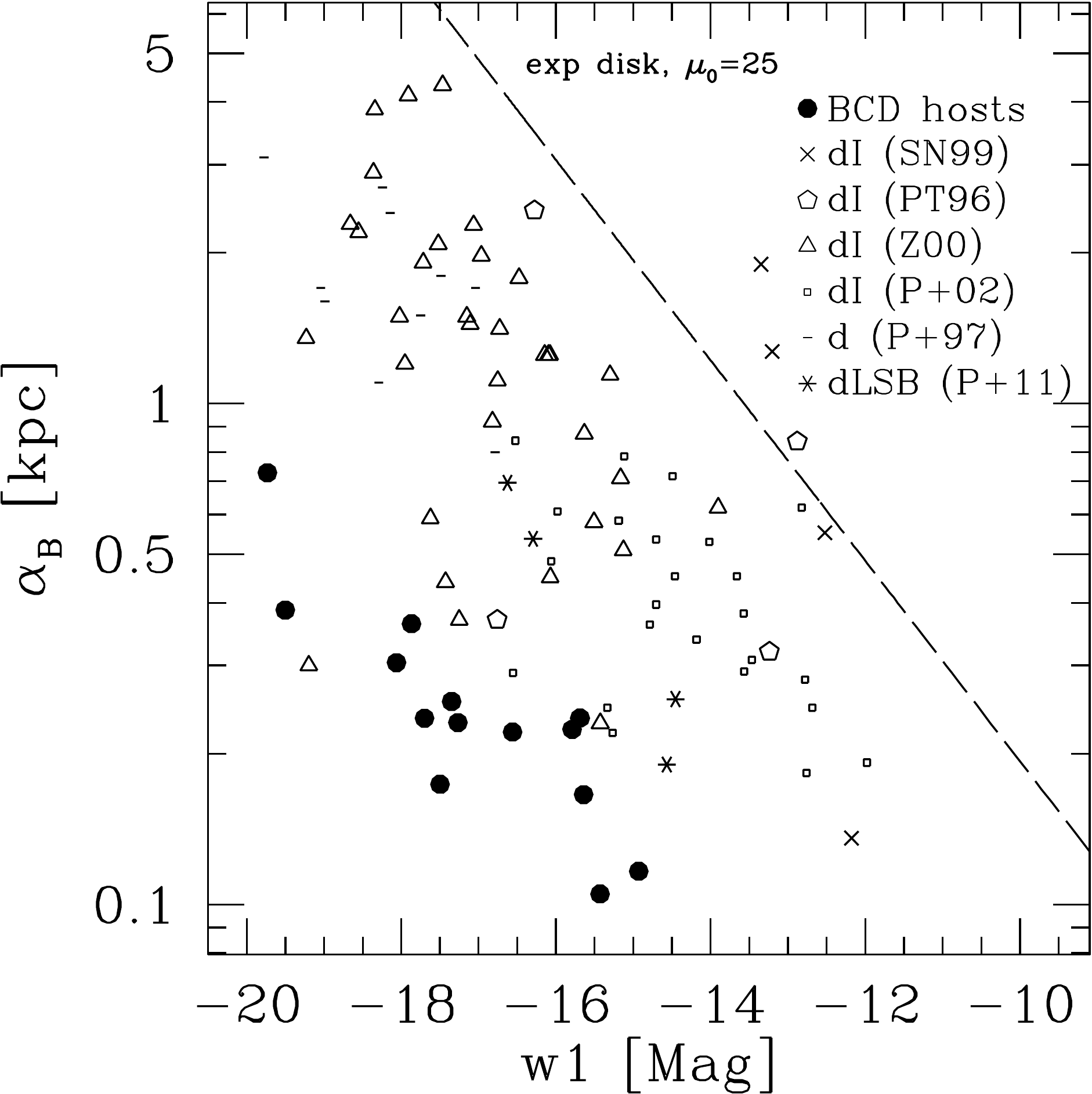}{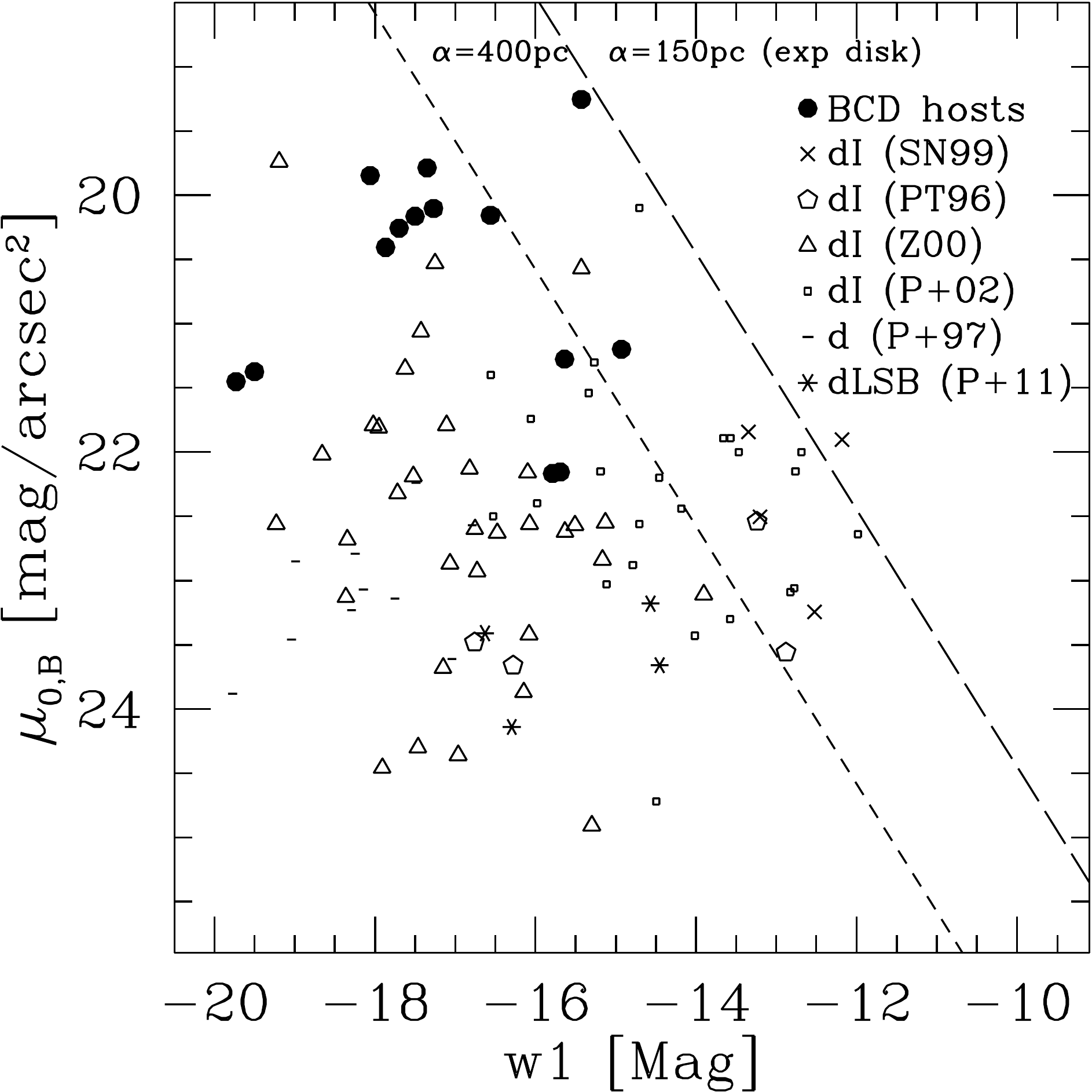}
\caption{Optical structural parameters ($B$ scale length $\alpha_B$
  and central surface brightness $\mu_{0,B}$) as a function of $M_{w1}$
  for the underlying host galaxies of BCDs and the comparison
  samples. 
  Diagonal lines indicate the relationships between parameters for
  exponential disks (of color $B-H=0$) with the given parameters.
  \label{wisegraph}  }
\end{figure*}

In order to more directly probe the relationship between structural
parameters and stellar mass of the underlying hosts of BCDs, we
consider the aperture magnitudes from the WISE catalog of photometry
(Wright \etal 2010), for all of our BCD sample and comparison
sample. In Figure \ref{wisegraph}, we show the relationship between
the $B$ structural parameters ($\alpha_B$ and $\mu_{0,B}$) and the
absolute luminosity in the $w1$ band of WISE.
Not all of the comparison sample dIs have been measured in WISE, owing
to their apparent faintness. For those galaxies which were
well-measured by WISE, we determined their absolute magnitude in $w1$
with the same distance used to determine $M_B$. The two graphs shown
in Figure \ref{wisegraph} are analogous to the optical structural
parameters shown in Figure \ref{opt}, but now we use $w1$ instead of
$M_B$. The $w1$ filter, at $3.4\mu$m, is more directly measuring the
stellar mass, while the $B$ filter measurement can be significantly
affected by the starbursts in the BCDs.

It is important to note that the trends apparent in the graphs of
optical structural parameters persist into the graphs with $w1$
luminosities. In fact, the trends may even be stronger when measured
this way. In the relationship between $\alpha_B$ and luminosity, when
we plot versus $w1$, the BCDs actually move further away from the
comparison sample than they were in $M_B$. When plotted in terms of
$M_B$, at a given scale length, a BCD host is on average $\sim 3$ mag
brighter than a dI galaxy. This offset increases to an average
separation of $\sim 4$ mag when considering $w1$ instead of
$M_B$. In the relationship between $\mu_{0,B}$ and luminosity, the
change from $M_B$ to $w1$ does not significantly change the position
of the BCD hosts relative to the dIs. The BCD hosts already had
significantly higher central surface brightnesses than the dIs, and no
luminosity shift can alter that. Overall, since the infrared
luminosity makes the BCD hosts even more distinct, we can be sure that
the unusual structural parameters of the BCDs are not just a result of
their current bursts of star formation. It is clear that the
underlying old stellar host is fundamentally different in BCDs.

\subsection{Color-color profiles}
\label{ccp}

\begin{figure*}[htb]
\figurenum{6}
\centering
\epsscale{0.80}
\plottwo{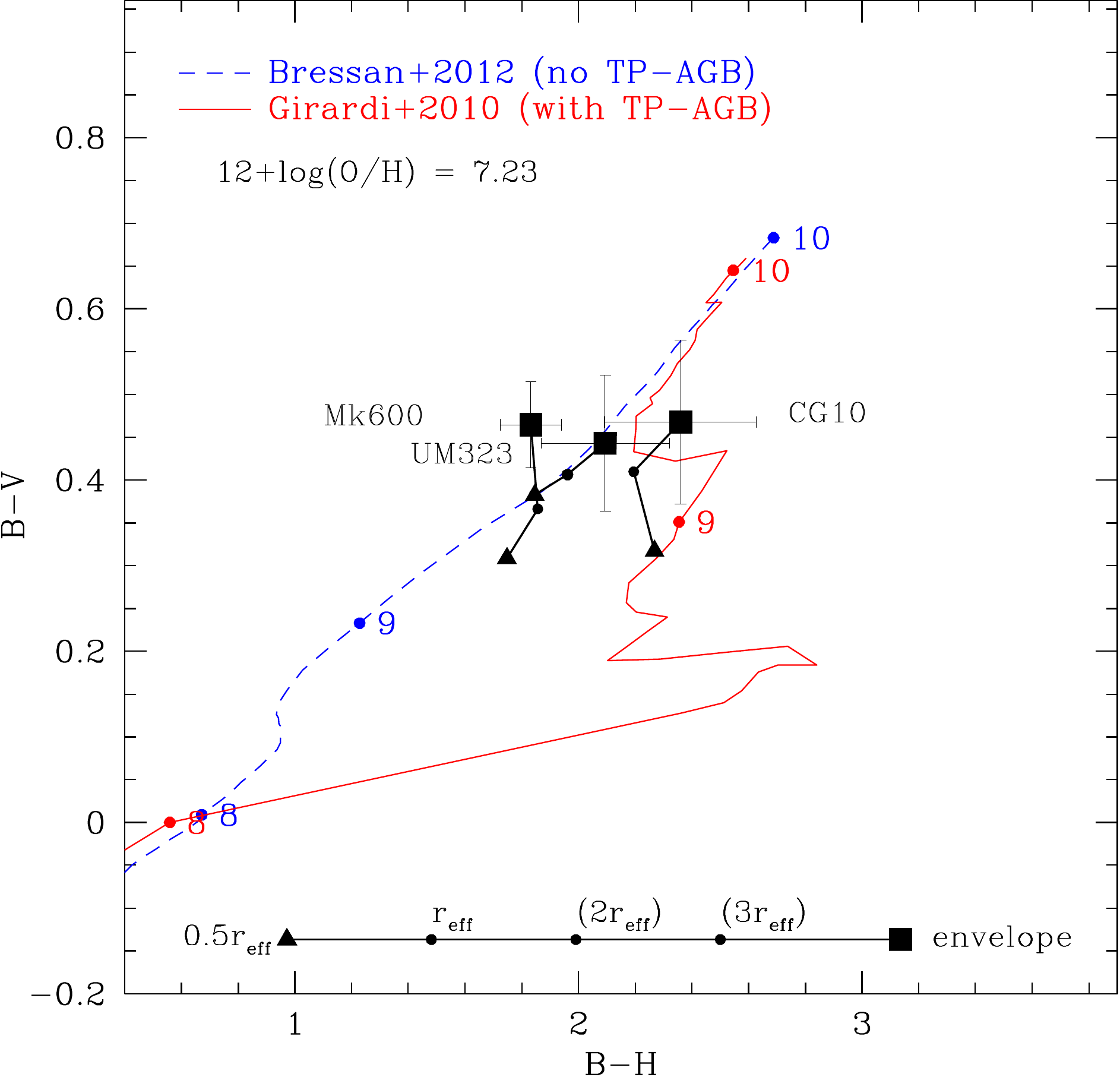}{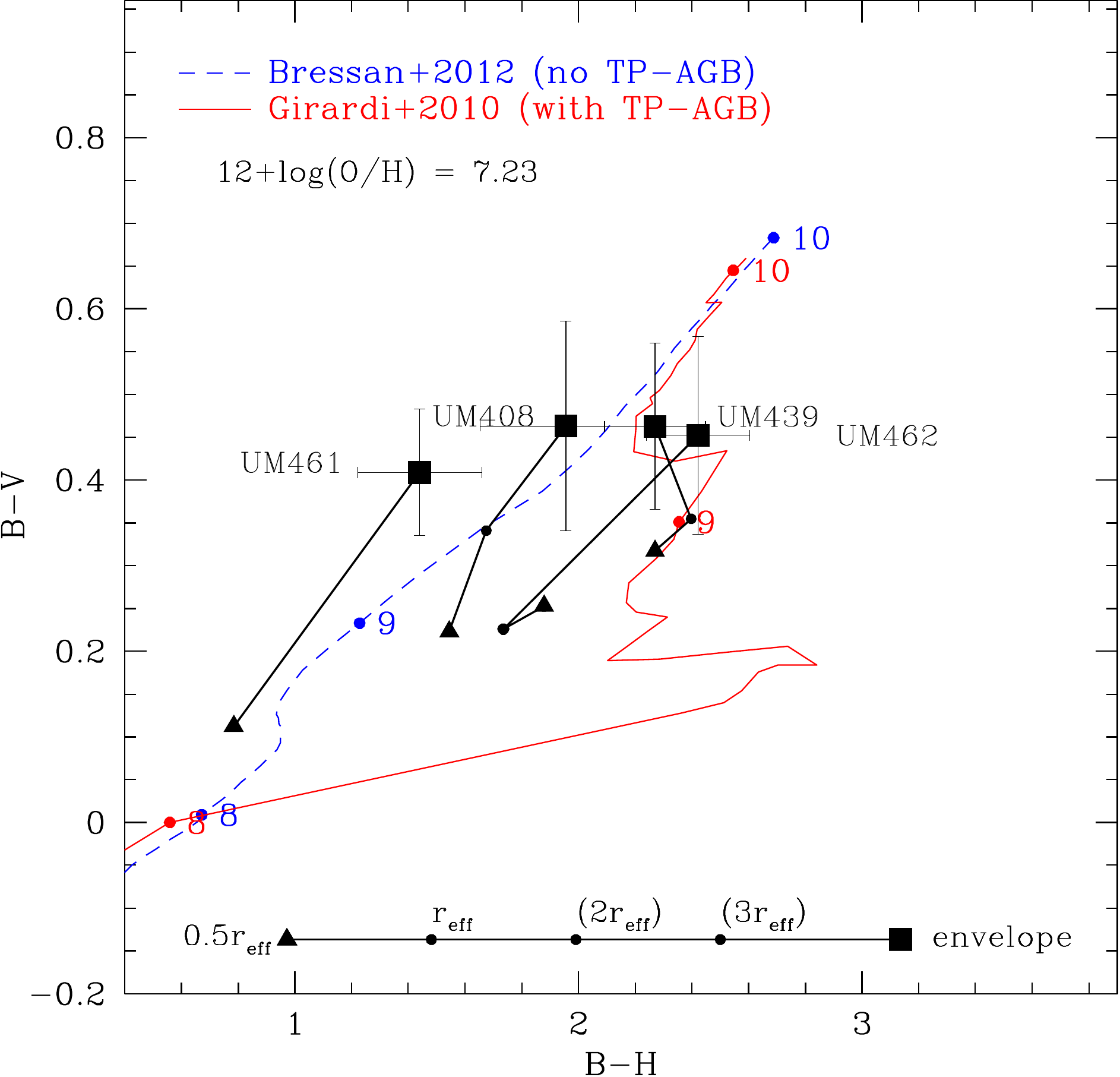}
\plottwo{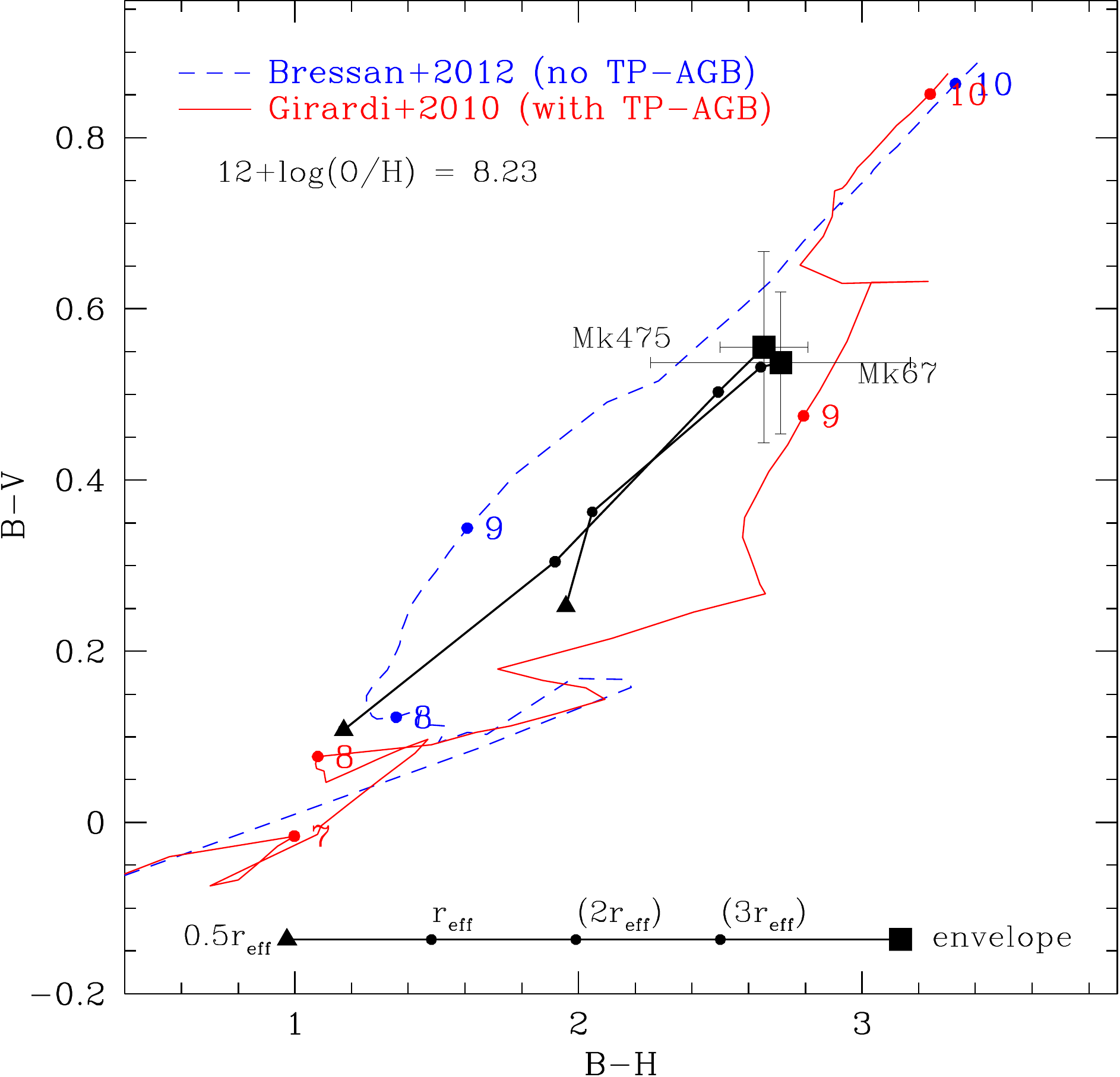}{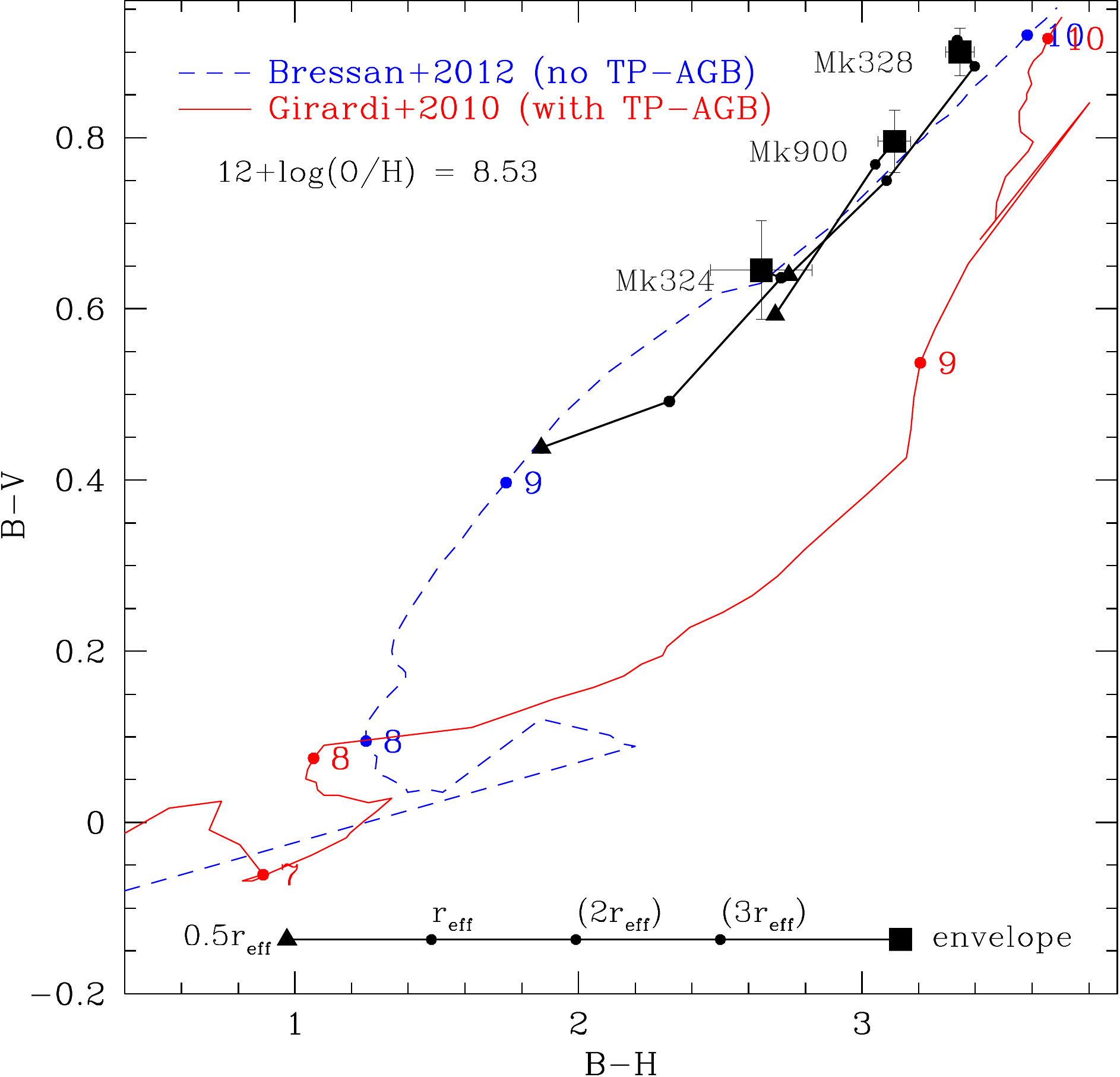}
\caption{Color-color diagrams showing $B-V$ and $B-H$ colors at
  different radii for the BCDs. Black lines connect the color profiles
  of the BCDs between half of their effective radius (black
  triangles), through one, two, or three effective radii (small black
  circles), out to the region used to fit the underlying host galaxies
  of the BCDs
  (solid squares). Also shown are SSP single burst models (blue and
  red lines) and the numbers show the logarithm of the age (in years)
  of the 
  population at that point. Different panels show different
  metallicity BCDs and models.
  \label{BVBH}  }
\end{figure*}

We use the {matched color profiles shown previously in Figure
  \ref{sb} to construct a modified surface color-color diagram} shown in
Figure \ref{BVBH}.
We divide our BCD sample into groups based on their metallicity, and
show the galaxies with metallicities $Z \approx 7.23$ in the
top two panels: those with $Z \approx 8.23$ in the bottom left panel,
and those with $Z \approx 8.53$ in the bottom right panel. For each
galaxy {we first plot a triangle at the $B-H$ and $B-V$ colors
  of its radial brightness profile at one half of the effective
  radius. Next we plot the colors at one effective radius, two
  effective radii, and so on, until we reach the region where the
  underlying host profile is being fit. The final square point for
  each galaxy shows the average $B-H$ and $B-V$ colors of the outer
  envelope where we are fitting the structure of the underlying host
  galaxy. The error bars on the final point show the standard
  deviation of all isophotes within the region where the exponential
  profile is fit.}
We do not plot the central
surface color since the $B$, $H$, and SDSS images often
have incompatible spatial resolutions. 

Also shown are evolutionary tracks from Simple Stellar
Population (SSP) models (Bressan \etal 2012, Girardi \etal 2010,
obtained at: \texttt{http://stev.oapd.inaf.it/cmd}). The Bressan \etal
(2012) models do not include contributions from Thermally-Pulsating
Asymptotic Giant Branch (TP-AGB) stars, while the Girardi \etal (2010)
models do include TP-AGB stars. Models are generated at metallicities
that match the gas-phase abundances of the BCDs and are shown in the 
appropriate panels. These $B-V$ vs $B-H$ evolutionary tracks do not
include any non-stellar effects in the light output of the BCDs,
such as dust absorption, hot dust emission, nebular emission lines,
nebular continuum,  etc. While dust can have significant effects on
galaxy colors, the low metallicities of these systems suggest that
they will not have very much dust. Most of our BCDs have strong
emission lines (notably H$\alpha$, but also [OIII] 5007 and others), which
may affect the broadband colors particularly in their inner
regions. Emission-line contamination should be negligible in the
underlying host colors as measured in the outermost isophotes. We take
these SSP model tracks as representative of stellar evolution in these
systems, but do not expect exact agreement with our observations.

The BCDs show colors indicative of younger stellar populations
near their centers, and colors consistent with older populations in
their outskirts. This old population in the outskirts has been
identified earlier, but
is confirmed more robustly with this comparison to actual SSP
models, {even though these simple models do not include
  non-stellar emission or absorption.}
The paths that individual BCDs take in color-color space all
show radial age gradients in the sense that the average age of the
stellar population increases with increasing radius throughout the
BCD. This radial increase in average stellar age is seen all the way
to the outskirts of the BCDs, which gives some constraints on
the formation and evolution of these galaxies. This universal age
gradient might suggest that new stars are formed in the (often
irregular) inner regions of BCDs, and gradually diffuse into the
outskirts over time. Alternatively, this age gradient could mean that
$\sim 10^{9.5}$ years ago (the typical age of the outskirts), star
formation took place across the entire BCD, and gradually became more
confined to the central regions like it is today, {perhaps
  as gas outflows quenched the ongoing star formation. It is also
  possible that most of the evolved stars were formed in an early
  star formation event and the younger average ages near the center
  are a result of recurrent starburst events producing young stars. }
Still, even at $0.5$
$r_{eff}$, the color-color diagram indicates relatively old average 
ages for the BCDs: the youngest, Mk 475, is around $10^8$ yr, but
most are around $10^{8.5} - 10^9$ yr. This underlying population age
is in stark contrast with 
the ages of the most recent star formation, {and it is
  important to note that there appears to be a significant presence of
  evolved stars even near the actively star-forming regions of these
  BCDs.}
The central regions of all
the BCDs have significant \Ha emission, which means O and B stars must
be present to ionize the neutral hydrogen gas. Those O and B stars
only live 30 Myr or less, so the central star formation is indeed very
recent. However, it is an open question whether or not the current
burst of star formation will create enough new young stars to
significantly lower the average age of the outer regions of the BCDs.

\subsection{Burst Strength}

In order to quantify the significance of the current star formation
event, we calculate the burst strength parameter for our BCDs. Burst
strength is usually parametrized in
terms of luminosity, and is the ratio between the light coming from
the recently formed stars and the overall total luminosity of the
galaxy. In the simplest scenario, the underlying host galaxy light
is well-described by an exponential profile, while the starburst
is simply an addition of light near the center. It is more complex to
determine the burst strength in real galaxies. Salzer \& Norton (1999)
observed a similar sample of BCDs and dIs, and derived surface
brightness profiles for both galaxy types. In the same manner as this
study, they 
fit the underlying host galaxy light with exponential profiles, and
integrated those fits to derive total luminosities of the underlying
host. However, as many groups have noted 
(e.g. McCall \etal 2012), the surface brightness profiles of dIs tend
to flatten near their centers, so the extrapolation of the exponential
profile tends to overestimate the observed central surface
brightness. In typical non-starbursting dIs, the integrated luminosity
of the exponential fit to the host is on average $\sim 0.7$ mag
brighter than the total observed luminosity of the galaxy (Salzer \&
Norton 1999). To meaningfully determine burst strength in our sample
of BCDs, we must remove this average offset to correct the burst
strength to its true value. {In fact, we find that in many
  cases, an extrapolation the exponential fits of the underlying
  hosts of the BCDs would predict a brighter
  central surface brightness than the observed value. Just as we
  parametrize both the dIs and BCD hosts with a single exponential
  function, we assume both will follow the same functional form. Even
  though the BCD hosts typically have 
  shorter exponential scale lengths than dIs of similar luminosity, we
  expect them both to similarly flatten in their centers.}
In this way, our
determination of the corrected burst strength of the BCDs will 
represent the excess luminosity contributed by the starburst, without
being affected by the centrally-flattening surface brightness profile
of the underlying host.

Following this method, we integrate the exponential profile fits to
the underlying hosts of our BCDs and apply the correction factor of
$0.7$ mag to account for the central flattening of the
underlying, older population of stars. We find that the average
corrected burst strength of our sample of BCDs is $\sim 0.8$ mag, with
a range of corrected burst strengths from $0.2$ mag to $1.5$ mag
across the sample. 
This means that the star formation in our BCDs is responsible
for average luminosity enhancements of $\sim 0.8$ mag or about a
factor of $2$ in luminosity.

Papaderos \etal (1996) also determined the burst strength of their
BCDs, and found results similar to ours. Their profile
decomposition scheme was substantially more complicated, involving
three-component fits to the surface brightness profiles. Even
with this difference in methodology, they found that the components
associated with the starburst contributed an average of $\sim 0.75$
mag on top of the underlying host galaxy. Our determination of burst
strength is less dependent on assumptions about the profiles of
BCDs, since we compare host light to the total observed luminosity of
the galaxy, rather than inter-comparing components of the same
profile.

In the context of our earlier plots of structural parameters, we
stress that this burst strength is too small to account for the
displacement of the BCD hosts from the dIs. While the BCD hosts are
offset from the dIs in terms of their exponential scale length when
measured in the luminosity direction (Figure \ref{opt}), this offset
is $\sim 2$ mag. Even if the $M_B$ values of the BCDs were corrected
to remove the light from the starburst and represent only the light
from the underlying host galaxy, the BCD hosts would still be distinct
from the dIs. Similarly, in terms of central surface brightness, no
shift in the luminosity of the BCD hosts can bring them into agreement
with the dIs.

\section{Discussion}\label{discussion}


In order to understand the evolutionary processes that affect dwarf
galaxies, we have carefully measured the structural parameters of the
underlying host galaxies in our sample of BCDs and compared them with
the structural parameters of a large number of dIs. We emphasize that
the structural parameters we have measured are, to the best of our
abilities, independent of the existence of the current
starburst. { The structures of the BCD host galaxies occupy an
  extreme part of parameter space defined by the structures of
  dIs. This parameter space covers one and a half orders of
  magnitude in scale length, and $\sim 7$ magnitudes (more than a
  factor of $600$) in central surface brightness. This broad parameter
  space is bounded at one extreme by low surface brightness dwarf
  galaxies (exceptionally faint $\mu_{0,B}$ and large $\alpha_B$ for
  their luminosities), and at the other extreme by BCD hosts
  (exceptionally bright $\mu_{0,B}$ and small $\alpha_B$ for their
  luminosities). The BCD hosts are similarly distinctive in the
  near-infrared structural parameter diagrams. }

The distinctive compactness of the underlying BCD hosts is similar to
the results of van Zee \etal (2001) who found that the neutral
hydrogen gas (HI) in BCDs is also more centrally concentrated than
other gas-rich dwarf galaxies. Van Zee \etal (2001) also found that
the rotation curves in BCDs were steeper than in comparable dwarf
galaxies, indicating a more compact mass distribution, as well. More
recently, Lelli \etal (2014) studied the HI in a sample of dwarf galaxies
and found that BCDs have stronger velocity gradients in their inner
regions than non-starbursting dwarf galaxies. Lelli \etal (2014) and
van Zee \etal (2001) both discuss the persistent nature of this
compact structure, even after the current burst of star
formation has faded. The unusually compact nature of the stars and gas
in these galaxies likely points to a difference in the underlying dark
matter distribution of BCDs. Certainly the compact distribution of
luminous matter seems to be a hallmark of BCDs; the dark matter
distributions are likely to be similarly compact. { It may be
  that this compact matter distribution is the reason BCDs can host
  such intense starbursts. While BCDs were first identified for their
  intense star formation activity, the strong starbursts may merely be
  symptoms of a compact matter distribution. In this way, the
  compact mass distribution is the fundamental parameter of BCDs, not
  their intense but transient star formation events.
}

This structural distinction of the BCD hosts is important both in
terms of identifying a well-defined and meaningful sample of
present-day BCDs, and also in understanding the past
and future evolution of present-day BCDs. {There are many
  possible definitions of BCDs that use various observational criteria
  ranging from spectral diagnostics to isophotal colors (Sargent \&
  Searle 1970, Thuan \& Martin 1981, Gil de Paz \& Madore 2005). } We
  suggest that the 
truly distinctive observable characteristic of BCDs compared with dIs
is their exceptionally concentrated underlying mass distribution, as
measured by the compact light distribution. This
 {is not a new definition or suggestion, as} compactness was part
of the original discovery and classification of BCDs by Sargent \&
Searle (1970). Thuan \& Martin (1981) loosely parametrized this
compactness by requiring BCDs to have optical sizes $\sim 1$
kpc. Shortly thereafter, this
classification got even broader with the Atlas of Virgo Cluster Dwarf
Galaxies (Sandage \& Binggeli 1984), where starbursting dwarf galaxies
with ``several knots and some low surface brightness fuzz'' were
cataloged as BCDs although the authors acknowledged that ``our BCD
objects would not fall into the extremes of this class''. The
BCD classification {continued to grow} increasingly broad with
time. Loose 
\& Thuan (1986) describe BCDs as ``low-luminosity galaxies
... undergoing intense bursts of star formation'', as do more recent
groups like Lelli \etal (2012) ``...low-mass galaxies that are
experiencing a starburst''. The frequently-cited definition in
Gil de Paz \etal (2003) parametrizes compactness via the peak surface
brightness, and requires $\mu_{B,\textrm{peak}} < $22 \magsec, rather
than actually measuring the compactness of the underlying light. Gil
de Paz \etal (2003), like many authors, determine these criteria
based on their sample of ``known'' BCDs. { Many
  other authors follow this definition and substitute small size for
  compactness, and some even classify BCDs purely
  spectroscopically. As the definition of a BCD has evolved,
  the importance of compactness has been increasingly overlooked.} Our 
conclusions from this study suggest that the compactness of
the underlying light is a defining characteristic of BCDs, and is just
as important to their classification as their blue color and their
low luminosity.

Even after the current starburst fades (if $M_B$ fades
by perhaps as much as $1$ magnitude), the BCD hosts will continue to
occupy this extreme area of parameter space. Fading of $M_B$ can
bring BCD hosts closer to the distribution of normal dIs in terms of
scale length, but a change in $M_B$ does not bring the BCD hosts
closer to the dIs in terms of central surface brightness. \emph{Major
structural changes would be required in the underlying light
distribution if BCDs could transform to or from dIs.} This type of
transformation would require mechanisms that could cause 
galaxy-wide structural changes in dwarf galaxies, in particular,
changes to the structure of the underlying mass distribution. Among
the most frequent mechanisms discussed for transforming the underlying
structure of non-interacting dwarf galaxies are substantial mass loss
from galactic winds (Mac Low \& Ferrara 1999) and infall of fresh
gas. In denser environments, interactions with other galaxies can also
have transformative effects (e.g., ram pressure and tidal stripping of
dwarf spheroids in cluster environments). Ongoing secular structural
evolution within the BCDs is not expected to change the underlying
structure dramatically enough, as the time scales for internal
evolution processes are longer than the other mechanisms
discussed. {Furthermore, the energy required to remove enough
  material from the central regions to cause global structural changes
  is prohibitively large.} 

Whether this underlying structural distinction is significant enough
to truly distinguish BCDs from dIs is a more complicated question
which requires additional detailed studies across all types of
dwarf galaxies. The broad continuum of structural properties of dwarf
galaxies (as seen in the broad parameter space populated by galaxies
in both panels of Figure \ref{opt}) further complicates the
classification process, and muddles the evolutionary pathways. One
class of dwarf galaxy which may be relevant to
these evolutionary connections is the ``Transition'' type dwarf
galaxy (TTD). TTDs have been studied by
many groups (Dellenbusch \etal 2008, Koleva \etal 2013,
and references therein), and share properties with both the gas-rich
classes of dwarf galaxies (dI and BCD) and the gas-poor classes (dE
and dSph). It has been suggested that TTDs may possibly represent a
short-lived transition period between classes. In some cases, TTDs may
be classified as BCDs (e.g., a galaxy with dE-like outer regions but a
starburst at its center) or vice versa. While star formation events
can alter the appearance of dwarf galaxies, our results suggest
that transitions between ``true'' BCDs and typical dIs would require a
significant amount of structural change to the underlying hosts. We
consider such transitions unlikely.

Given the apparent permanence of the underlying compact global
structure of BCD hosts, and the short gas depletion time scales of the
current star formation in BCDs, there must 
be galaxies which have had a BCD-like phase in their past but which
are not forming stars as actively today. Indeed the galaxies we have
identified in Table \ref{similar} with similar structural parameters
to the BCD hosts may be examples of galaxies which, if starbursting,
would be classified as BCDs. 
Other groups have searched for BCDs
before or after their burst phase. S\'{a}nchez Almeida
\etal (2009) has identified a population of dwarf 
galaxies in SDSS which may be BCDs during quiescence (QBCDs). QBCDs
have some properties in common with BCDs (i.e., similar structure
and color of the outskirts) but are not necessarily 
undergoing a significant burst of star formation. While the sample
of 21,000 QBCDs used by S\'{a}nchez Almeida \etal may include some examples
of BCD-like galaxies in quiescence, they note that the gas-phase
abundances of the QBCDs are $\sim0.35$ dex more metal rich than BCDs on
average, which makes them an unlikely population to transform
simply into BCDs with the addition of a starburst. However, the less
likely scenario of pristine gas infall might be able to lower their
metallicities and trigger a starburst to bring them out of
quiescence.

Next we consider the impact of the duty cycle of star formation in
BCDs, as today's BCDs cannot sustain their starbursts 
indefinitely. Lee \etal (2009) used a volume-limited survey of 261
nearby galaxies (within 11 Mpc) to determine that only $\sim 6$\%
of dwarf galaxies are currently experiencing global star formation
events. In fact, two of our BCDs are nearby enough to be inside the
Lee \etal (2009) volume (Mk 36 \& Mk 475), and are among the three
largest H$\alpha$ equivalent width systems in their entire
sample. Even if BCDs have a significantly greater duty cycle of star 
formation than normal dwarf galaxies, there should be many more
non-bursting BCDs than bursting BCDs at any given time. Depending on
what triggers the bursts in the BCDs, the pre-burst and post-burst
BCDs may have observable signatures, as well. For example, a
post-burst BCD would have a unique stellar population for a while as
its starburst population becomes intermediate aged. These starburst
events themselves may also last longer than is commonly
assumed. McQuinn \etal 2010 studied twenty dwarf galaxies and found
that most of the starburst durations were between 450-650 Myrs,
considerably longer than the lifetimes of O and B stars. Still, we
expect the underlying host galaxy structure to change on a slower time
scale than that of the starburst, so we could look for galaxies
with similarly compact underlying old stellar populations but without
current significant star formation (e.g., the types of galaxies listed
in Table \ref{similar}). In principle, surveying for non-bursting
dwarf galaxies with compact structural parameters would allow for the
determination of the BCD duty cycle. However, carrying out such a
study in practice would be very difficult, as it would require
detailed observations of a large volume-limited sample.


In addition to classifications of dwarf galaxies based on the
structural parameters of their underlying host galaxies, we briefly
consider 
what can be inferred about their evolutionary status by studying their
stellar populations. The detailed SED-modeling of the stellar
populations and star formation histories of this sample of BCDs will
be explored in a subsequent paper, but for now we consider
the color gradients found in the BCDs and shown in Figure
\ref{BVBH}. We find agreement with other groups that the outer regions
of BCDs indeed show colors consistent with evolved stellar
populations, but it is revealing to also see how the stellar
population changes radially across the galaxy. Marlowe \etal 1999
produced similar plots (their Fig. 7) which showed that the cores of
``Blue Amorphous Galaxies'' typically have 
colors consistent with ages of $\sim 10^8$ years while the outer
regions of their underlying host galaxies have colors consistent with
ages of $\sim 10^{10}$ 
years. Comparing an individual galaxy's age range (or, analogously,
color gradient) from its center to its outskirts may give insight not
only about its particular star formation history (SFH) but also about
its preferred modes of star formation. These details will be discussed
in the context of detailed SFH and stellar mass estimates from our
subsequent SED analysis (Janowiecki \etal in preparation).

\section{Acknowledgments}

We are grateful to S. Salim, L. van Zee, and H. Evans for useful
feedback on this work. S.J. also thanks J. Hargis for many useful
conversations during the course of this work.

Funding for SDSS-III has been provided by the Alfred P. Sloan
Foundation, the Participating Institutions, the National Science
Foundation, and the U.S. Department of Energy Office of Science. The
SDSS-III web site is http://www.sdss3.org/.

SDSS-III is managed by the Astrophysical Research Consortium for the
Participating Institutions of the SDSS-III Collaboration including the
University of Arizona, the Brazilian Participation Group, Brookhaven
National Laboratory, University of Cambridge, Carnegie Mellon
University, University of Florida, the French Participation Group, the
German Participation Group, Harvard University, the Instituto de
Astrofisica de Canarias, the Michigan State/Notre Dame/JINA
Participation Group, Johns Hopkins University, Lawrence Berkeley
National Laboratory, Max Planck Institute for Astrophysics, Max Planck
Institute for Extraterrestrial Physics, New Mexico State University,
New York University, Ohio State University, Pennsylvania State
University, University of Portsmouth, Princeton University, the
Spanish Participation Group, University of Tokyo, University of Utah,
Vanderbilt University, University of Virginia, University of
Washington, and Yale University. 

This research has made extensive use of NASA's Astrophysics Data
System Bibliographic Services. This research has also made use of the
NASA/IPAC Extragalactic Database (NED), which is operated by the Jet
Propulsion Laboratory, California Institute of Technology, under
contract with the National Aeronautics and Space Administration.

This publication makes use of data products from the Wide-field
Infrared Survey Explorer, which is a joint project of the University
of California, Los Angeles, and the Jet Propulsion
Laboratory/California Institute of Technology, and NEOWISE, which is a
project of the Jet Propulsion Laboratory/California Institute of
Technology. WISE and NEOWISE are funded by the National Aeronautics
and Space Administration.

{\it Facility:} \facility{WIYN}

\section{Appendix A: WHIRC reduction}

Observing extended sources against the high sky background in ground
based near-infrared imaging is a great challenge. Near-infrared point
source observations are 
somewhat more straightforward, but surface photometry requires
particular efforts. However, since our target galaxies ($\sim$$1'$) are
smaller than the full WHIRC field of view ($\sim$$3'$), we can still
employ relatively simply observing techniques with plenty of
well-sampled sky around our targets. This appendix fully describes
our reduction process, which is generalizable to any ground-based
near-infrared observations of extended sources smaller than the
detector field of view. It also describes the process of removing some
particular artifacts and features of WHIRC. Much of the process comes
from Dick Joyce's invaluable manual
(\texttt{www.noao.edu/kpno/manuals/whirc/}) and the reduction guide to 
NEWFIRM (\texttt{www.noao.edu/staff/med/newfirm/}), and from a very
useful conversation with Janice Lee. 

The WHIRC team provides a script (\textsc{wprep}) that carries out
necessary steps on every image from an observing run. In our case, it
first applies a linearity correction, since WHIRC and most NIR detectors
experience modest gain changes as the pixel wells fill up ($\sim$$4\%$
drop at half well capacity). Next, it applies a rough World Coordinate
System (WCS) and trims unnecessary reference columns from one side of
the images.

Standard flat field images are not useful on NIR imagers because of
the high background from the instrument. Instead, it was recommended
to take images of the flat field screen with the flat field lamps both
on and off. We took 10 lamps-on and 10
lamps-off in each filter. By subtracting the (average-combined)
lamps-off flat from the lamps-on flat in each filter, a true
sensitivity correction is obtained. However, we found that our dark
sky flats (discussed later) were a much better sensitivity correction
to our observations. We did, however, use the lamps-on and lamps-off
dome flats to create a bad pixel mask (BPM) with the
\textsc{ccdmask}. Sets of dark exposures were also taken at each
exposure time of our science frames, and are median-combined. These
darks are necessary to create the dark sky flats.

\begin{deluxetable*}{cccl}
\tablewidth{5in}
\tablecaption{Photometric Calibrations in JHK \label{JHK}}
\tablehead{\colhead{Term} & \colhead{value} & \colhead{error} & \colhead{Description}}
\startdata
$J$    & - & - & calibrated $J$ magnitude \\
$H$    & - & - & calibrated $H$ magnitude \\
$K$    & - & - & calibrated $K$ magnitude \\
\hline
$m_J$  & - & - & observed instrumental magnitude in $J$ filter \\
$m_H$  & - & - & observed instrumental magnitude in $H$ filter \\
$m_K$  & - & - & observed instrumental magnitude in $K$ filter \\
\hline
$X$    & - & - & airmass of observation \\
\hline
$k_J$  & 0.061 & - & standard KPNO airmass extinction coefficient in $J$ \\
$k_J$  & 0.031 & - & standard KPNO airmass extinction coefficient in $H$ \\
$k_J$  & 0.024 & - & standard KPNO airmass extinction coefficient in $K$ \\
\hline
$\xi_{J}$  & 23.124 & 0.032 & best-fit photometric zero-point in $J$ \\
$\xi_{JH}$ & -0.036 & 0.036 & best-fit photometric zero-point in $J-H$ \\
$\xi_{JK}$ &  0.586 & 0.032 & best-fit photometric zero-point in $J-K$ \\
\hline
$\epsilon_{J}$ & 0.0238 & 0.0672 & best-fit photometric color term in $J$ equation \\
$\mu_{JH}$     & 0.9590 & 0.0658 & best-fit photometric color term in $J-H$ equation \\
$\mu_{JK}$     & 1.0653 & 0.0552 & best-fit photometric color term in $J-K$ equation \\
\enddata
\end{deluxetable*}

One of the major customizations to our reduction process facilitates
the removal of an intermittent dark ``palm-print'' feature on the
first image of most dither sequences. Our observing program used
4-point dither patterns, and in many cases the first pointing had the
dark ``palm-print'' feature imposed on it. This feature was later
determined to be related to the temperature change of the detector
when it transitions from idle mode (and is continuously reading out)
to exposing mode (no longer reading out, but integrating), and is a
short-term transient feature that appeared only in the first dither
points. Our iterative sky subtraction is able to successfully model
and remove it from the affected images.

As a result of this complication, our sky subtraction method is a
three step iterative process, involving three separate sky
subtractions. The first sky subtraction is the crudest, involving a
single median combination (using only the central 1000x1000 pixels) and
subtraction for each set of observations in
a particular filter of a particular target. We also generated crude
dark sky flats by median-combining all of the science observations
from a particular exposure time and filter in a given night,
subtracting the appropriate dark frame, and used those rough
dark sky flats to flatten the images. This first
sky-subtraction is used to identify which images are affected by the
dark ``palm-print'' feature, and get a rough look at the data, as a
raw un-sky-subtracted image is very difficult to interpret.

Once the images affected by the ``palm-print'' are identified, we
begin the second-pass sky subtraction. We use the same method of
median combining and subtracting a single sky image for each filter
and target, and a simple dark sky flattening for each filter, but this
time the images affected by the ``palm-print'' are left out
of the median combinations. We do still subtract the median-sky images
from the affected images, though, and are able to assess the
consistency and stability of the dark feature, as well as the
stability of the sky throughout each dither sequence.

Next we want to combine all of these sky subtracted images for each
target. We tried
using NEWFIRM's \textsc{nfwcs} task at this point to refine and
improve the WCS on the images, but WHIRC's small field of view and the
scarcity of 2MASS stars in the frame made it fail on most
images. Instead, we used the USNO-B2 catalog (Monet \etal 2003) to
check and 
improve the WCS on all of the images. Once the WCS was suitable, we
used \textsc{mscimage} to re-project all of the images for a particular
target and filter onto the same pixel scale so that they were
aligned. We applied this same re-projection to the original BPMs as
well, and used them on the final images. These reprojected
sky-subtracted images are then median-combined (using the sky level
from the central region as the zero-level in \textsc{imcombine}) to
create our first stacked images of each target.

The added depth of these images allows us to detect and mask objects
too faint to be seen in individual frames, to further improve the sky
subtraction. We used the \textsc{acesegment} task to automatically
detect and generate a mask for all of the objects in the combined
image. These masks are then reprojected using \textsc{mscimage} back
to each of the original images from that dither sequence. Using these
detailed masks we generated improved dark sky flats for each filter.

Finally, we begin the final-pass sky subtraction using these detailed
masks on every raw image. This final sky subtraction is more
sophisticated than the previous sky subtractions since we create a
custom sky image for each individual science image. In most cases we
created the sky image from a masked
weighted median combination of the (usually) 7 nearest (most
contemporaneous?) images of the same dither sequence. A combination of
the detailed object masks and the bad pixel mask were used on each of
the raw images, and the images with
the ``palm-print'' are excluded from the sky image creation. We used
the second-pass sky subtraction to assess sky stability and determine
whether 5, 7, or 9 images should be used in the median combine. We
also weighted each image in the median combination by $1/\sqrt{n+1}$
where $n$ is the number of images separating the target image from the
sky image in the dither sequence. This weighting method increases the
contribution from nearby images and decreases the contribution from
more distant images. After
subtracting the sky image from each science frame, we flatten the
images with the improved dark sky flats created using the detailed
masks.

Now we must remove the dark ``palm-print'' feature from the affected
first-images. The sky subtraction on these images is satisfactory,
since only the adjacent non-affected images were used to create the
sky images. We median combine all sky-subtracted images in a
particular filter that contain the dark feature in order to make a
template model of the feature. This template is scaled to precisely
match each affected image, by matching median levels in selected regions both
on and off the feature. After the template has been scaled to the
individual image, the template is subtracted from the image, and the
dark ``palm-print'' feature is removed.

Before the final stacking of dither sequences, we also checked the
photometric stability throughout the night by measuring the
brightnesses of 2MASS
stars in each individual frame. We used this photometry to check for
absolute zeropoint variations throughout the night and also be sure
that the photometry was stable throughout individual dither
sequences. If 
standard stars are being observed to calibrate the observations, this
photometric verification is useful. When we did experience
non-photometric conditions, the 2MASS stars in the final stacked image
were used to determine a post-calibration zeropoint.

Just before doing the final image stacking, we checked that all of the
images have excellent WCS and re-projected them to be mutually
aligned. Additionally, we calculate the original sky level of each
image (before any sky subtraction was applied) and store it as a
header keyword in order to later restore the Poisson image
statistics. We use \textsc{imcombine} to average and ccdclip the
images, using both the median of the central 1000x1000 pixels as the
zeropoint (which was usually very close to zero) and the offsets from
the WCS. After this image is combined, we add a constant value to it
in order to make the final sky level the same as the average
throughout the sequence.

At this point, the images have been fully reduced and are ready to be
measured. The combination of the iterative sky subtraction and object
masking (and the dark ``palm-print'' feature removal) makes these
reduced images suitable for surface photometry. An extensive detailed
description of the parameters used for each task is available upon
request, but it was too detailed to present in this appendix.

In addition to the per-image calibrations from 2MASS, we calibrated
many of our observations using JHK standard stars from Hunt \etal
(1998). Presented in the table below is a sample of our derived
zero-points and color terms, and the values we used for airmass
extinction. These values come from observations of 7 standard stars
in JHK filters on the night of November 5th 2008. The full set of
photometric equations for calibrated $J$ magnitudes, and $J-H$ and
$J-K$ colors are given 
below. Descriptions and values of the adopted and best-fit terms are
given in Table \ref{JHK}.

\vspace{0.5cm}

$J = m_J - X k_J + \epsilon_J (m_J - m_H) + \xi_J$

\vspace{0.5cm}

$J-H = (m_J - X k_J) - (m_H - X k_H) + \mu_{JH} (m_J - m_H) + \xi_{JH}$

\vspace{0.5cm}

$J-K = (m_J - X k_J) - (m_K - X k_K) + \mu_{JK} (m_J - m_K) + \xi_{JK}$

\vspace{0.5cm}

Sample images and detailed descriptions of observing and reducing
WHIRC data are given on the website
(\texttt{www.noao.edu/kpno/manuals/whirc/WHIRC.html}) and were
invaluable in our data reduction process.

\end{document}